\documentclass{aa}
\usepackage{natbib}
\usepackage{graphicx}
\usepackage{multicol}
\usepackage{subcaption}
\usepackage{amsmath}	
\usepackage{txfonts}

\newcommand\HI{$\textrm{H}\scriptstyle\mathrm{I}$}

\newcommand{\ve}[1]{\mathbf{q1}}

\newcommand{\f}{\frac}

\newcommand{\be}{\begin{equation}}      
\newcommand{\ee}{\end{equation}}      
      
\newcommand{\bef}{\begin{figure}}      
\newcommand{\eef}{\end{figure}}      
\newcommand{\bea}{\begin{eqnarray}}    
\newcommand{\eea}{\end{eqnarray}}

\newcommand{\av}[1]{\ensuremath{\left\langle q1 \right\rangle}}

\newcommand{\tve}[1]{\tilde{\boldsymbol{q1}}}

\def\bse{\begin{subequations}}
\def\ese{\end{subequations}}

\def\lsim{\raise 0.4ex\hbox{$<$}\kern -0.8em\lower 0.62ex\hbox{$\sim$}} 
\def\gsim{\raise 0.4ex\hbox{$>$}\kern -0.7em\lower 0.62ex\hbox{$\sim$}}

\def\f0N{f_0^{(N)}}
\def\bec{\begin{center}}
\def\eec{\end{center}}

\begin{document}
\titlerunning{ Dark Matter Disc Model in Dwarf Galaxies}
\authorrunning{Sylos Labini et al}

   \title{Exploring the Dark Matter Disc Model in Dwarf Galaxies: Insights from the LITTLE THINGS Sample}


   \author{Francesco Sylos Labini
          \inst{1,2}\fnmsep\thanks{sylos@cref.it}
          \and
          Roberto Capuzzo-Dolcetta \inst{3}
          \and 
          Giordano De Marzo \inst{1,3,4,5}
          \and 
          Matteo Straccamore \inst{1,6} 
          }

   \institute{Centro Ricerche Enrico Fermi,  I-00184, Roma, Italia
         \and
            INFN Unit\`{a} Roma 1, Dipartimento di Fisica, Universit\`{a} di  Roma Sapienza, I-00185 Roma, Italia
         \and
Dipartimento di Fisica, Sapienza, Universit\`{a} di  Roma,  I-00185, Roma, Italia
         \and
Complexity Science Hub Vienna, Josefstaedter Strasse 39, 1080, Vienna, Austria
         \and
University of Konstanz, Universit\"{a}tstra$\beta$e 10, 78457, Konstanz, Germany
         \and
         Sony CSL - Rome, Joint Initiative CREF-SONY,  00184, Roma, Italia
             }

   \date{Received xxxx; accepted yyyy}

 
  \abstract
   {We conducted an analysis of the velocity field of dwarf galaxies in the LITTLE THINGS sample, focusing on deriving 2D velocity maps that encompass both the transverse and radial velocity fields. 
Within the range of radial distances where velocity anisotropies are sufficiently small for the disc to be considered rotationally supported, and where the warped geometry of the disc can be neglected, we reconstructed the rotation curve while taking into account the effect of the asymmetric drift.
To fit the rotation curves, we employed the standard halo model and the dark matter disc (DMD) model, which assumes that dark matter is primarily confined to the galactic discs and can be traced by the distribution of \HI{}. Interestingly, our analysis revealed that the fits from the DMD model are statistically comparable to those obtained using the standard halo model, but the inferred masses of the galaxies in the DMD model are approximately 10 to 100 times smaller than the masses inferred in the standard halo model.
In the DMD model, the inner slope of the rotation curve is directly related to   a linear combination of the surface density profiles of the stellar and gas components, which generally exhibit a flat core. Consequently, the observation of a linear relationship between the rotation curve and the radius in the disc central regions is consistent with the framework of the DMD model.
}
\maketitle
%

\section{Introduction}
In recent decades, the study of neutral hydrogen (\HI{}) using high-resolution radio interferometric surveys has greatly advanced our understanding of the kinematic properties of galaxies. Two notable surveys, The \HI{} Nearby Galaxy Survey (THINGS) \citep{Walter_etal_2008}  and the Local Irregulars That Trace Luminosity Extremes, THINGS (LITTLE THINGS) \citep{Hunter_etal_2012} survey, have been instrumental in this field. These surveys have provided us with the means to measure high-resolution \HI{}\;  rotation curves, which are crucial for unraveling the mass distributions of galaxies.

The THINGS sample of galaxies was carefully selected to encompass a wide range of physical properties, spanning from low-mass, metal-poor, quiescent dwarf galaxies to massive spiral galaxies. On the other hand, the LITTLE THINGS sample primarily consists of late-type dwarf galaxies. These dwarf galaxies show a relatively simple structure, lacking prominent bulge and spiral components. Consequently, they offer a valuable opportunity to study the distribution and quantity of dark matter (DM) since this mass component significantly influences their dynamics. The \HI{} observations from these surveys have provided substantial observational constraints on the central distribution of DM in galaxies.

Motivated by cosmological simulations, the assumption commonly adopted  is that 
disc galaxies are quasi-flattened systems primarily governed by rotation surrounded by nearly spherical halos with an almost isotropic velocity dispersion.  Various halo models have been employed to interpret the observed kinematic properties of these galaxies.  The standard one is the Navarro-Frenk-White (NFW) model \citep{Navarro_etal_1996} that has been found to satisfactorily describe non-linear structures emerging from cosmological Lambda cold dark matter ($\Lambda$CDM) simulations. The NFW model is  commonly referred to as a \textit{cusp-like} halo model, as it assumes a density power-law behavior, $\rho(r) \sim r^{\alpha}$ with $\alpha \approx -1$, in the central region of the halos \citep{Moore_1994,Navarro_etal_1996,Navarro_etal_1997,Navarro_etal_2004,Navarro_etal_2010,Moore_etal_1999,Power_etal_2003,Diemand_etal_2008,Stadel_etal_2009,Ishiyama_etal_2013}. 
{  A \textit{core-like} halo model with an almost constant density $\rho(r) \sim \rho_0$ for small values of $r$, has also been considered in the analysis of the observed rotation curves (see, e.g., \cite{Begeman_etal_1991,Oh_etal_2015,Mancera_Pina_etal_2022} ).}

The understanding of the behavior of the density distribution in the central regions of dwarf galaxies depends on the ability to infer different distributions of DM by studying their  rotation curves.  Actually, in many cases these rotation curves exhibit a linear increase in the circular velocity towards their centers, i.e. $v_c\propto r$ for small values of $r$, indicating the presence of a flat core in the mass density \citep{Moore_1994,deBlok_etal_1996,deBlok_etal_2001,deBlok+Bosma_2002,Oh_etal_2008,Oh_etal_2011}.This clear discrepancy between the central DM distribution within 1 kpc in galaxies observed in $\Lambda$CDM simulations and observations, known as the "cusp/core" problem, has been one aspect contributing to the small-scale crisis in $\Lambda$CDM cosmology (see, e.g., \cite{Kroupa_2012,Pawlowski+Kroupa_2013,Dabringhausen+Kroupa_2103}).
{   Indeed, the presence of cusps is a fundamental feature of current cosmological simulations, confirmed repeatedly by significant increases in resolution and computing power over the years. The origin of the cusp is tied to the physical processes underlying the formation of non-linear structures. The density profile of halo structures, characterized by a cusp in the inner regions  ($\sim r^{-1}$) and a slower decay in the outer regions ($\sim r^{-3}$),  emerges from a bottom-up aggregation process. In this scenario, non-linear structures form hierarchically through a gradual dynamical process.
In contrast, when non-linear structures form via a top-down, monolithic collapse, the inner region tends to exhibit a nearly flat core \citep{Benhaiem+SylosLabini+Joyce_2019,SylosLabini_etal_2020}. 

Thus, in principle, the distinction between a cored profile and one with a cusp is crucial, as it may indicate fundamentally different underlying dynamical evolution processes. However, this distinction becomes less clear when additional physical effects are considered in simulations. For instance, \cite{Read_etal_2016} demonstrated that stellar feedback can heat  DM, resulting in a "coreNFW" density profile. They concluded that standard cosmological simulations, when accounting for such feedback, provide an excellent match to the rotation curves of dwarf galaxies. Similarly, \cite{Lazar_etal_2020} showed that, by incorporating galaxy formation physics, DM haloes would exhibit density profiles that are both cored and cuspy. For this reason they  introduced the core-Einasto profile, a three-parameter density model, to better characterize DM haloes. 

Such cored profiles have been used to fit galaxy rotation curves: for instance, \cite{Ou_etal_2024} fitted the Milky Way's rotation curve using two different mass models for the DM halo: a generalized NFW profile and an Einasto profile \citep{Einasto_1965,Retana-Montenegro_etal_2012}. Both models include an additional free parameter compared to the standard NFW profile, making them more suitable for fitting a declining rotation curve as it is the one of our own Galaxy \citep{Eilers_2019,Wang_etal_2023,Ou_etal_2024}.}

{  To measure the rotation curves of dwarf galaxies at small radii, \citet{Oh_etal_2008,Oh_etal_2011} used high-resolution dark matter (DM) density profiles of seven dwarf galaxies from the THINGS survey, supplemented with data from the Spitzer Infrared Nearby Galaxies Survey (SINGS) by \citet{Kennicutt_etal_2003}}. These studies showed that the average inner density slope  for these galaxies is approximately  $\alpha = -0.29 \pm 0.07$ which is consistent with the value $\alpha= -0.2 \pm 0.2$ derived earlier for a larger number of Low Surface Brightness (LSB) galaxies by \cite{deBlok+Bosma_2002}. This significant observational data set provides a strong empirical evidence in support of the existence of a core-like distribution of DM around the centers of dwarf irregular  galaxies.

To overcome the limitation in the number of galaxies probed by the THINGS survey, \cite{Oh_etal_2015} expanded the investigation by including a larger sample of dwarf galaxies. This was done by the use of data from the LITTLE THINGS survey \citep{Hunter_etal_2012}, which contains 41 nearby gas-rich dwarf galaxies using the NRAO26 very large array. The LITTLE THINGS observations have a fine angular resolution of approximately 6" as well as a fine velocity resolution of less than or equal to 2.6 km s$^{-1}$. In the catalog, these high-resolution observations were complemented with data in various other wavelength bands, including H$\alpha$, optical (U, B, V), near-infrared, archival Spitzer infrared, and GALEX ultraviolet images \citep{Hunter+Elmegreen_2006}. Follow-up observations were also conducted using ALMA and Herschel.

The availability of these high-quality, multi-wavelength, datasets significantly reduced observational uncertainties, which could have concealed central cusps in low-resolution data. From the initial sample of 41 galaxies, \cite{Oh_etal_2015} selected a subset of 26 dwarf galaxies (three of which were also part of THINGS) that exhibited regular rotation patterns in their velocity fields. They measured the rotation curves using the rotating Tilted Ring Model (TRM) \citep{Warner_etal_1973,Rogstad_etal_1974} distinguishing the contribution from baryons and from DM. This decomposition allowed them to estimate the relative contributions to the total mass budget.

The majority of galaxies in the LITTLE THINGS sample display rotation curves linearly increasing with radius in their inner regions.  The average value of the density slopes for these 26 LITTLE THINGS dwarf galaxies was found to be $\alpha  = -0.32 \pm 0.24$, which is consistent with previous findings for low surface brightness galaxies and for the 7 THINGS dwarf galaxies. This result significantly deviates from the predicted cusp-like DM distribution
($\alpha \simeq -1$) indicating a different mass distribution than a cuspy one. 

{  Similar results have been obtained also by 
\cite{Iorio_etal_2017} who analyzed a subsample of 17 galaxies from the LITTLE THINGS survey. They directly fit 3D models (two spatial dimensions and one spectral dimension) using the software $^\text{3D}$BAROLO \citep{DiTeordoro_Fraternali_2015}, a code based on the TRM approach that performs full 3D numerical minimization on the entire data cube. A key advantage of this approach is that the data cube models are convolved with the instrumental response, ensuring that the final results are not affected by beam smearing. The resulting rotation curves are consistent with those of \cite{Oh_etal_2015} within the errors for half of the sample galaxies. The differences observed are primarily due to variations in the best-fitting orientation angles and/or discrepancies in the asymmetric-drift correction terms.
}

{  More recently,  \cite{Mancera_Pina_etal_2022} investigated how significantly mass models are affected when the possibile flared geometry of gas discs is taken into account. Accurate computations of scale heights require detailed kinematic modeling of interferometric \HI{} and CO data, along with robust bulge-disc decomposition. They found that the effect of gas flaring on rotation curve decomposition may play a role only in the smallest, gas-dominated dwarf galaxies. For most galaxies, however, the effect is minor and can be safely ignored.
} 


In this study, we examine the LITTLE THINGS survey to estimate the mass of galaxies by employing a disc-based model for the distribution of dark matter (DM), in contrast to the standard assumption of a spherical halo DM distribution. This modeling strategy is inspired by evidence suggesting a correlation between DM and the distribution of \HI{} gas in disc galaxies \citep{Bosma_1978, Bosma_1981}. Previous investigations on nearby disc galaxies have revealed that the ratio of the total surface density of the disc (derived from rotation curve measurements) to the surface density of gas alone (obtained from \HI{} observational data) remains relatively constant beyond approximately one-third of the optical radius. This correlation between DM and gas in disc galaxies implies that rotation curves at larger radii may be scaled versions of those determined from the \HI{} distribution. Several studies have indicated that gas surface density can effectively represent the distribution of DM in galaxies. Authors such as \cite{Sancisi_1999, Hoekstra_etal_2001, Hessman+Ziebart_2011, Swaters_etal_2012} have explored the DM-\HI{} relationship. 
We recall that, within the framework of the Bosma effect, \citet{Pfenniger_etal_1994} proposed the idea that DM could reside in the galactic disc in the form of very cold molecular hydrogen (H$_2$). In the Milky Way, extensive clouds of cold dust (i.e., $T \approx 10-20$ K) and dark gas, invisible in \HI{} and CO but detected via $\gamma$-rays, were observed in the solar neighborhood by \citet{Grenier_etal_2005}. Dark gas was also found by the Planck mission \citep{Planck_2011,Casandjian_etal_2022}. However, H$_2$ in less excited and colder regions than those traced by CO has yet to be directly inferred as below 20 K, CO molecules freeze onto dust grains, preventing them from emitting rotational lines. Nonetheless, regions colder than 20 K are not uncommon, leaving the possibility of undetected H$_2$ in the outer gas discs of galaxies, where stellar radiation decreases rapidly. Unfortunately, the direct detection of such clouds in the outer discs of galaxies remains a challenging observational task \citep{Combes+Pfenniger_1997}. Recently, \cite{SylosLabini_etal_2024_Mass} discovered that the rotation curves of 16 nearby disc galaxies in the THINGS sample can be accurately described by both the NFW halo model and the DMD model with similar levels of precision. Furthermore, \cite{SylosLabini_etal_2023_MW} demonstrated that the rotation curve of the Milky Way can be effectively modeled using the DMD. While the presence of a correlation does not imply causation, its identification may be associated with a fundamental characteristic of disc galaxies, such as the distribution of DM within the disc itself, rather than in a spherical halo encompassing the galaxy, as commonly assumed but not definitively proven.  Therefore, we believe it is valuable to conduct a more comprehensive investigation of this phenomenon than has been previously undertaken, by extending the findings of \cite{SylosLabini_etal_2024_Mass} regarding the galaxies in the THINGS sample to the dwarf galaxies in the LITTLE THINGS sample.

In order to fit the DMD properly, we will create our own estimation of the rotation curves for the galaxies in the LITTLE THINGS sample. For this purpose, we will utilize the velocity ring model (VRM), as recently introduced by \cite{SylosLabini_etal_2023_VRM}. This method operates under the assumption that the galactic disc is flat, meaning it is not warped, and facilitates the reconstruction of coarse-grained 2D velocity maps that cover both the transversal and radial velocity fields. The assumption of a flat disc generally holds true in the inner regions of galactic discs which are the focus of our study. Indeed, warps have been directly observed in the outer regions of edge-on disc galaxies \citep{Sancisi_1976, Reshetnikov+Combes_1998, Schwarzkopf+Dettmar_2001, Garcia-Ruiz_etal_2002, Sanchez-Saavedra_etal_2003}, typically emerge starting at the optical radius. Therefore, these maps enable the examination of spatial anisotropies in the 2D velocity field within the inner disc and allow for the quantification of velocity anisotropies. 

{  
It is important to stress that  the TRM and the VRM are based on two distinct assumptions. The TRM, in its simplest implementation, assumes the radial velocity to be zero, i.e., $v_R=0$,  while allowing for a warped disc geometry, whereas the VRM assumes a flat disc and permits $v_R \ne 0$. We have adopted the VRM in this study because it enables the reconstruction of coarse-grained 2D velocity component maps, which, in turn, allow for the characterization of spatial anisotropies. This approach is particularly suitable for the dwarf galaxies in the LITTLE THINGS sample, as they exhibit anisotropic 2D line-of-sight velocity maps that are difficult to reconcile with a symmetric geometric deformation of the disk, such as a warp. In addition, this choice is justified a-posteriori by the observation that the results for the transverse velocity obtained with the VRM are very similar to the rotation velocity derived using the TRM. Additionally, the warps in these galaxies are generally small, as indicated by the limited variation in orientation angles. This is due to the fact that dwarf galaxies, which make up our sample, have relatively limited spatial extents. Actually, warps are typically originated by tidal interactions with neighboring galaxies, and so are more commonly found in the outermost regions of large disk galaxies.}

The paper is structured as follows:
In Section \ref{sect:sample}, we provide an overview of the main characteristics of the galaxies included in the LITTLE THINGS survey, which form the focus of our study.
Section \ref{sect:methods} introduces the methods used to estimate the rotation curves of our sample galaxies and to correct for the effects of asymmetric drift.
In Section \ref{sect:massmodels}, we discuss the DMD mass model employed in our analysis.
Section \ref{sect:results_kinematics} presents and discusses the results for specific galaxies in our sample, including their rotation curves and other relevant findings. Results for the remaining galaxies are provided in the Appendix.
{  In Section \ref{sect:results_mass_models}, we analyze the results of the fits using the two mass models considered in this study.}
Finally, Section \ref{sect:conclusion} summarizes the main conclusions derived from our analysis.

\section{The sample}
\label{sect:sample}
The LITTLE THINGS  survey comprises 37 dwarf irregular and 4 blue compact dwarf galaxies known to contain  \HI{}{}. 
{  These are nearby galaxies with distances ranging from 1 Mpc to 10 Mpc, as listed in Table 1 of \cite{Hunter_etal_2012},   exhibiting a wide range of properties.} More information about the galaxies included in the LITTLE THINGS sample can be found in \cite{Hunter_etal_2012,Oh_etal_2015}. 
{  Note that for the inclination angle, we have adopted the mean value reported by \cite{Oh_etal_2015}. The determination of the inclination angle by \cite{Iorio_etal_2017}  not always 
agree with that by \cite{Iorio_etal_2017}: we will comment in more details on this issue when discussing individual galaxies. 
On the other hand, the position angle (P.A.) is assumed to be constant and is therefore irrelevant for our analysis
(see below and \cite{SylosLabini_etal_2023_VRM} for more details). }

Despite the presence of significant velocity anisotropies in the outer regions of the discs (see discussion in Sect.\ref{sect:results_kinematics}), the sample galaxies display an approximately regular rotation pattern in their 2D \HI{}{} velocity fields. Consequently, reliable rotation curves can be derived for the inner regions, but a thorough investigation of the outer regions is required. {  The LITTLE THINGS \HI{}{} data have a linear resolution of approximately $6''$, corresponding to the 25 to 200 pc range for the sample galaxies, with an average of 100 pc.} This resolution is sufficient to resolve the inner 1 kpc region of the galaxies. Furthermore, the high-resolution \HI{}{} data, with a velocity resolution of less than or equal to 2.6 km/s, enables accurate determination of the underlying kinematics of the sample galaxies.

The LITTLE THINGS dataset also includes a wealth of ancillary data in multiple wavelengths, which prove valuable for investigating star formation, interstellar medium, dust, molecular clouds, and other aspects \cite{Hunter_etal_2012}. In particular, the separation of the baryonic contribution from the total rotation curve is achieved by utilizing Spitzer archival IRAC 3.6$\mu$m data and supplementary optical color information \cite{Hunter+Elmegreen_2006}. These images are less susceptible to dust effects compared to shorter wavelength maps and trace the old stellar populations that constitute the predominant fraction of the stellar mass in galaxies \citep{Walter_etal_2007,Oh_etal_2015}.

For this study, we utilize a sub-sample of high-resolution \HI{}{} data from LITTLE THINGS, consisting of 22 nearby dwarf galaxies, the same sub-sample examined by \cite{Oh_etal_2015} but  four galaxies Haro29, Haro36, UGC8508 and NGC1569 which present a too irregular line-of-sight (LOS) velocity ($v_{los}$) map. Each galaxy's stellar and  \HI{}{} mass are reported in Tab.\ref{table1}.

{  
The mass of the gaseous components in galaxies can be reliably measured from \HI{}{} observations without requiring questionable assumptions, whereas the stellar mass depends on the specific hypothesis adopted. However, since the majority of the baryons in this sample are estimated to be in the form of gaseous components, the final value of the baryonic mass remains robust.
\cite{Oh_etal_2015} calculated the mass of the gaseous component using total integrated \HI{}{} intensity maps (moment 0), corrected for the warp as determined by tilted-ring models. 
The obtained value of the mass was then scaled up by a factor of 1.4 to account for helium and metals. The contribution of molecular hydrogen (H\textsubscript{2}) was neglected, as the low metallicities in dwarf galaxies typically result in only a small fraction of the gaseous component being in the form of H\textsubscript{2}.
Similarly, \cite{Hunter_etal_2012} measured the total \HI{}{} mass by summing the flux over the primary-beam-corrected integrated \HI{}{} map. 
The relative difference between \cite{Hunter_etal_2012}  measures and those of \cite{Oh_etal_2015} is on average 20\%  (similar differences are found with the values of \cite{Iorio_etal_2017}).  These differences are mainly due to the
different methods used to estimate the total mass and we take the value of this difference as the error on the gas mass.

To derive mass models of the stellar components of the galaxies, \cite{Oh_etal_2015} utilized {\it Spitzer} IRAC 3.6 $\mu$m images, which trace the old stellar populations that dominate late-type dwarf galaxies. 
The surface brightness profiles of the stellar components of the galaxies were recovered by applying the derived tilted-ring parameters to the {\it Spitzer} IRAC 3.6 $\mu$m images.

To estimate the stellar mass in galaxies, it is necessary to assume a specific value for the stellar mass-to-light ratio, $\Upsilon_*$, which typically introduces the greatest uncertainty when converting the luminosity profile into the mass density profile.
 \cite{Oh_etal_2015} employed an empirical relation between galaxy optical colors and $\Upsilon_*$ values in the 3.6 $\mu$m band, derived from stellar population synthesis models.
In addition, they report the stellar masses of the sample galaxies, which were derived using a spectral energy distribution fitting technique. The difference between these estimations is, on average, 20\%, and we refer to this as the error in the stellar mass estimation.

Values of the gaseous and stellar mass components and their error are reported in Tab.\ref{table1}.
}


\begin{table*}
\begin{center}
\begin{tabular}{ | c | c | c | c | c | c | c | c | c | c | }
\hline
Name           & $M_{g} (10^7 M_\odot)$   & $M_{s} (10^7 M_\odot) $   & $M_{\text{bar}} (10^7 M_\odot)$  & $M_{\text{dmd}} (10^9 M_\odot)$ & $M_{\text{200}} (10^{10} M_\odot)$& $\frac{M_{\text{dmd}}}{M_{\text{bar}}}$ & $\frac{M_{\text{200}}}{M_{\text{bar}}}$   & $\chi^2_{dmd}$ & $\chi^2_{nfw}$      \\ 
(1)                &            (2)                          &               (3)                        &                   (4)                               &                         (5)                         &                              (6)                       &                                (7)                           &                                 (8)                           &            (9)          &(10)             \\ 
\hline
CVnIdwA     & $2.9    \pm 0.6    $             & $0.41    \pm 0.1$                & $3.3     \pm 0.7$                            & $0.24    \pm 0.06$                         &$0.11  \pm 0.02$                       &7.3         &  33                              &   0.3    & 0.2             \\  
DDO43        & $23     \pm 5       $             & ---                                       & $23      \pm 5   $                            & $1.8       \pm 0.4$                          &$0.6   \pm 0.1$                         &8.2         &  26                              &   1.0    & 0.3             \\  
DDO46        & $22     \pm 4       $             & ---                                       & $22       \pm 4  $                            & $7.6       \pm 1.6$                          & $2.2   \pm 0.4$                         &34          &  95                              &   2.2    &  1.4            \\  
DDO47        & $47     \pm 10     $             & ---                                       & $47       \pm 10 $                           & $7.6       \pm 1.5$                          & $27.0  \pm 5.0$                        &19          &  575                            &   2.4    &  0.5            \\  
DDO50        & $130   \pm 26     $             & $10    \pm 2.0$                  & $140     \pm 30 $                           & $1.8        \pm 0.7$                          & $0.2 \pm 0.1$                          & 1.3 	 &  1.4                             &   1.6    & 2.0        \\  
DDO52        & $33.4  \pm 6       $             & $7.2    \pm 1.4$                  & $41       \pm 7  $                            & $3.5       \pm 0.9$                          & $15.0  \pm 5$                          &11.3        & 360                             &   1.5    &  1.2       \\  
DDO53        & $7.0    \pm 1.4    $             & $1.0    \pm 0.4$                  & $8.0      \pm 1.8$                          & $1.0        \pm 0.2$                          & $2.0  \pm 1.0  $                       &13		 & 250                             &    0.6   &  0.5           \\  
DDO70        & $3.8    \pm 0.8    $             & $1.2    \pm 0.2$                  & $4.0      \pm 1.0$                          &  $0.34     \pm 0.07$                        & $1.7 \pm 0.6  $                       & 8.5         & 350                             &    2.3   & 1.6            \\  
DDO87        & $29.1  \pm 6       $             & $6.2    \pm 1.2$                  & $35.3    \pm 7.2$                          &  $2.1       \pm 0.5$                          & $4.7  \pm 2.0$                         & 6.2         & 133                             &   0.9  &  0.8       \\  
DDO101      & $3.5    \pm 0.7    $             & $5.8    \pm 1.2$                  & $9.3      \pm 1.9$                          &  $1.3       \pm 0.3$                          & $5.5  \pm 2.0$                         & 13.4         & 765                           &   1.4    &  2.5         \\  
DDO126      & $16.4  \pm 3.2    $             & $2.3    \pm 0.5$                  & $18.7    \pm 3.7$                          &  $ 1.1      \pm 0.2$                          & $3.7  \pm 1.0$                          &  5.8         & 195                           &    1.7   &   3.0         \\  
DDO133      & $13     \pm 2.6    $             & $2.6    \pm 0.5$                  & $15.5      \pm 3.1$                         &  $ 1.3      \pm 0.2$                          & $2.2  \pm 0.7   $                      &  8.1         & 142                           &   0.7  &   0.4     \\  
DDO154      & $13     \pm 2.6    $             & $3.5    \pm 0.7$                  & $16.5     \pm 3.3$                          &  $ 4.0     \pm 1.0$                          & $1.5  \pm 0.3$                         &  24          & 90                              &   1.8    & 1.5          \\  
DDO168      & $26     \pm 5       $             & $5.1    \pm 1.0$                  & $31        \pm 6.0$                          &  $ 1.6     \pm 0.4$                          & $2.0  \pm 0.7$                         & 5.2          & 65 $ $                        &  1.3     & 0.8      \\  
DDO210      & $0.14   \pm 0.03 $             & $0.04  \pm 0.01$                & $0.18     \pm 0.04$                        &  $ 0.14   \pm 0.04$                        & $0.04 \pm 0.01$                       & 80       & 170 $ $                       &   1.3   &  0.3  \\  
DDO216      & $0.49   \pm0.1    $             & $1.6    \pm 0.3$                  & $2.1      \pm 0.4$                           & $  0.14  \pm 0.03$                         & $0.1   \pm 0.04$                       &  6.5        &  38$ $                          &  1.2    & 0.5    \\  
F564-V3      & $4.4     \pm 0.8   $             & ---                                        & $4.4      \pm 0.8$                           &  $ 1.0    \pm 0.2$                          & $1.0   \pm 0.2$                         &  22.6       &  206                           &  2.0    & 0.7    \\  
IC10            & $1.7     \pm 0.34 $             & $12     \pm 2.4$                   & $13.7   \pm 3.0$                           & $  0.8    \pm 0.3$                            & $0.8   \pm 0.3$                         & 5.6          &  44$ $                         &   0.3   & 0.2     \\  
IC1613        & $5.9     \pm 1.2   $             & $1.9    \pm 0.4$                   & $7.9      \pm 1.6$                          & $ 0.21     \pm 0.04$                       & $0.9  \pm 0.2$                          &2.6           &  $114  $                      &   0.3  &  0.3      \\  
NGC2366   & $110     \pm 42    $             & $11     \pm 2.2$                   & $121     \pm 45 $                          &  $ 2.8       \pm 0. 6$                       & $1.0    \pm 0.2$                        &2.3          &    8.4                           & 1.0    &  1.0   \\  
NGC3738   & $13       \pm 2.6   $             & $13     \pm 2.6$                   & $26      \pm 5.0$                          &  $ 489       \pm 100$                        & $14.6 \pm 4$                          &20              & 30                             &  1.2   & 2.1    \\  
WLM           & $8.0     \pm 1.6   $              & $1.2    \pm 0.24$                 & $9.2      \pm1.8$                         &   $ 1.0     \pm 0.2$                           & $1.2    \pm 0.3$                       &10.5        &  $130.4  $                    &    1.2 &   0.5         \\  
\hline
\end{tabular}
\end{center}
\caption{  
\label{table1} 
The columns represent:
(1) Name of the galaxy;
(2) the gas mass  $M_{g}$   (see discussion in Sect.\ref{sect:sample}) 
(3) the stellar mass $M_{s}$   (see discussion in Sect.\ref{sect:sample}) 
(4) the total baryonic mass        $M_{\text{bar}}=M_{s}+M_{g}$;
(5) the DMD mass $M_{\text{dmd}}$ (see Eq.\ref {eq:massdmd});
(6) the NFW virial mass  $M_{\text{200}}$;  
(7) the ratio the DMD mass and the total to baryon  mass; 
(8) the ratio the NFW virial mass and the total to baryon  mass; 
(9) value of the $\chi^2_{dmd}$;
(10) value of the $\chi^2_{nfw}$ 
}
\end{table*}


\section{Methods}
\label{sect:methods}

First, we present the Velocity Ring Model (VRM) to illustrate its application in the study. The VRM is a method used to reconstruct the 2D velocity component fields by assuming a flat galactic disc. It allows for the analysis of the spatial distribution and amplitude of spatial velocity anisotropies in both radial and tangential directions.

Next, we discuss the estimation of asymmetric drift that is necessary to compute the {  circular} velocity from kinematic quantities, assuming the galaxy is in a steady state described by the Jeans equations.

\subsection{The velocity ring model}
\label{vrm}

Observations of the LOS component of emitter velocities in external galaxies allow for the determination of their kinematics, though certain assumptions must be made during the analysis. 
{  Indeed, the  Doppler effect only allows to determine a component of the velocity vector, $v_{los}$,  of each star and the reconstruction of the full vector is thus a challenging task. 
The simplest assumption made by the TRM (Tilted Ring Model) method is that radial velocities, $v_R$, are zero \citep{Warner_etal_1973, Rogstad_etal_1974}, while allowing for galactic discs whose geometry may be deformed by a warp. The projection of a disc galaxy onto the sky is characterized by at least two orientation angles: the inclination angle,  $i$, which defines the overall tilt of the disc relative to the observer, and the P.A., which is the angle between the galactic major axis and an arbitrary reference direction, such as the North Galactic Pole. For a warped disc, one or both of these angles vary with the radial distance from the galactic center instead of remaining constant.
This variation in orientation angles with radius is, at least partially, degenerate with the presence of non-circular motions, such as radial velocities. While it is possible within the TRM framework to consider more general cases where 
$v_R \ne 0$, methods based on the TRM often perform the initial fit under the assumption of $v_R=0$. This can erroneously lead to interpret real radial motions as spurious warps so that, in general, methods employing the TRM struggle to disentangle radial velocities from warp effects due to this degeneracy. Consequently, the TRM may underestimate the radial motion of the disc, as it attributes distortions of the kinematic major axis primarily to warps \citep{Schmidt_etal_2016, DiTeodoro+Peek_2021, Wang+Lilly_2023}. While radial flows have been detected using the TRM for over two decades \citep[e.g.,][]{Fraternali_etal_2001}, it remains uncertain whether their full extent has been accurately measured.%
}

{  As mentioned above, assumptions must be made to reconstruct the full velocity vector from a scalar quantity. Those underlying the TRM are not the only possible choice. }
Recently, the velocity ring model (VRM) was introduced with the aim of characterizing both the transversal and the radial velocity components \citep{SylosLabini_etal_2023_VRM}. This model shares the assumption of a flat galactic disc with other existing methods \citep{Barnes+Sellwood_2003,Spekkens+Sellwood_2007,Sellwood+Sanchez_2010,Sellwood_etal_2021}. However, the VRM offers the ability to characterize non-axisymmetric and heterogeneous velocity fields of any form. Within this framework, both radial and transverse velocity components can exhibit arbitrarily complex spatial anisotropic patterns. The VRM method is able to reconstruct coarse-grained two-dimensional maps, allowing for the investigation of spatial anisotropies in angular sectors at various distances from the galactic center. Let us briefly illustrate the VRM method. 

\begin{figure}
 \centering
\includegraphics[width=7.0cm,angle=0]{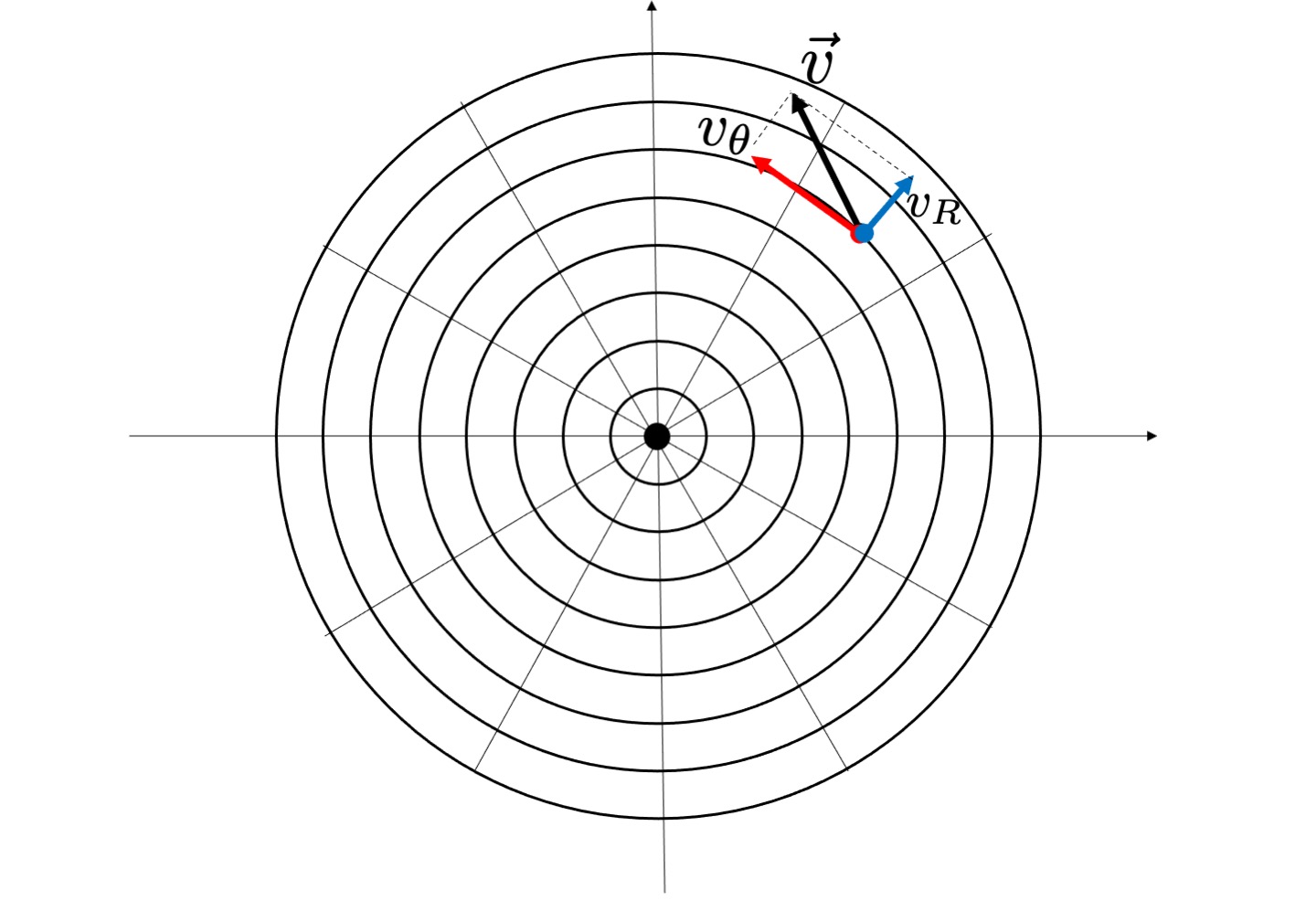}
\caption{  
 Illustrative representation of the VRM: the LOS velocity is decomposed, in the polar coordinates in the plane a galaxy, into a radial and transversal component. The projected image of a galaxy is divided in $N_r$ radial shells and $N_a$ arcs: in the example $N_r=8$ and $N_a=12$. 
   } 
\label{fig:VRM}
\end{figure}

\begin{figure}
\quad
\begin{subfigure}[a]{0.5\textwidth}
\centering
\includegraphics[width=7.0cm,angle=0]{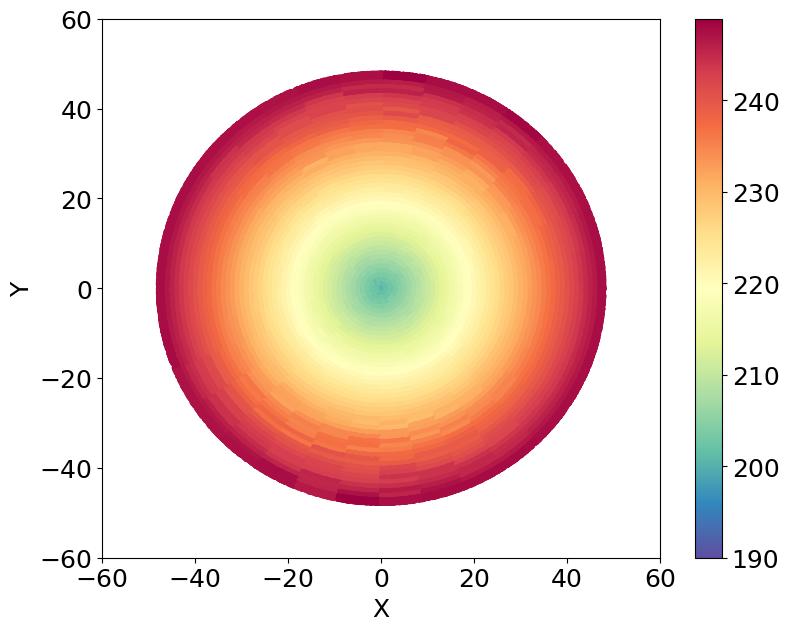}
\caption{}
\end{subfigure}
\begin{subfigure}[a]{0.55\textwidth}
\centering
\includegraphics[width=7.0cm,angle=0]{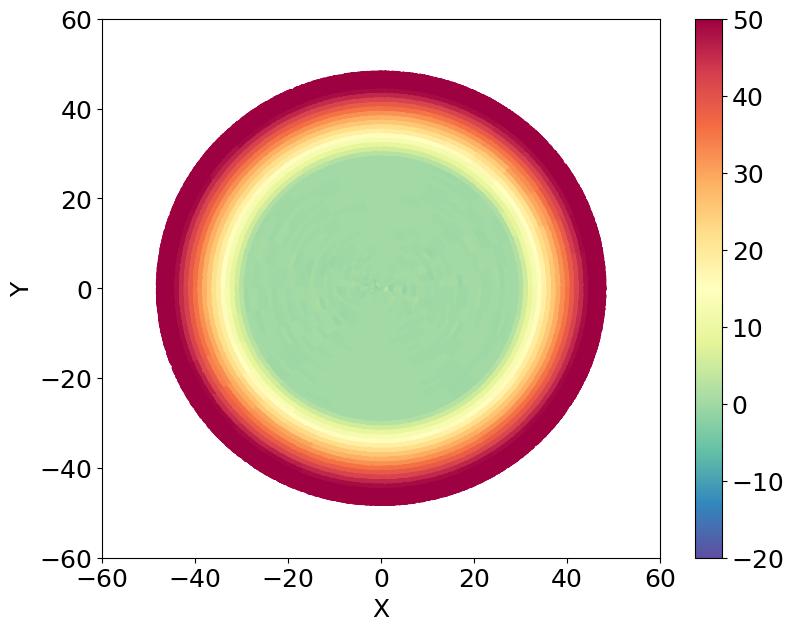}
\caption{}
\end{subfigure}
\caption{  
        VRM-reconstructed coarse-grained map of the transversal  $v_\theta$ (panel (a)) and of the radial velocity component (panel (b)) for a simple toy disc model 
        with an isotropic velocity distribution (see text for more details). 
} 
\label{fig:VRM2}
\end{figure}

The map of an external galaxy consists of the angular coordinates $(r,\phi)$ in the projected image of the galaxy onto the plane of the sky, along with the LOS  component of the velocity $v_{\text{los}}(r,\phi)$. These can be transformed in
\be
\label{eq:vlos} 
v_{\text{los}}  = \left[v_\theta  \cos(\theta) + v_R  \sin(\theta) \right] \sin(i)  \;, 
\ee
where the transversal and radial velocity components, $v_\theta=v_\theta(R, \theta)$ and $v_R=v_R(R, \theta)$, refer to the velocity of the galaxy in the plane of the galaxy, where $R$ and $\theta$ are the polar coordinates  in the plane of the galaxy and $i$ is the inclination angle (the other orientation angle, the P.A., enters into the transformation from  $(r,\phi)$ to $(R, \theta$)). In the intrinsic coordinates of the galaxy, assuming a flat disc, it is possible to divide the disc into $N_r$ rings, all having the same inclination angle and P.A. 
{  The velocity field within each ring can be decomposed into a radial component, $v_R = v_R(R)$ , and a transverse component, $v_\theta = v_\theta(R)$. By inverting Eq. \ref{eq:vlos}, the VRM determines the values of $v_R(R)$ and $v_\theta(R)$ in each ring
 (see Fig.\ref{fig:VRM}). In its simplest implementation, the VRM provides the average values of $v_R$ and $v_\theta $ in concentric rings. If each ring is further divided into 
$N_a$ arcs (see below), the VRM provides the values of these two velocity components for each circular sector.

To present a simple example of the coarse-grained maps of the transversal $v_\theta$ and radial velocity components, we show in Fig. \ref{fig:VRM2} the results for a toy disc model. In this model, the $v_\theta$ linearly increases from 200 km/s at the center to 250 km/s at the outermost shells, while $v_R$ is zero up to half of the disc radius and then increases linearly to 66 km/s. In this case, the velocity field is isotropic, so the VRM-reconstructed coarse-grained velocity component maps exhibit no angular anisotropy, only a radial variation. We refer the interested reader to \cite{SylosLabini_etal_2023_VRM} for a more detailed discussion of the VRM methods and its results to toy disc models and real galactic images. 
}

%

{  The VRM maps can be further refined by } splitting each ring into $N_a$ arcs, each characterized by a different radial and transversal velocity, it is possible to characterize velocity anisotropies, i.e.  the dependency of the two velocity component  on both the radial distance $R$  the angular coordinate $\theta$, meaning that $v_\theta=v_\theta(R, \theta)$, $v_R=v_R(R, \theta)$. The VRM divides the galaxy image into a grid of $N_{\text{cells}}$ cells, determined by the product of the number of rings and the number of arcs, i.e. $N_{\text{cells}}=N_r \times N_a$. Within the $i^{th}$ cell, the VRM provides the transverse velocity component, $v_\theta^i$, and the radial velocity component, $v_R^i$. It is then possible to analyze the convergence of the VRM method by studying the convergence of several moments that measure the velocity field in a given region  by varying the resolution of the coarse-grained map (see \cite{SylosLabini_etal_2023_VRM} for details). 

{   
On one hand, the size of each of the $N_{\text{cells}}$ must be larger than the beam size $\sigma$, and on the other hand, there should be enough cells to perform a reliable statistical analysis of the anisotropy structures in the coarse-grained velocity component 2D maps.
The first requirement is satisfied as long as the cell size exceeds the beam size of $25" \times 25"$ , to which the original map is smoothed. Conversely, the optimal number of cells can be determined by analyzing the convergence of various moments (see \cite{SylosLabini_etal_2023_VRM} for details) that characterize the velocity field within a given region, as the resolution of the coarse-grained map is varied. This process involves, for instance, changing the number of arcs while keeping the number of rings fixed. The convergence of these moments ensuring that the reconstructed velocity field remains stable and independent of the resolution employed by the VRM, is critical for evaluating the method's reliability. It demonstrates that the noise introduced by the VRM reconstruction, which is inevitably present, does not dominate over the signal.
In what follows we will report results for $N_r=10$ and $N_a=8$: this corresponds to the galaxy with the lowest resolution for which cells are of the order than, or larger than, the beam size. }

In principle, the degeneracy between warps and radial flows  in external galaxies makes difficult to discriminate radial flows from geometric deformations.  However, the possible warped geometry of a disc galaxy is typically more pronounced in the outermost regions that may be affected by tidal forces from external perturbers: indeed some edge-on disc galaxies show warps starting at the optical radius \citep{Sancisi_1976, Reshetnikov+Combes_1998, Schwarzkopf+Dettmar_2001, Garcia-Ruiz_etal_2002, Sanchez-Saavedra_etal_2003}. For this reason, in order to control that the key assumption of the VRM that the disc is flat is a good working hypothesis, we monitor the behavior of the orientation angles determined by the TRM as a function of the radial distance. Only when they present a smooth radial change, do we interpret the behaviors as indicating the presence of a warp; otherwise, fluctuations in the values of the orientation angles all over the disc are interpreted as being due to velocity anisotropies in the VRM framework. In practice, when the orientation angles smoothly vary with scale, we use the {  circular} velocity, averaged over rings, determined by the TRM; However, we find that, in general, the correction due to the warp  for the galaxies in this sample is small in the inner regions of the  discs and thus $v_\theta(R)$ is generally very similar the {  circular} velocity determined by the TRM. In a forthcoming paper \citep{SylosLabini_etal_2024_Warp}, we will present a method to correct the VRM determinations in case the disc is warped.

\subsection{Correction for asymmetric drift}
\label{AsymDrift}

Under the assumption that the galaxy is in stationary state and collisions can be neglected, we use the Jeans equations to relate the observed kinematic properties to the gravitational field of the system \citep{Binney_Tremaine_2008}. In particular, for an axisymmetric system in cylindrical $R, z, \theta$ coordinates,  we have that  
\bea
\label{eq:jeans1}
\frac{\partial \rho \overline{v^2_R}} { \partial R} + 
\frac{ \partial \rho \overline{v_R v_z }}{ \partial z} + 
\rho \left( \frac{\overline{v^2_R} - \overline{v^2_\theta}}{R} + 
\frac{ \partial \Phi}{ \partial R} \right) =0 \;,
\eea 
where {  $\rho$ is mass density in the disc, $v_R$, $v_\theta$ and $v_z$ are respectively the radial, tangential and vertical component of the velocity is cylindrical  coordinates and } 
$\overline{v^2_R}, \overline{v_R v_z }$ and $ \overline{v^2_\theta}$ are the moments of the velocity distribution and  $\Phi$ is the {  total} system's gravitational potential. {  Note that hereafter, we will refer to the mean values and fluctuations (i.e., anisotropies) of the radial and tangential velocity components, while disregarding variations in the vertical velocity component, as it is not observable in external galaxies.}

By defining   the {  circular} velocity as 
\be
\label{eq:vc2}
v_c^2 = R \frac{ \partial \Phi}{ \partial R} \;
\ee
we can rewrite Eq.\ref{eq:jeans1} as
\be
\label{eq:jeans1b}
v_c^2 = \overline{v^2_\theta} - \overline{v^2_R} \left( 1 + \frac{R} { \rho} \frac{\partial \rho}  { \partial R}  \right) - R \frac{\partial  \overline{v^2_R} }  { \partial R} +
- \frac{R} { \rho}  \frac{ \partial \rho \overline{v_R v_z }}{ \partial z} \;.
\ee
We neglect hereafter the terms with $ \overline{v_Rv_z}$ as they are generally considered subdominant: while for the case of the Milky Way it is now possible to directly measure these terms, and measure that they are indeed negligible \citep{Eilers_2019,Wang_etal_2023,Ou_etal_2024} for the case of external galaxies this is a reasonable  working hypothesis.  Neglecting these mixed terms we find 
\be
\label{eq:jeans1c}
v_c^2 \approx  \overline{v^2_\theta} - \overline{v^2_R} \left( 1 + \frac{R} { \rho} \frac{\partial \rho}  { \partial R}  \right) - R \frac{\partial  \overline{v^2_R} }  { \partial R} \;.
\ee
From the very definition of dispersion we can write that 
\be
\label{eq:disep}
 \overline{v^2_\theta}  =  \overline{v_\theta}^2 +  \sigma^2_{v_\theta}  
\ee
and
\be
 \overline{v^2_R}  =  \overline{v_R}^2 +  \sigma^2_{v_R} \approx \sigma^2_{v_R}\;.
 \ee
We assume $\overline{v_R} \approx \overline{v_z} =0$: while we do not have direct observations of the vertical component we can measure the radial one by means of the VRM method  and limit the analysis to the range where it is fluctuating around zero.

Considering Eq.\ref{eq:disep}, we can rewrite  Eq.\ref{eq:jeans1c} as 
\be
\label{eq:jeans1d}
v_c^2 \approx \left( \overline{v_\theta}^2 + \sigma^2_{v_\theta}\right) - \sigma^2_{v_R} 
\left( 1 + \frac{R} { \rho } \frac{\partial \rho }  { \partial R}   +  \frac{R} { \sigma^2_{v_R}  } \frac{\partial   \sigma^2_{v_R} }  { \partial R} \right) \;.
\ee
By assuming $\sigma^2_{v_\theta} \approx  \sigma^2_{v_R} \approx \sigma^2$, where $\sigma^2$ is given by 2nd moment of the observed velocity field (moment 2), 
Eq.\ref{eq:jeans1d} becomes
\be
\label{eq:jeans1g}
v_c^2 \approx  \overline{v_\theta}^2 - 
R  \sigma^2  \left( \frac{1} { \rho } \frac{\partial \rho }  { \partial R}   +  \frac{1} { \sigma^2  } \frac{\partial   \sigma^2 }  { \partial R} \right) = 
 \overline{v_\theta}^2 +  \sigma_{D}^2 
\ee
where we defined the asymmetric drift as 
\be
\label{AD1}
 \sigma_{D}^2 =  - R \sigma^2 \left( \frac{1} { \rho } \frac{\partial \rho }  { \partial R}   +  \frac{1} { \sigma^2   } \frac{\partial   \sigma^2 }  { \partial R} \right)   \;. 
 \ee 
In Eqs.\ref{eq:jeans1g}-\ref{AD1} the kinematic terms can be measured from the data while density  $\rho(R)$ depends on the mass model adopted as we are going to discuss in the next section. 

{  
Note that  the density $\rho$ in Eq.\ref{eq:jeans1} and thus in  Eq.\ref{AD1} corresponds to the density within the disc.Therefore in the NFW case, 
$\rho$ refers exclusively to the baryonic component (i.e., the stellar and gaseous components), which is dominated by the gas mass. Conversely, in the DMD model, where DM is also assumed to reside in the disc, 
$\rho$  represents the combined mass density of the baryonic and DM components. For a more detailed discussion of this point, see \cite{SylosLabini_2024}.  
}


\section{Mass models}
\label{sect:massmodels}

The  dynamics of the sample galaxies is determined by the combination of two components: the baryonic matter and the DM. The study conducted by \cite{Oh_etal_2015} focused on the consideration of the halo model  proposed by \cite{Navarro_etal_1996}  (NFW) and the spherical pseudo-isothermal halo model introduced by \cite{Begeman_etal_1991}. The NFW halo model represents a cusp-like halo, whereas the pseudo-isothermal halo exhibits a core-like behavior. In both cases, the rotationally supported disc of the galaxy is surrounded by a spherical halo consisting of DM. The difference lies in the density behavior of the halo for small radii: the NFW model follows a density profile of $\rho(r) \sim r^{-1}$, while the pseudo-isothermal halo maintains a constant density, $\rho(r) \sim \text{const}$. In this work, we will make fits of the NFW halo model and compare them with those obtained by \cite{Oh_etal_2015}. In this model, the {  circular} velocity   can be written as \citep{Navarro_etal_1996}
\be
\label{nfw_fit} 
v_c^2(R) = v_{s}^2(R) + v_{g}^2(R) + v_{h}^2(R)
\ee
where $v_g(R)$ and $v_s(R)$ are the {  circular} velocity due to the  gas and stellar components and and $v_h(R)$ is the {  circular} velocity of the spherical halo that is determined by the two free parameters of the model, i.e., $r_s$ and $\rho_0$, that define the NFW density profile
\be
\label{eq:nfw}
\rho_{h}= \frac{\rho_0}{\frac{r}{r_s}\left(1 + \frac{r}{r_s}\right)^2}\;,
\ee
where $r$ is the 3D radius.

{  The {  circular} velocity of the $i^{th}$  mass component in Eq.\ref{nfw_fit}  is defined as the equilibrium rotational velocity that this component would induce on a test particle in the plane of the galaxy, assuming it were isolated and unaffected by external influences. These velocities in the plane are calculated based on the observed stellar mass density distributions and the assumed  DM  distribution. While the {  circular} velocity of the halo, $v_h(R)$, can be computed analytically, the {  circular} velocities of the disc components, $v_g$ (gas) and $v_s$  (stars), must be determined through simulations.  As the analytical calculation of the centrifugal force for a disc is not straightforward we have adopted a numerical strategy: we generate a numerical realization of a thin disc model with the observed surface brightness of the stellar (or gas) component and the same mass, then numerically compute the resulting centripetal acceleration (the derivation of the masses of the gaseous and stellar components was discussed in Sect.\ref{sect:sample}).}

The total mass of the baryonic disc plus the halo is 
\be
\label{eq:nfwmass}
M_\text{nfw} = M_{s} + M_{g} + M_{h} = M_{\text{bar}} +  M_{200} 
\ee
where the baryonic mass $M_{\text{bar}}$ is the sum of the stellar and gas components, i.e., 
\be
\label{eq:massbar}
M_{\text{bar}} = M_{s} + M_{g} \;,
\ee
and 
$M_{200}$ is the mass of the halo in a sphere of radius $r_{200}$, defined as the virial radius:  as the mass of the halo grows linearly at large distances, it is necessary to consider a cut-off that corresponds to the virial radius $r_{200}$ where density fluctuations due to the halo are 200 times larger than the average mass density of the universe \citep{Navarro_etal_1996}. The virial  radius $r_{200}$ depends on $r_s$ and $\rho_0$.

The second model we will consider is the DMD model, which assumes that DM is confined to the galaxy's disc. The basis for this model can be traced back to the observed correlation between DM and the distribution of neutral hydrogen gas. In this scenario, it is assumed that the distribution of \HI{} gas serves as a proxy for the DM distribution, allowing us to express the {  circular} velocity as
\be
\label{dmd_fit} 
v_c^2(R) = \gamma_{s} v_{s}^2(R) + \gamma_{g} v_{g}^2(R) 
\ee
where  $\gamma_s, \gamma_g$ are the two free parameters of the fit: their meaning is how much DM  is associated, respectively, with the stellar and gas component. In Eq.\ref{dmd_fit} the velocity $v_s$ represents the rotation velocity induced by the stellar component on a test particle in the galaxy's plane, assuming it is in isolation without external influences. Similarly, $v_g$ represents the rotation velocity induced by the gas component on the same test particle under the same conditions. These are computed by (i) estimating the surface density profiles of the stellar and gas components (the former is reported in \cite{Oh_etal_2015}); (ii) generating artificial discs with such profiles, (iii) computing the gravitational 
potential and (iv) making its radial derivative on the plane (i.e., Eq.\ref{eq:vc2}).

In this case the total mass of the baryonic and dark component is 
\be
\label{eq:massdmd}
M_{\text{dmd}} =  \gamma_{s}  M_{s} + \gamma_{g} M_{g}   \;. 
\ee
Given that both the surface density profiles of the gas and star components  decay approximately exponentially, in this case, to estimate the total mass,  there is no need to consider a cut-off in distance. 
\bigskip

Note that the  density of the disc $\rho(R)$ that enters into {  the Jeans equation (Eq.\ref{eq:jeans1}), and thus in the expression of the {  circular} velocity (Eq.\ref{eq:jeans1g}) and in Eq.\ref{AD1} } is the baryonic density for the case of NFW model,
{  which, up to a prefactor that is not relevant in Eq.\ref{eq:jeans1}, is} 
\be
\label{eq:rho_dmd} 
\rho_\text{}(R) \sim  \Sigma_{\text{bar}}(R)  =  \Sigma_s(R) + \Sigma_{\text{g}}(R)
\ee 
where $\Sigma_s(R)$ and $ \Sigma_g (R) $ are respectively  the stellar and \HI\;  surface density: the first was measured by \citep{Oh_etal_2015} while the second is measured directly from the data of the \HI{} (mom0).
{  In particular, the gas surface density is computed by assuming a constant inclination angle across the entire disc. To remain consistent with the mass determination discussed in Sect.\ref{sect:sample}, the gas surface density was scaled up by a factor of 1.4 to account for helium and metals, while the contribution of molecular hydrogen was neglected.}

The (quasi) spherical halo is assumed be in equilibrium with an (quasi) isotropic velocity dispersion.

{  For the DMD model, the density of the disc entering the Jeans equation (Eq.\ref{eq:jeans1}) is given (again, up to an irrelevant prefactor) by}
\be
\label{rho_dmd} 
\rho_\text{}(R) \sim \Sigma_{\text{dmd}}(R) =  \gamma_s \Sigma_s(R) +  \gamma_g \Sigma_{g}(R) \;.
\ee 
{  Specifically, it is a linear combination of the stellar and gas surface densities, rescaled by the two free parameters of the DMD model, $\gamma_{s}$  and $\gamma_{g}$ .  This ensures that Eq.\ref{rho_dmd} remains consistent with the mass model described in Eq.\ref{dmd_fit}.}

In the analysis of the individual galaxies in our sample, 
we have fitted $\Sigma_{\text{dmd}}(R)$   with the function 
\be
\label{sigma} 
\Sigma_{\text{dmd}}(R) = \frac{\Sigma_0}{1+ \left( \frac{R}{R_g} \right)^\beta}
\ee
that corresponds to a cored profile. In general, for the galaxies in the present sample we find   $\beta$ is in the range $[1,4]$.
We make polynomial fit to $\sigma$ and we then insert  the results of the two fits  in Eqs.\ref{eq:jeans1g}-\ref{AD1} to  obtain the  asymmetric drift 
and the {  circular} velocity $v_c$.

We stress that from a physical perspective, the distinction between the NFW  and DMD  models lies in the distribution and dynamical nature DM. In the NFW model, DM is assumed to be distributed in a (quasi) spherical halo surrounding the galaxy. The DM in this model is supported by a (quasi) isotropic velocity dispersion, meaning that the random motions of DM particles play a significant role in supporting the gravitational potential. On the other hand, the DMD model posits that DM is primarily confined to the galactic disc and follows the distribution of neutral hydrogen. In this model, the DM is supported by rotation, similar to the visible matter in the disc. This fundamental difference in the dynamical nature of DM between the NFW and DMD models arises from distinct evolutionary paths. The NFW model assumes a hierarchical structure formation scenario, where DM halos form through gravitational collapse and subsequent merging of smaller structures. The DMD model, as discussed in \cite{SylosLabini_etal_2020} and related references, proposes an alternative scenario where  galaxy formation proceeds through a top-down collapse, leading to its confinement within the disc.


 \begin{figure*}
 \quad
 \begin{subfigure}[a]{0.3\textwidth}
 \centering
\includegraphics[width=5.0cm,angle=0]{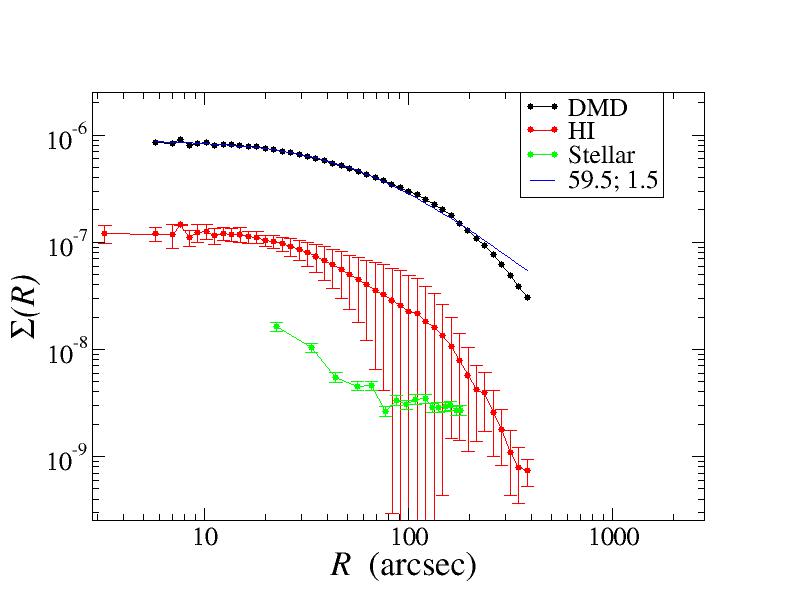}
\caption{}
\end{subfigure}
\quad
\begin{subfigure}[a]{0.3\textwidth}
\centering
\includegraphics[width=5.0cm,angle=0]{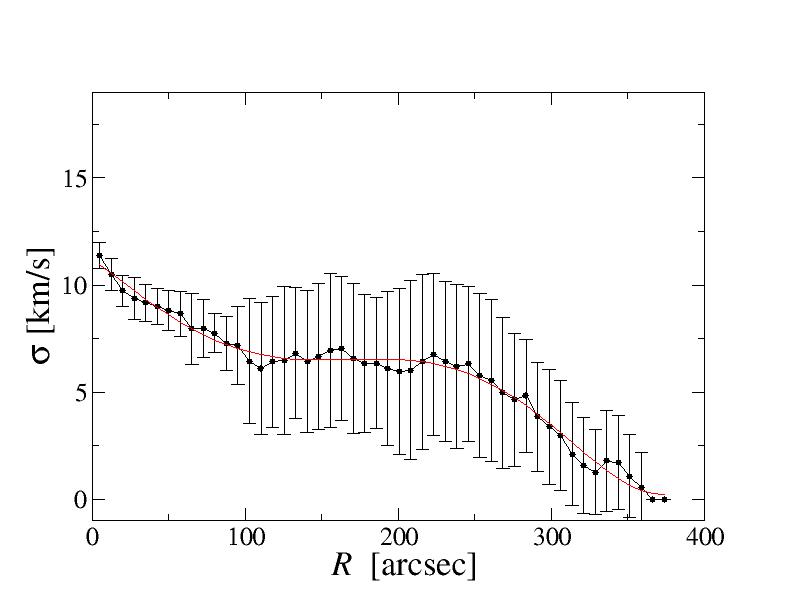}
\caption{}
\end{subfigure}
\quad
\begin{subfigure}[a]{0.3\textwidth}
\centering
\includegraphics[width=5.0cm,angle=0]{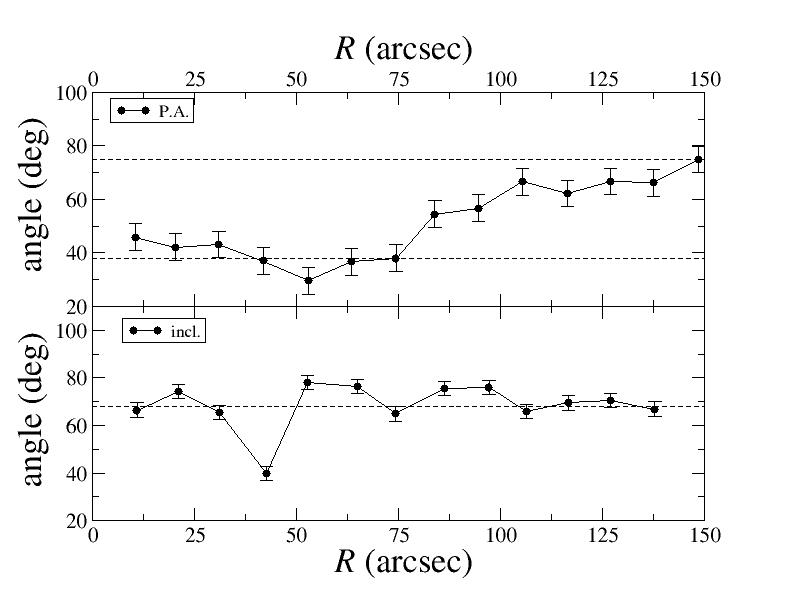}
\caption{}
\end{subfigure}
\quad
\caption{Profiles of  CVnIdwA:
    (a) {  total DMD, \HI{}\ and stellar surface density profiles.} 
    The parameters $R_g, \beta$ of the function in Eq.\ref{sigma} (blu line) are reported in the labels. 
    (b)  {  Velocity dispersion profile $\sigma(R)$ averaged in rings with a 6$^{th}$ degrees polynomial fit (red line).}
    (c)  Orientation angles from the TRM as a function of the radial distance: position angle (upper panel),  inclination angle (bottom panel)	
} 
\label{fig:CVnIdwA2}
\end{figure*}

\begin{figure*}
\quad
\begin{subfigure}[a]{0.3\textwidth}
\centering
\includegraphics[width=5.0cm,angle=0]{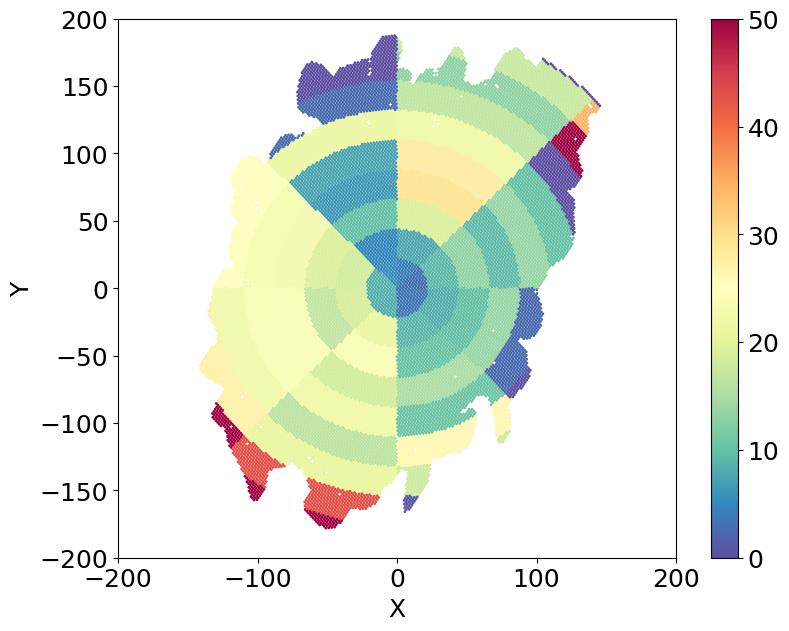}
\caption{}
\end{subfigure}
\quad
\begin{subfigure}[a]{0.3\textwidth}
\centering
\includegraphics[width=5.0cm,angle=0]{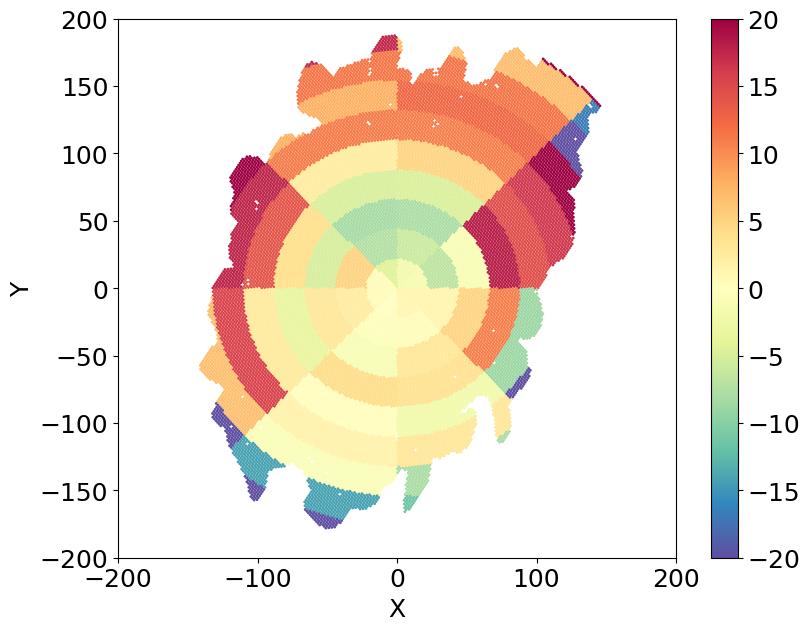}
\caption{}
\end{subfigure}
\quad
\begin{subfigure}[a]{0.3\textwidth}
\centering
\includegraphics[width=5.0cm,angle=0]{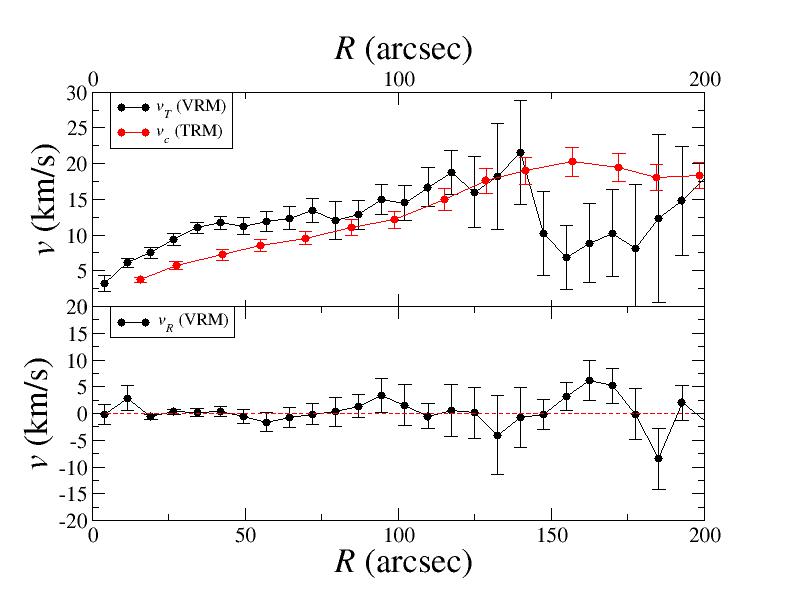}
\caption{}
\end{subfigure}
\quad
\caption{  
VRM maps of CVnIdwA:
(a)  transversal velocity component map with resolution $N_r=10$ and $N_a=8$ and 
(b)  radial velocity  component map with resolution $N_r=10$ and $N_a=8$).
(c)  Upper panel: transversal velocity profile as measured by the VRM without corrections (black dots), and determined by the TRM (red dots).
Bottom panel: radial velocity profile.  
} 
\label{fig:CVnIdwA3}
\end{figure*}

\begin{figure*}
\quad
\begin{subfigure}[a]{0.3\textwidth}
\centering
\includegraphics[width=5.0cm,angle=0]{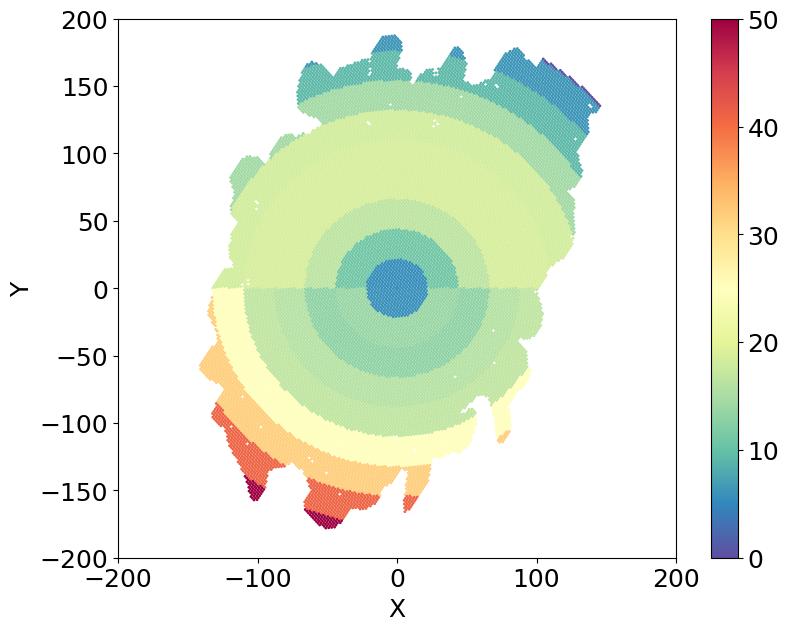}
\caption{}
\end{subfigure}
\quad
\begin{subfigure}[a]{0.3\textwidth}
\centering
\includegraphics[width=5.0cm,angle=0]{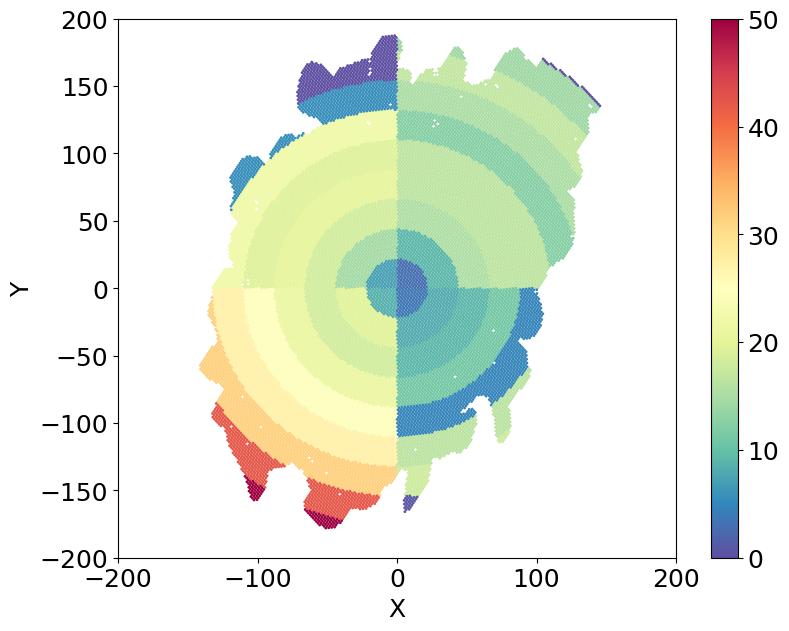}
\caption{}
\end{subfigure}
\quad
\begin{subfigure}[a]{0.3\textwidth}
\centering
\includegraphics[width=5.0cm,angle=0]{fig_e_CVnIdwA_VT_R10A8.pdf.jpg}
\caption{}
\end{subfigure}\\
\quad
\begin{subfigure}[a]{0.33\textwidth}
\centering
\includegraphics[width=5.0cm,angle=0]{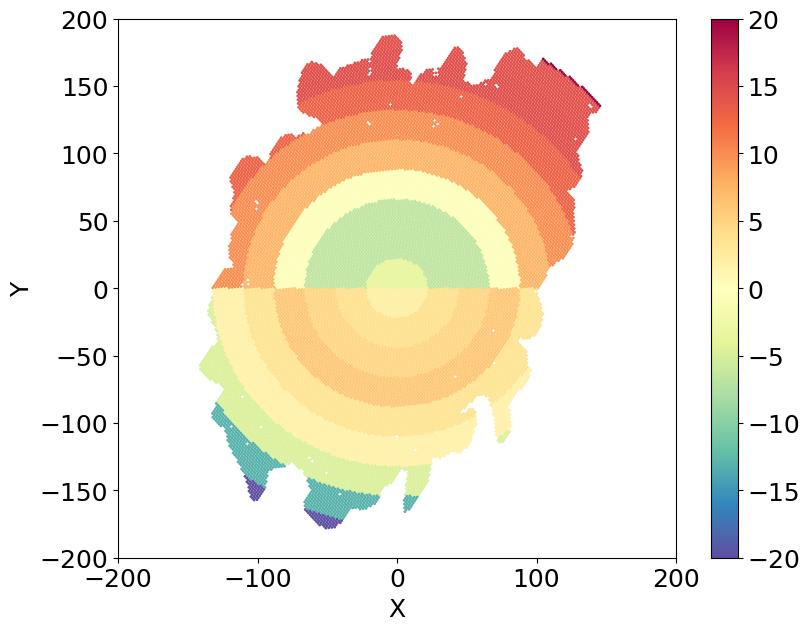}
\caption{}
\end{subfigure}
\quad
\begin{subfigure}[a]{0.29\textwidth}
\centering
\includegraphics[width=5.0cm,angle=0]{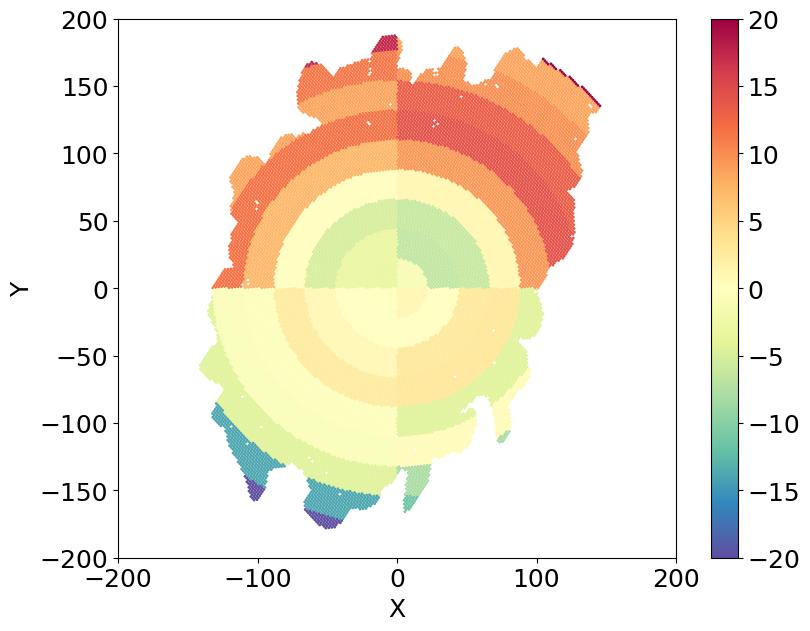}
\caption{}
\end{subfigure}
\quad
\begin{subfigure}[a]{0.3\textwidth}
\centering
\includegraphics[width=5.0cm,angle=0]{fig_e_CVnIdwA_VR_R10A8.pdf.jpg}
\caption{}
\end{subfigure}
\quad
\caption{  
VRM maps of CVnIdwA with varying resolution:
(a,b,c)  transversal velocity component map with resolution $N_r=10$ and $N_a=2,4,8$; 
(d,e,f)  radial velocity component map with resolution $N_r=10$ and $N_a=2,4,8$ 
} 
\label{fig:CVnIdwA3b}
\end{figure*}

\begin{figure}
\centering 
\includegraphics[width=7.0cm,angle=0]{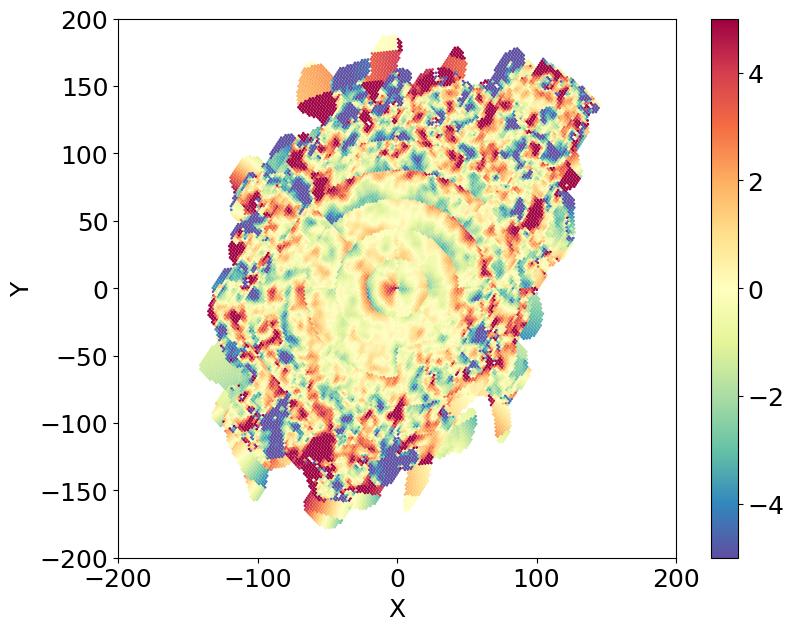}
\caption{  Residual field map of CVnIdwA withwith resolution $N_r=10$ and $N_a=8$.} 
\label{fig:CVnIdwA3c}
\end{figure}


\section{Kinematical Analysis}
\label{sect:results_kinematics}

In this section, we present the  results of the analysis obtained for one specific galaxy in our sample (CVnIdwA), showcasing the rotation curves and other significant findings.  For a comprehensive overview of the results obtained for all other  galaxies in our sample, including rotation curves and additional analyses, we direct readers to the Appendix, where detailed information is provided.

  {Let us start with the determination of the \HI{}, stellar surface, and total DMD density profiles shown in panel (a) of Fig.~\ref{fig:CVnIdwA2}. The \HI{} surface density profile was derived based on the analysis of the  intensity (mom0) map. The sky-plane coordinates  ($r,\phi)$ provided by observations were transformed into galaxy-plane coordinates ($R,\theta)$, assuming a constant inclination angle throughout the disc:}
\bea
&&
\tan(\theta) = \frac{\tan(\phi-\phi_0)} {\cos(i)}  
\\ && \nonumber 
R = r \frac{\cos ( \phi - \phi_0 ) } {\cos(\theta)} 
\eea 
  {where $i$ is the inclination angle and $\phi_0$ is the P.A. (see \cite{SylosLabini_etal_2023_VRM} for more details).   The \HI{}{} density profile was computed in concentric circular rings of radius $R$ and thickness $\Delta R$.} 

  {The  stellar surface density profile is that measured by \citet{Oh_etal_2015} and thus has a different sampling, i.e. $\Delta R$, than that of the \HI{}  one.  
Thus, in order to numerically compute their sum in Eq.\ref{eq:rho_dmd} the stellar profile was approximated by an analytical fitting function. 
}
  {
Finally, the DMD surface brightness profile was computed from Eq.\ref{rho_dmd}  once the fit with the DMD mass model has been performed and the two free parameters $\gamma_s$ and  $\gamma_g$ determined (see Sect.\ref{sect:results_mass_models}). In the panel (a) of Fig.~\ref{fig:CVnIdwA2} it is also shown the best of the DMD profile  fit with Eq.~\ref{sigma} (solid line). In this case, we find that the exponent is  $\beta \approx 1.9$, which is a typical result for the galaxies in this sample.}

  { The velocity dispersion $\sigma$  is directly estimated from the moment-2 maps using a single value for the inclination angle as for the estimation of the \HI{}{} surface density profile (see panel (b) of Fig.~\ref{fig:CVnIdwA2} where the behavior of the velocity dispersion, $\sigma(R)$, is shown along with the sixth-degree polynomial best-fit function).  We note that the values estimated for the galaxies in our sample,  typically $\leq 15$~km~s$^{-1}$, are consistent with those found in other dwarfs \citep{Oh_etal_2015,Iorio_etal_2017,DiTeodoro+Peek_2021,ManceraPina_etal_2021}. For CVnIdwA the velocity dispersion shows a slow decay up to $R_d \approx 200'' \,$ (= 1.7 kpc,  assuming the distance to the galaxy is $D = 3.6$  Mpc; see \citealt{Hunter_etal_2012}) and a faster decay beyond that radius. 
}  

  { The TRM analysis performed by \cite{Oh_etal_2015} provides two orientation angles: the P.A. (upper part of panel (c) of Fig.~\ref{fig:CVnIdwA2}) and the inclination angle (bottom part of panel (c) of Fig.~\ref{fig:CVnIdwA2}). The  P.A. exhibits an approximately constant behavior for $R < R_d$. Beyond this radius, it clearly increases, with a change of about $20^\circ$. On the other hand, the inclination angle remains approximately constant but shows noticeable fluctuations. The radial trend of the  P.A. hints at the presence of a warp in the outer regions of the galaxy, while the noisy behavior of the inclination angle suggests the presence of velocity gradients.
}

{  
As mentioned above, the advantage of using the VRM lies in its ability to reliably estimate the spatial anisotropies of the velocity components.
Results of the  VRM analysis are shown in Fig.\ref{fig:CVnIdwA3}:  the transversal velocity (panel (a)  of Fig.\ref{fig:CVnIdwA3}) presents small spatial anisotropies in the inner disc, $R<R_d\approx 150"$, while 
larger ones are seen in the outer regions where, however, the signal may be affected by the presence of a warp. A similar situation occurs for the radial velocity map (panel (b)  of Fig.\ref{fig:CVnIdwA3}). Note that we  used a resolution of  10  rings and  8 arcs so that the total number of cells is $N_{\text{cells}}=80$ with each cell having a size of approximately $\sim 700$ arcsec$^2$, compared to the beam size of $\sim 625$   arcsec$^2$.  We have verified that the anisotropy maps converge when the number of  cells changes. This is illustrated in Fig.\ref{fig:CVnIdwA3b} which shows the maps of the transverse and radial velocity components for different numbers of arcs $N_a=2, 4, 8$ and same number of rings $N_r=10$.  By visually inspecting the maps, it can be seen that the anisotropy structures remain approximately consistent for $N_a=2, 4, 8$ while the signal becomes noticeably noisier for $N_a=16, 32$ (not shown). 

The transversal velocity profile, averaged over rings is close to the {  circular} velocity  $v_c(R)$ measured by the TRM by \cite{Oh_etal_2015} (upper part of panel (c)  of Fig.\ref{fig:CVnIdwA3}): the difference, at large radial distances, can be attributed to the effect of the warp which is not taken into account by the VRM. The radial profile (bottom part of panel (c)  of Fig.\ref{fig:CVnIdwA3}) shows  large fluctuations (compared to $v_t(R)$) beyond $\approx R_d$: from these behaviors we can conclude that for $R<R_d$ the disc is in rotational equilibrium; beyond $R_d$ perturbations from a  steady state can be relevant. 

As mentioned above, the best VRM  is determined by minimizing the residuals, i.e. the differences between the observed and modeled velocities, with respect to the free parameters of the model in Eq.~\ref{eq:vlos}. The number of free parameters is given by $N_{\text{par}} = 2 N_a N_r = 2 N_{\text{cells}}$, where $N_{\text{cells}}$ represents the number of cells, each with two free parameters: the radial and tangential velocity components (see \cite{SylosLabini_etal_2023_VRM} for more details).  For CVnIdwA3, we find that the residuals are small: in the inner disc, they are $< 2 \, \mathrm{km/s}$, while in the outer disc, they are $< 4 \, \mathrm{km/s}$ (see Fig.~\ref{fig:CVnIdwA3c}), corresponding respectively to $\sim 10 \%$ and $\sim 20\%$ of the peak velocity. This residual map is representative of all other galaxies whose residual maps are not shown.
}


\section{Mass estimation}
\label{sect:results_mass_models}

{  
In this section we provide the fitting results using both the DMD and NFW  models for all the galaxies in our sample. 

We determined the tangential velocity, $v_\theta(R)$, averaged in rings and measured through the VRM, to estimate the circular velocity, which is then used to perform fits with theoretical mass models. As mentioned above, we generally find that $v_\theta(R)$, within the range of distances of interest (i.e., in the inner disc), agrees well with the circular velocity determined by the TRM.
Differences are typically observed at larger radii, but these are expected due to the different assumptions.

The second quantity required to compute the circular velocity (see Eq.\ref{eq:jeans1g}) is the asymmetric drift correction, $\sigma^2_D$, which is estimated as explained in Sect.~\ref{AsymDrift}. This correction is close to the one determined by \citet{Oh_etal_2015}, but some differences can be detected due to the different methods used to determine it. 

In summary, both $v_\theta(R)$ and $\sigma^2_D$ were measured differently from the corresponding quantities in \citet{Oh_etal_2015}. Nevertheless, given the limited range of scales over which the fits were performed, there is, in most cases, {  good agreement} between our estimates of $v_c(R)$ and theirs.  
{  This is also the case when comparing with the results of \citet{Iorio_etal_2017} and \citet{Mancera_Pina_etal_2022}. In particular, the latter authors find that \HI{} flaring can impact the shape of the rotation curve in the outer regions of galaxies and, consequently, can alter the values of the best-fit parameters of the considered mass models for some LITTLE THINGS dwarfs, such as CVnIdwA. However, we note that such changes are limited in amplitude and, although consistent with the error bars of the VRM results, they primarily impact the outermost regions of galaxies, where differences due to the methods used to recover the velocity from the LOS maps, i.e., VRM or TRM, become more significant. Indeed, as discussed above, radial motions are typically detected by the VRM precisely in these external regions.}%

It should be noted that, in some cases, we limit the fit to a smaller radius than that considered by \citet{Oh_etal_2015} because the VRM indicates that, in the outer parts of the discs, the velocity field is generally more perturbed. For this reason, we focus on the range of distances where the tangential velocity component is larger than the radial one. Additionally, the radial range over which the fits are performed typically spans less than half a decade, and in many cases only a third of a decade (e.g., for CVnIdwA3 the best fit with the mass models has been done in the range 0-2.0 kpc, i.e. $R<R_d$). As a result, even small differences in the data points of $v_c$ can lead to significant variations in the derived model parameters.
}
  {For this reason, from a quantitative perspective, differences do exist in some cases in the estimated parameters of the NFW model (the ones that can be compared), and consequently in the virial mass, $M_{200}$.  }

{  
The best-fit parameters for both models are obtained by minimizing the $\chi^2$ value between the model circular velocity and the observed one. The errors on the two parameters of each model --- $\gamma_s$ and $\gamma_g$ for the DMD model, and $R_s$ and $\rho_0$ for the NFW model --- are derived from the diagonal elements of the covariance matrix. From these errors, we can easily compute the errors on the total mass, respectively $M_{\text{dmd}}$ and $M_{200}$, using error propagation.
%
%
Results for all galaxies are reported in Tab.\ref{table1}. 
}

\begin{figure}
\includegraphics[width=8cm,angle=0]{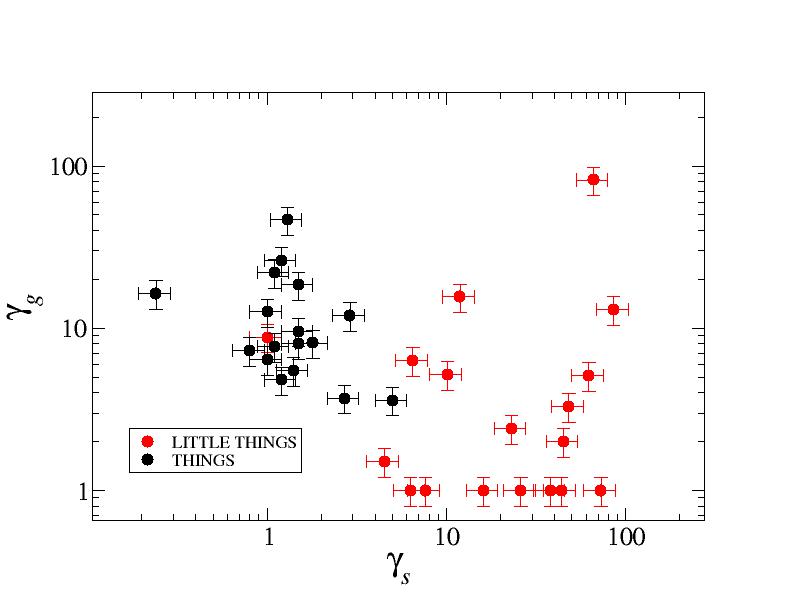}
\caption{  Parameters space $\gamma_s-\gamma_g$ for the THINGS galaxies and the LITTLE THINGS sample.} 
\label{fig:gammaGgammaS}
\end{figure}

{  It is interesting to compare results of the DMD mass model for disc and dwarf galaxies.
Figure \ref{fig:gammaGgammaS} illustrates the parameter space $\gamma_s-\gamma_g$ for the disc galaxies in the THINGS sample (from \cite{SylosLabini_etal_2024_Mass}) and the dwarf galaxies  in the LITTLE THINGS sample. It is observed that the galaxies in the LITTLE THINGS sample generally have larger values of $\gamma_s$, implying that the DM associated with the stellar component is more significant compared to the THINGS sample.   As a result, the DM associated with the  stellar component becomes much more important for the dynamics of these disc galaxies compared to the dwarf galaxies in the THINGS sample.}

 \begin{figure*}
     \quad
    \begin{subfigure}[a]{0.3\textwidth}
    \centering
	\includegraphics[width=5.0cm,angle=0]{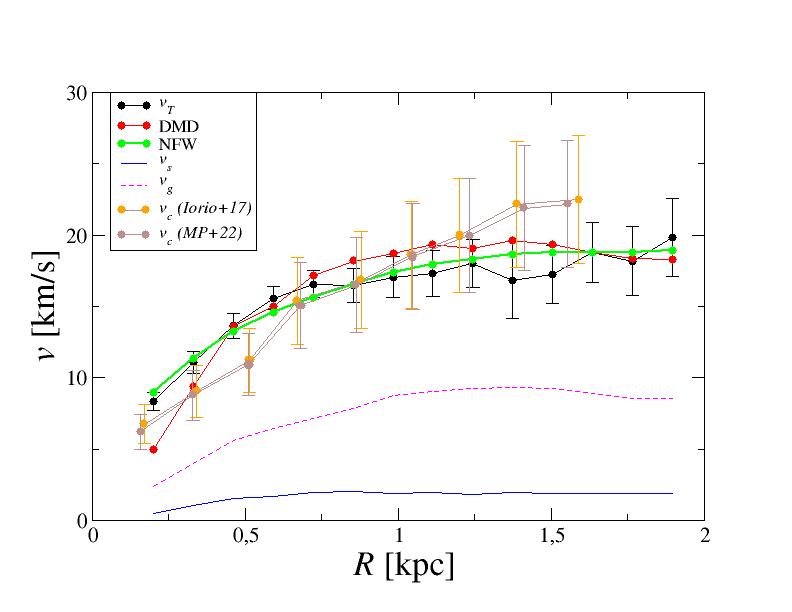}
     \caption{CVnIdwA}
    \end{subfigure}
     \quad
    \begin{subfigure}[a]{0.3\textwidth}
    \centering
    \includegraphics[width=5.0cm,angle=0]{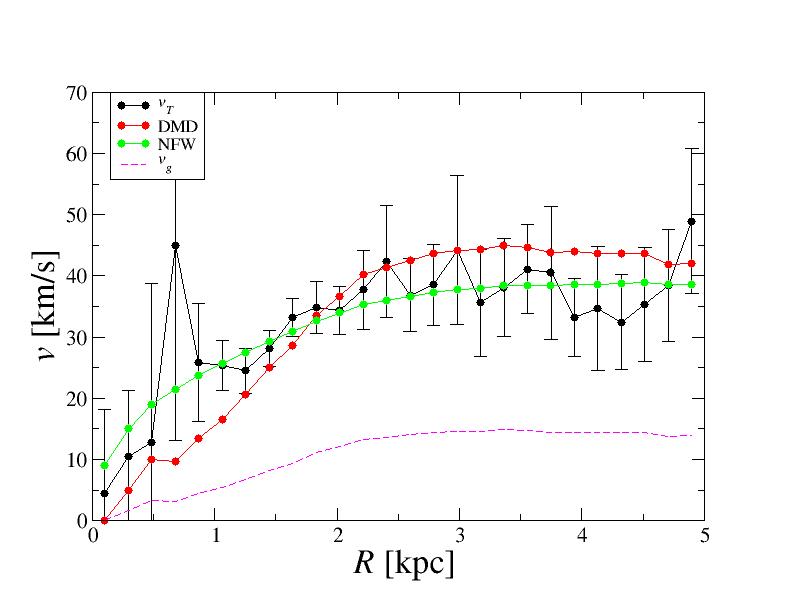}     
    \caption{DDO43}
    \end{subfigure}
     \quad
     \quad
    \begin{subfigure}[a]{0.3\textwidth}
    \centering
    \includegraphics[width=5.0cm,angle=0]{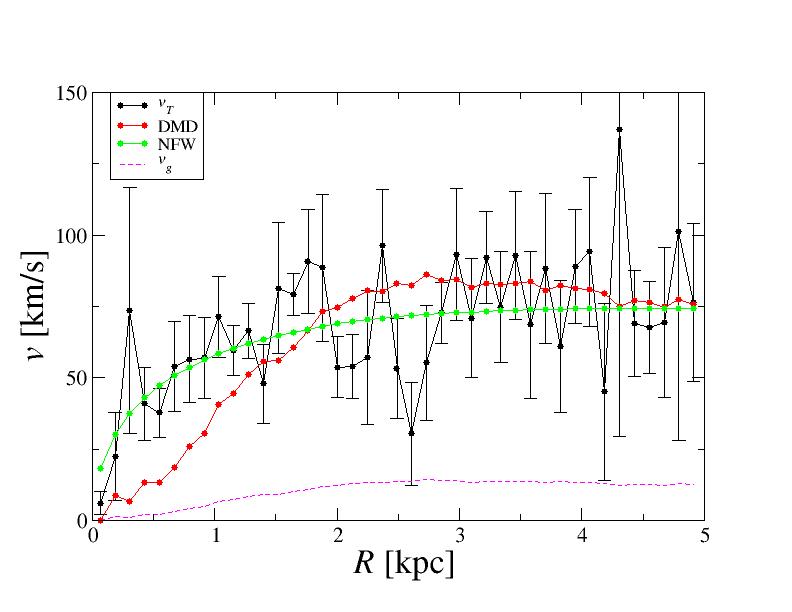}     
    \caption{DDO46}
    \end{subfigure}
    \quad
    \begin{subfigure}[a]{0.3\textwidth}
    \centering
    \includegraphics[width=5.0cm,angle=0]{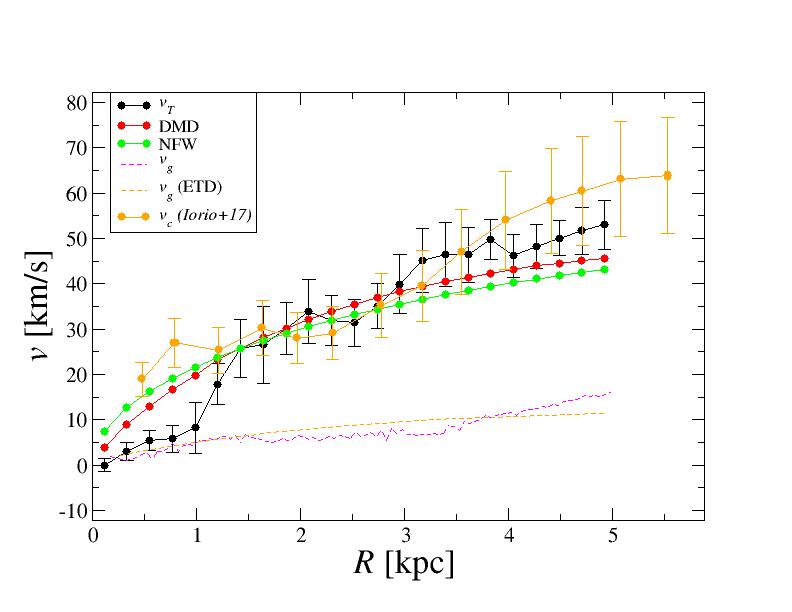}     
    \caption{DDO47}
    \end{subfigure}
   \quad
    \begin{subfigure}[a]{0.3\textwidth}
    \centering
    \includegraphics[width=5.0cm,angle=0]{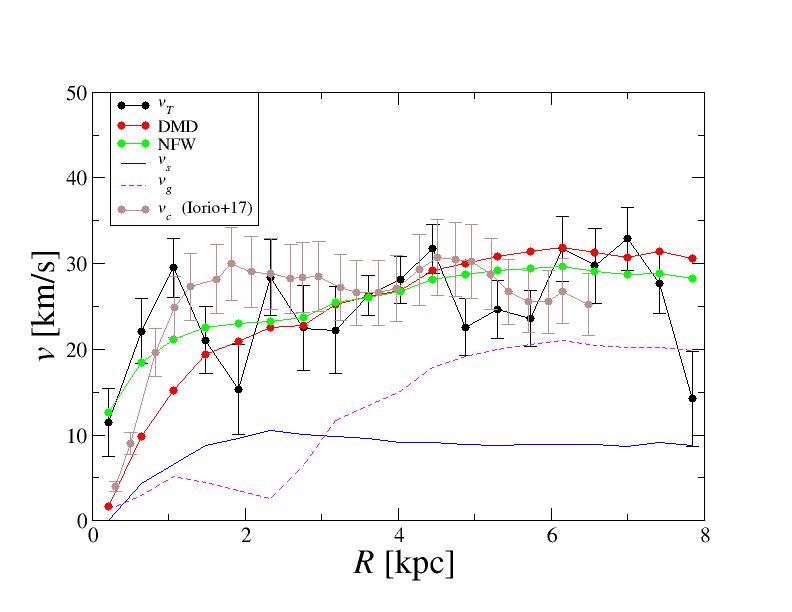}     
    \caption{DDO50}
    \end{subfigure}
    \quad
    \begin{subfigure}[a]{0.3\textwidth}
    \centering
    \includegraphics[width=5.0cm,angle=0]{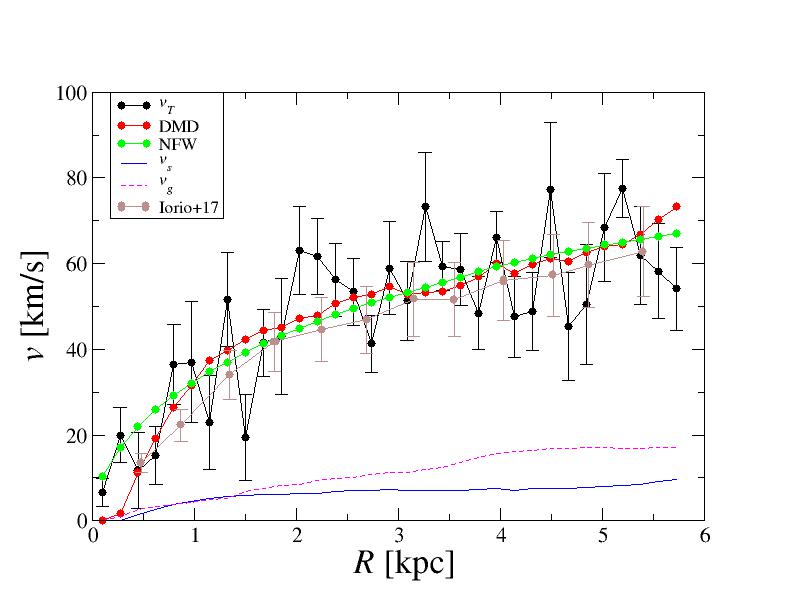}    
     \caption{DDO52}
    \end{subfigure}
   \quad
    \begin{subfigure}[a]{0.3\textwidth}
    \centering
    \includegraphics[width=5.0cm,angle=0]{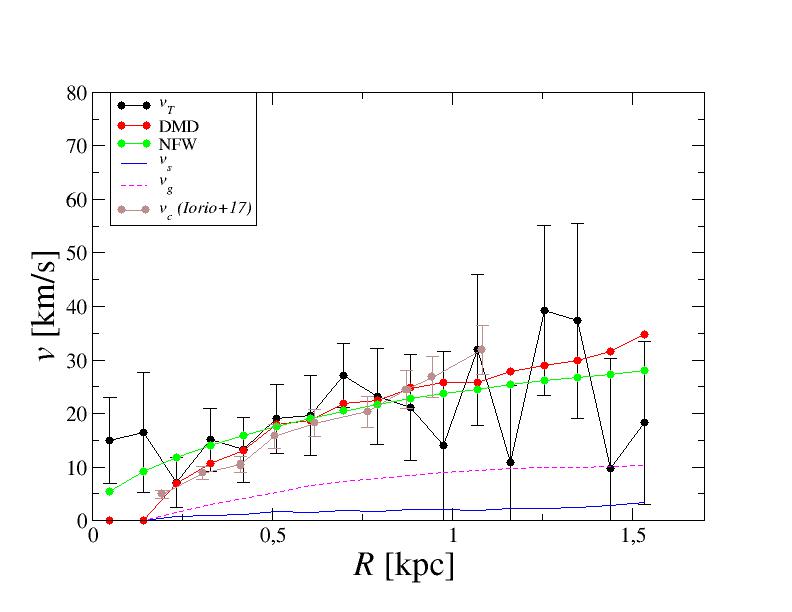}     
    \caption{DDO53}
    \end{subfigure}
   \quad
    \begin{subfigure}[a]{0.3\textwidth}
    \centering
    \includegraphics[width=5.0cm,angle=0]{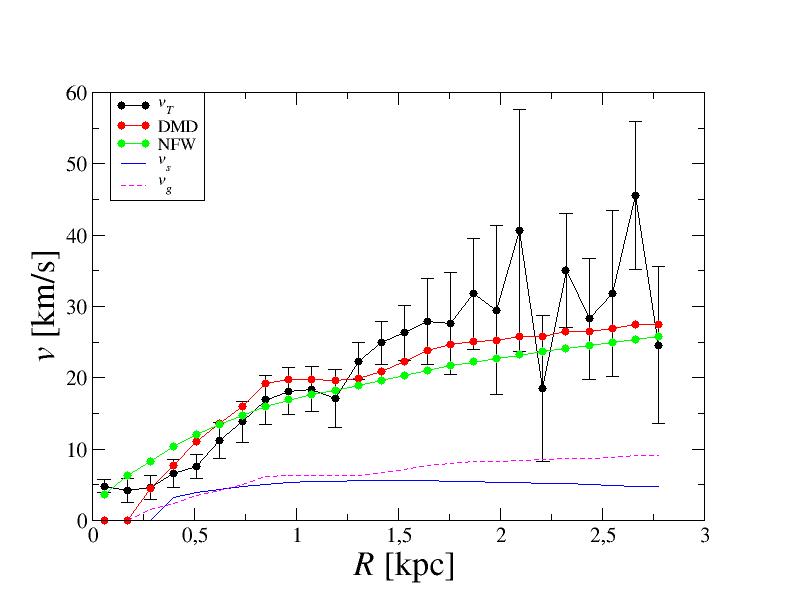}     
    \caption{DDO70}
    \end{subfigure}
   \quad
    \begin{subfigure}[a]{0.3\textwidth}
    \centering
    \includegraphics[width=5.0cm,angle=0]{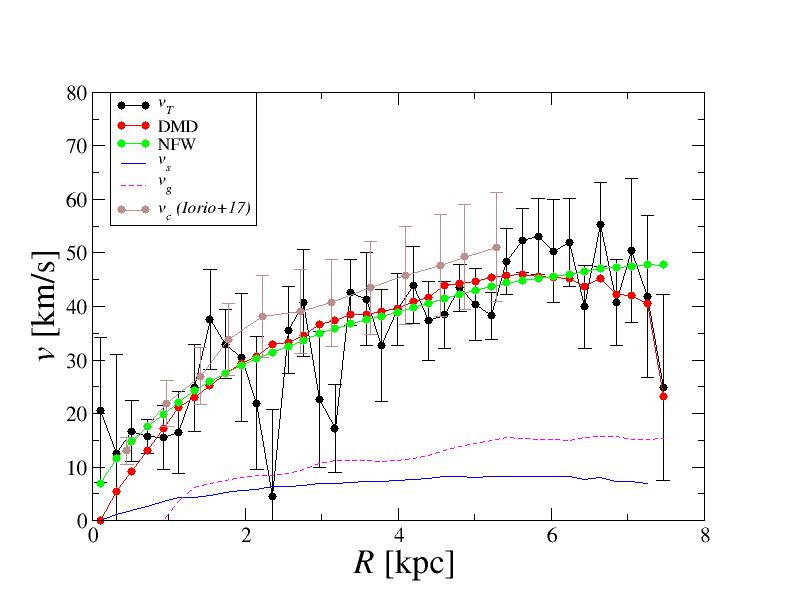}     
    \caption{DDO87}
    \end{subfigure}
       \quad
    \begin{subfigure}[a]{0.3\textwidth}
    \centering
    \includegraphics[width=5.0cm,angle=0]{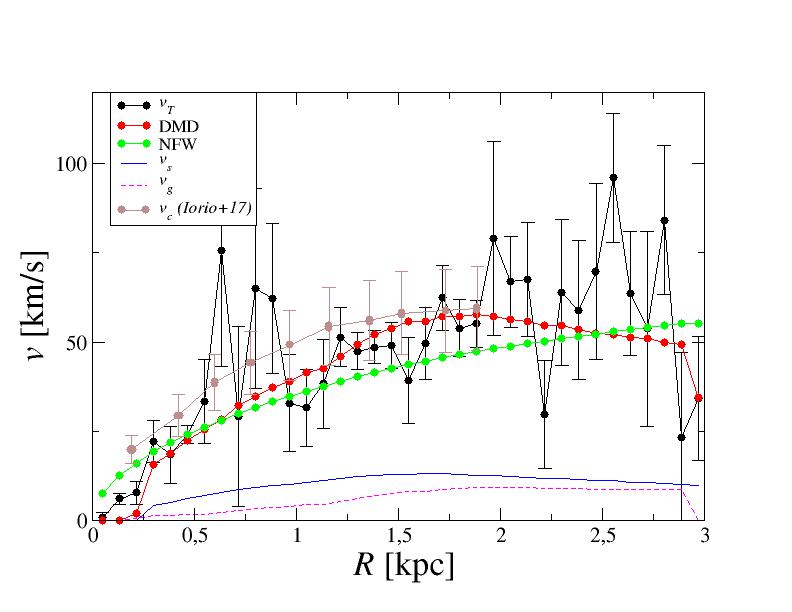}     
    \caption{DDO101}
    \end{subfigure}
   \quad
    \begin{subfigure}[a]{0.35\textwidth}
    \centering
    \includegraphics[width=5.0cm,angle=0]{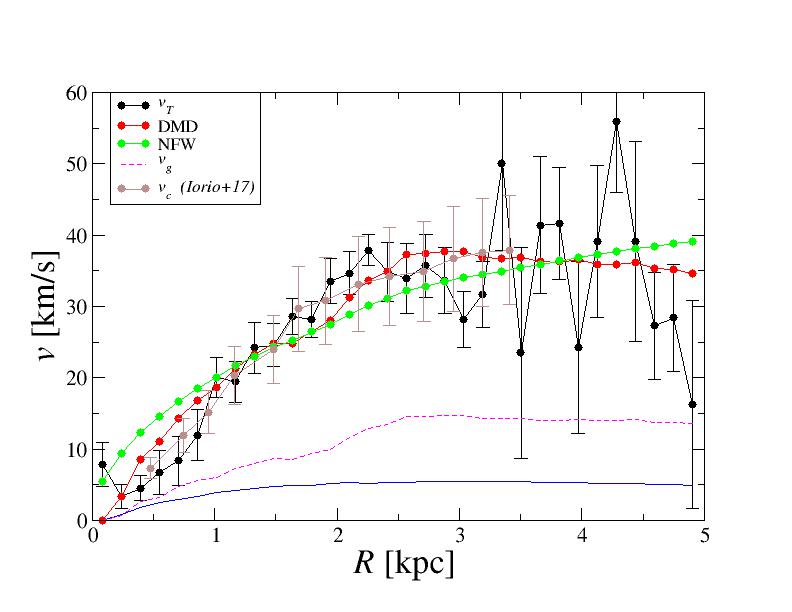}     
    \caption{DDO126}
    \end{subfigure}
       \quad
    \begin{subfigure}[a]{0.32\textwidth}
    \centering
    \includegraphics[width=5.0cm,angle=0]{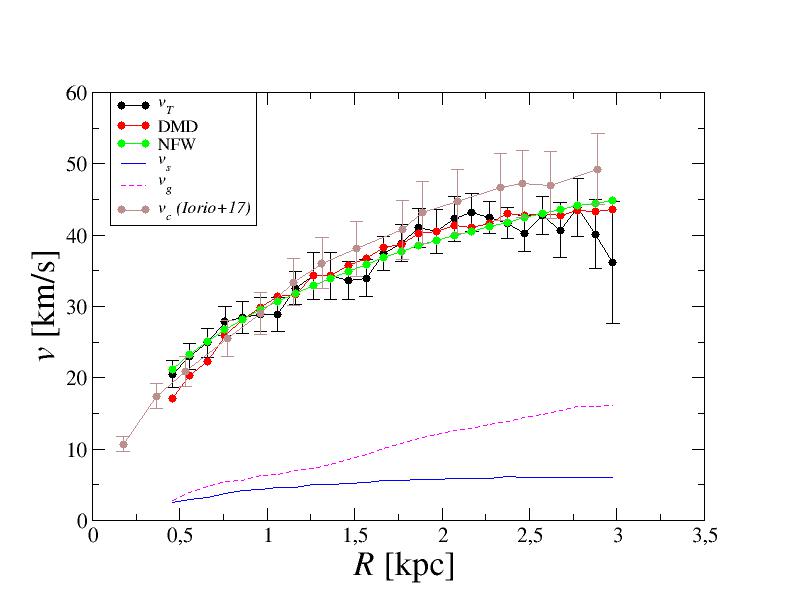}     
    \caption{DDO133}
    \end{subfigure}
   \quad
    \caption{  Best fit with the DMD and NFW models for some of the galaxies in our sample. 
    The {  circular} velocity due to the stellar and gas components is also shown. 
    {  For comparison, in some cases, we have reported the asymmetric drift-corrected rotation curves measured by \citet{Iorio_etal_2017} (Iorio+17) and \citet{Mancera_Pina_etal_2022} (MP+22).      } } 
    \label{fig:fit1}
    \end{figure*}

 \begin{figure*}
     \begin{subfigure}[a]{0.3\textwidth}
    \centering
    \includegraphics[width=5.0cm,angle=0]{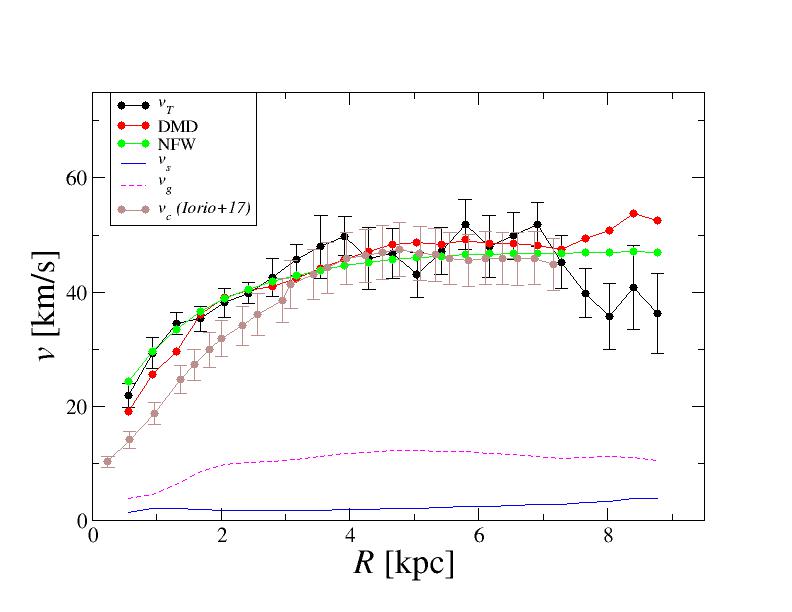}     
    \caption{DDO154}
    \end{subfigure}
   \quad
    \begin{subfigure}[a]{0.3\textwidth}
    \centering
    \includegraphics[width=5.0cm,angle=0]{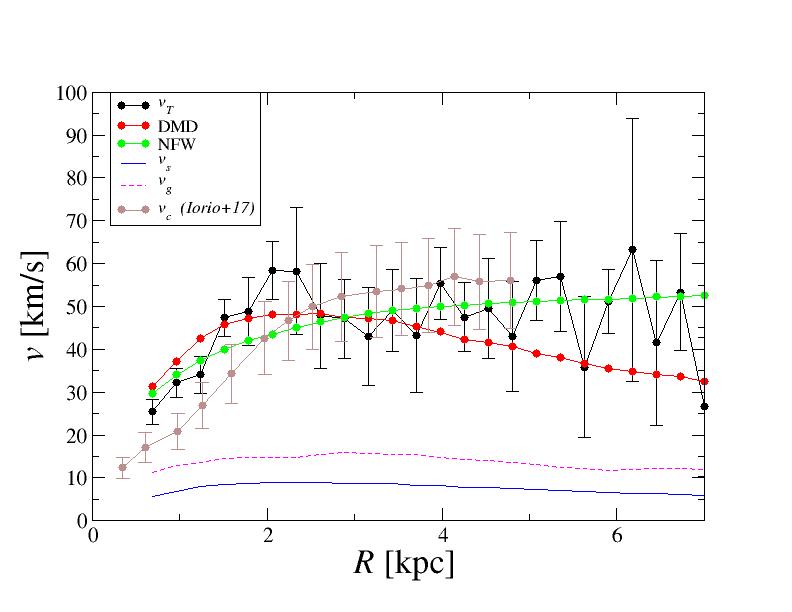}     
    \caption{DDO168}
    \end{subfigure}
       \quad
    \begin{subfigure}[a]{0.3\textwidth}
    \centering
    \includegraphics[width=5.0cm,angle=0]{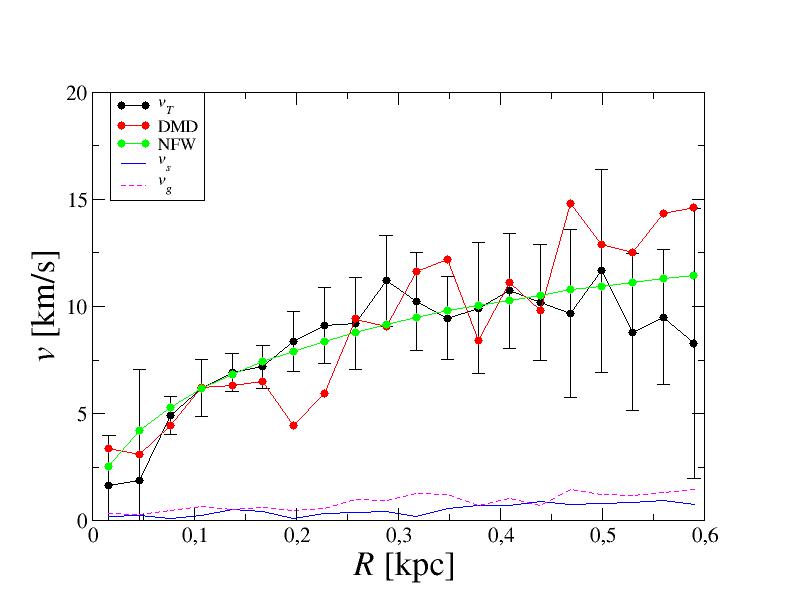}     
    \caption{DDO210}
    \end{subfigure}
     \quad
    \begin{subfigure}[a]{0.3\textwidth}
    \centering
    \includegraphics[width=5.0cm,angle=0]{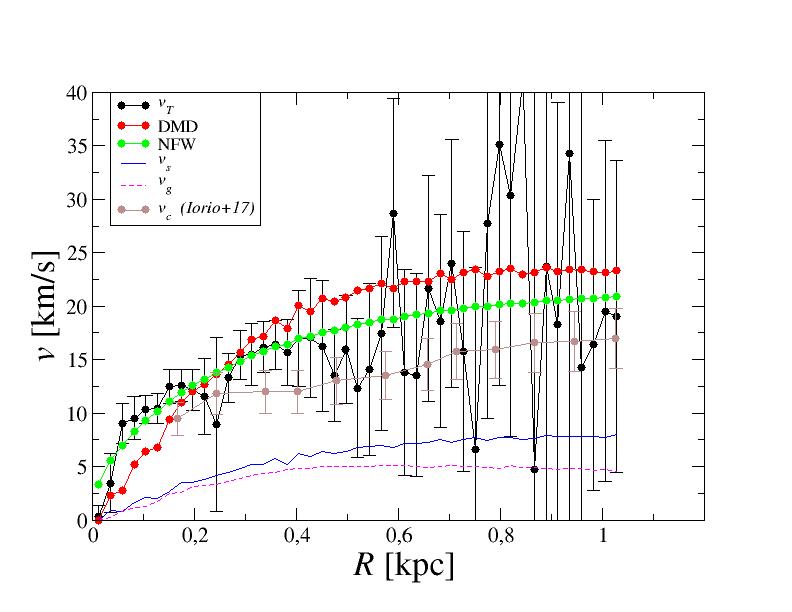}     
    \caption{DDO216}
    \end{subfigure}  
    \quad
    \begin{subfigure}[a]{0.3\textwidth}
    \centering
    \includegraphics[width=5.0cm,angle=0]{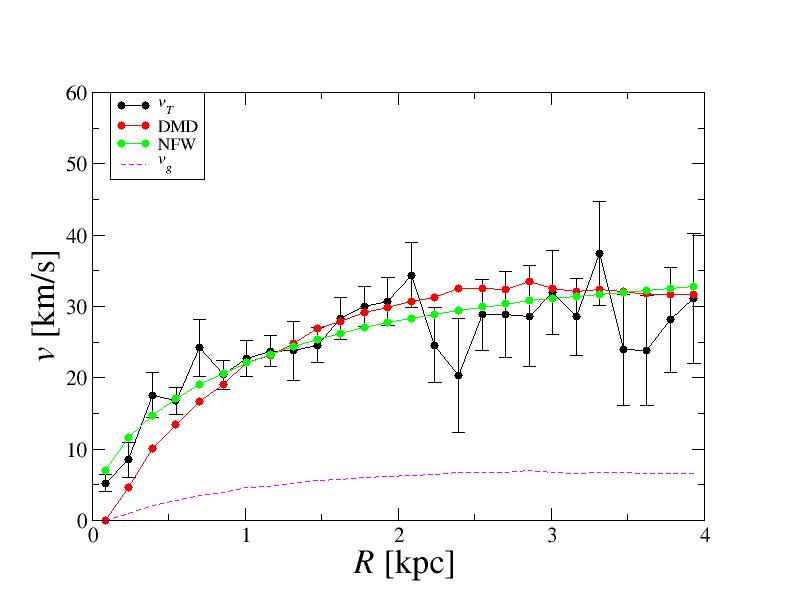}     
    \caption{F564-V3}
    \end{subfigure}  
    \quad
    \begin{subfigure}[a]{0.3\textwidth}
    \centering
    \includegraphics[width=5.0cm,angle=0]{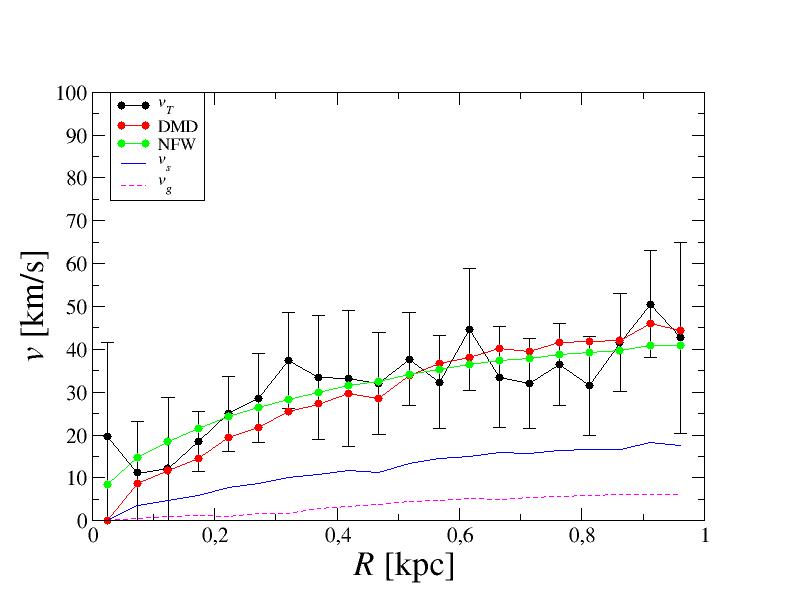}     
    \caption{IC10}
   \end{subfigure}  
        \quad
    \begin{subfigure}[a]{0.3\textwidth}
    \centering
    \includegraphics[width=5.0cm,angle=0]{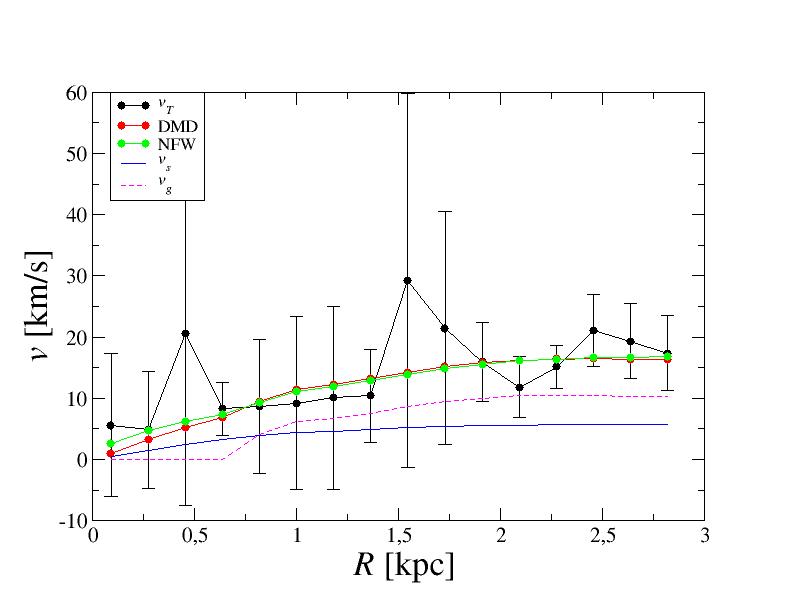}     
     \caption{IC1613}
    \end{subfigure}   
    \quad
     \begin{subfigure}[a]{0.3\textwidth}
     \centering
     \includegraphics[width=5.0cm,angle=0]{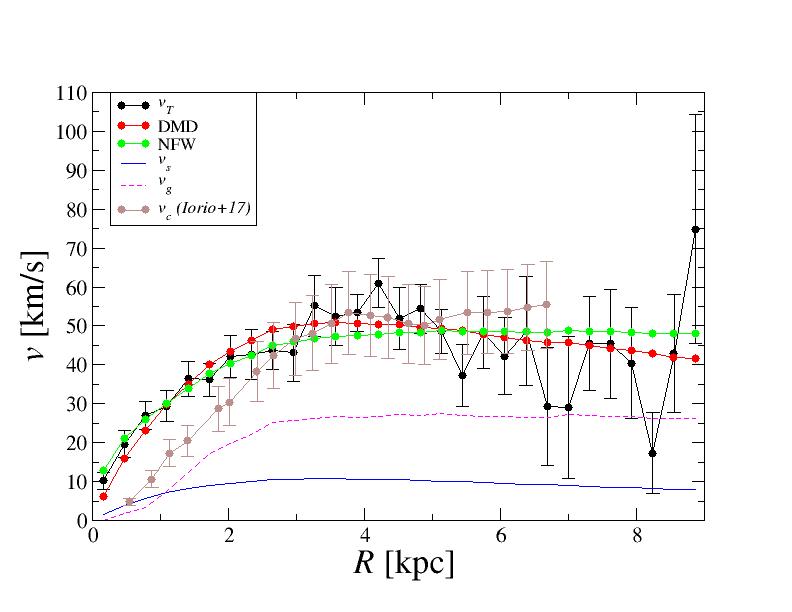}    
      \caption{NGC2366}
      \end{subfigure}  
      \quad
    \begin{subfigure}[a]{0.3\textwidth}
    \centering
    \includegraphics[width=5.0cm,angle=0]{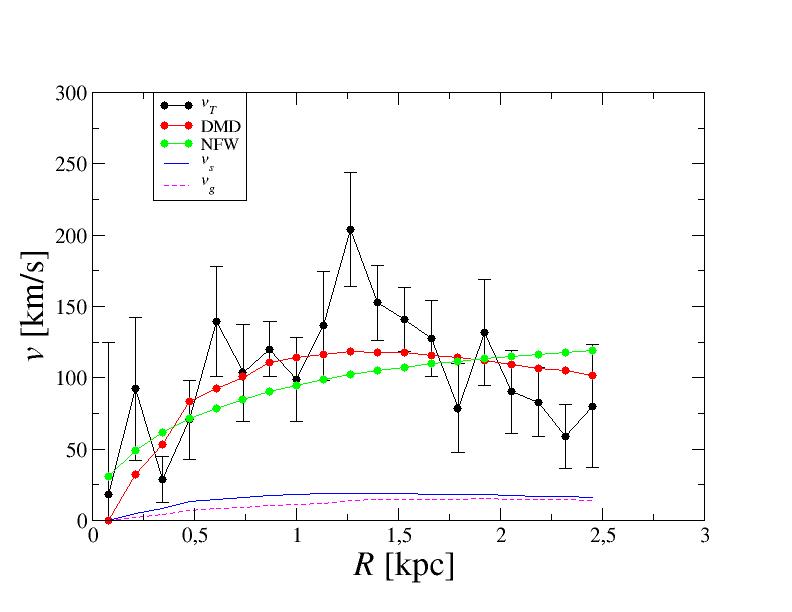}     
    \caption{NGC3738}
    \end{subfigure}     
    \quad
     \begin{subfigure}[a]{0.3\textwidth}
    \centering
    \includegraphics[width=5.0cm,angle=0]{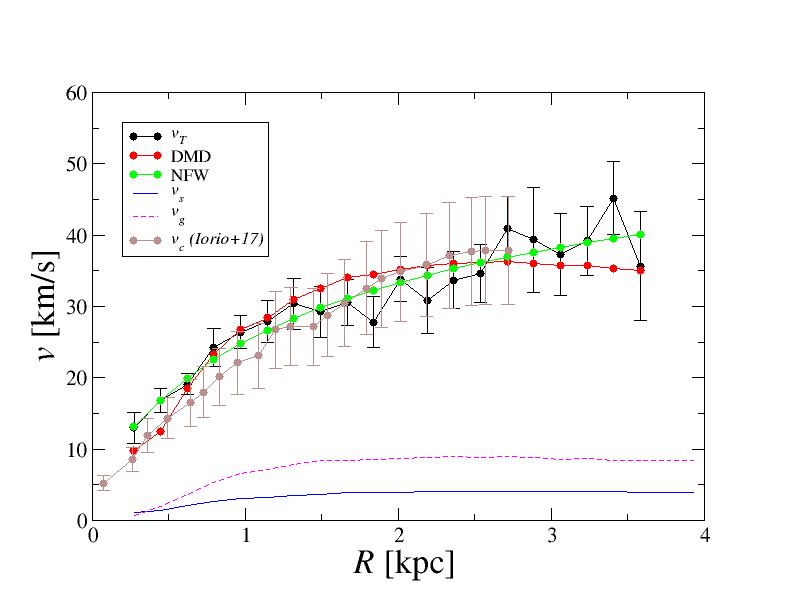}     
    \caption{WLM}
    \end{subfigure}
        \quad
    \caption{As Fig.\ref{fig:fit1} but for the remaining galaxies in the sample. 
      } 
    \label{fig:fit2}
    \end{figure*}

\section{Conclusions} 
\label{sect:conclusion}

The velocity fields of the 22 late-type dwarf  galaxies in the sub-sample of the LITTLE THINGS survey  \citep{Hunter_etal_2012}  that we have considered show highly perturbed characteristics. 
{ 
This is evident from the simple visual inspection of the observed 2D map of the line-of-sight velocity, $v_{\text{los}}$. 
Indeed, in almost all cases, the kinematic axis of the galaxy, i.e. the axis along which
 $v_{\text{los}}$ 
 has the largest velocity gradient
 changes orientation with radius, so the velocity map is not symmetrical with respect to the major axis of the galaxy's projected image, as would be expected for a flat, rotating disc.  
These irregularities in the observed velocity field are usually interpreted as resulting from local deformations of the disc, which, instead of being flat, is warped. While this is certainly a possible explanation for the observed maps, it may neglect the effects of non-circular motions. Warps are typically observed in the outermost regions of disc galaxies, whereas in the inner regions, they tend to be absent. Given the relatively small spatial extent of the dwarf galaxies in the subsample considered, it is thus possible that non-circular motions, rather than warps, are the primary source of the radial change in the orientation of the galaxy's kinematic axis. 

For this reason we have analyzed these galaxies} using the Velocity Ring Model \citep{SylosLabini_etal_2023_VRM}, which, under the assumption of a flat disc, allows for the reconstruction not only of the radial behaviors, averaged in rings, of the transversal and radial velocity components but also their coarse-grained spatial maps that can trace the distribution of anisotropies. While the Tilted Ring Model finds that some of the galaxies exhibit clear warping, thereby invalidating the assumption of a flat disc to describe their geometry, many lack the typical signature of a warp that manifests as a smooth radial dependence of the orientation angles with radial distance in the outermost regions of the galaxy. {  Rather, they show orientation angles characterized by significant fluctuations, which, in the VRM framework, correspond to velocity anisotropies.} Consequently, through examining the inner parts of the disc, we may infer that velocity anisotropies are generally significant, indicating the perturbed nature of the velocity fields in this sample of dwarf galaxies.

We then performed the best fit using two different mass models to the transversal velocity of these galaxies in the inner regions of the disc, where the assumption that they are rotationally supported appears to be well satisfied and where the geometric deformations due to a possible warp are generally negligible. 

The dark matter disc (DMD) model assumes,  consistently with the Bosma effect \citep{Bosma_1981},  that the distribution of DM is confined to the disc and follows the same exponential decay as the gas.  In the DMD model, two free parameters, $\gamma_s$ and $\gamma_g$, are introduced that correspond to how much DM is associated respectively with the stellar and gas components. This model was shown to well describe both the velocity field of the Milky Way \citep{SylosLabini_etal_2023_MW} and of the disc disc galaxies from the THINGS sample \citep{SylosLabini_etal_2024_Mass}. 

\begin{figure}
\includegraphics[width=8cm,angle=0]{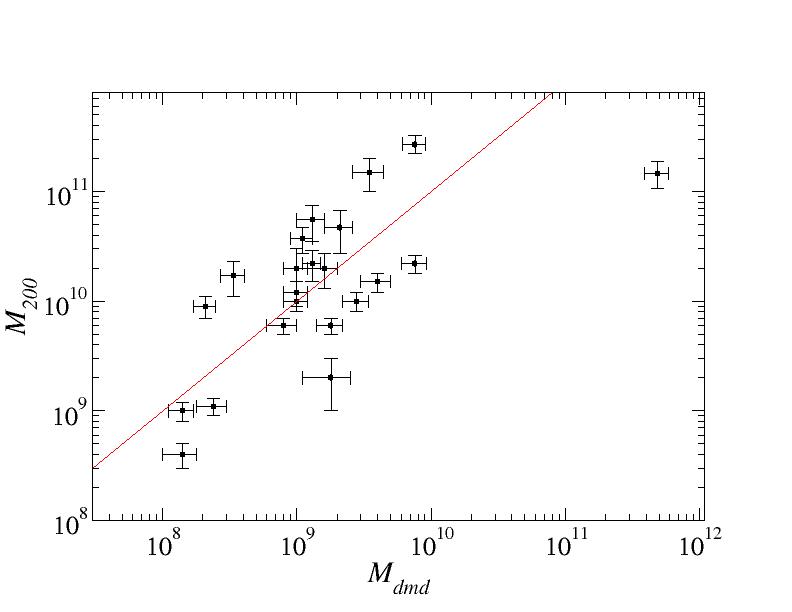}
\caption{  The virial mass estimated from the NFW mass model fit versus the disc mass obtained from the DMD model fit.
 The reference line corresponds to $M_{200} = 10 \times M_{dmd}$} 
\label{fig:M200Mdmd}
\end{figure}

The results of the DMD fits for the 22 dwarf galaxies in the sub-sample of the LITTLE THINGS survey that exhibit rotationally dominated inner regions \citep{Oh_etal_2015} are presented in Tab.\ref{table1}. It is worth noting that the total mass of the disc is at most 40 times that of the baryonic components. Tab.\ref{table1}  also includes the virial mass obtained by fitting the same rotation curves with a standard Navarro-Frenk-White (NFW) halo model \citep{Navarro_etal_1996}. One can observe that the virial mass is hundreds to thousands of times larger than the baryonic mass and much larger than the total mass in the DMD model. This result is not surprising since in the DMD model, the DM is confined to the disc and not spread out in a large spherical region around the baryonic disc. 
{ 
Indeed, the potential of a disc is stronger than that of a halo, so with less mass, it can provide the same contribution to the circular speed \citep{Binney_Tremaine_2008,Mancera_Pina_etal_2022}. Additionally, in the DMD model, the radius of the disc is determined by that of the neutral hydrogen (\HI{}) distribution, which is much smaller than the virial radius in the NFW model (i.e., the scale at which the density of the system is 200 times the average density of the universe). This difference makes the total mass in the DMD model much smaller than that required by the NFW model (see Fig.\ref{fig:M200Mdmd}).
}

One of the notable outcomes of the  DMD model is its ability to naturally address the core/cusp problem. Indeed, the standard DM halo profile predicts a density distribution that follows $\rho(r) \sim r^{-1}$ as $r$ approaches zero, leading to a cuspy central density profile. However, numerous observations have indicated that the rotational velocity profiles of galaxies correspond to a flat density profile in the central regions.
In the DMD model the mass density profile can be computed as the weighted sum of the stellar and \HI{}  surface density profiles, where the weights are the two parameters of the DMD fit, i.e.  $\gamma_s$ and $\gamma_g$. In general, both profiles are flat at small radii and so is the total  mass density profile, i.e. $\rho(R) \sim   \propto r^{-\alpha} \sim $ const., with $\alpha = 0$. In this situation it is expected that the rotational velocity increase linearly in function of the radius $v_c \propto r^{(2-\alpha)/2} \sim r$. This aligns with the findings of  \cite{Oh_etal_2015} and supports the notion that the DMD model can reproduce the observed linear rotational velocity profiles in the central regions of galaxies. Thus,  the DMD model provides a consistent explanation for the observed behavior of the rotational velocities, which is in contrast to the cuspy density profiles predicted by the standard NFW halo model.

Note that, the scales over which dwarf galaxy discs extend, typically not more than 5 kpc, are not directly relevant to cosmology. Therefore, estimating the DM  content for these objects does not have a direct impact on cosmological parameters.  The key question posed by the DMD model is to quantify the degree of flatness in the distribution of DM. In particular, it seeks to determine whether the DM distribution resembles a closer-to-spherical system or a flatter axially symmetric disc. This question cannot be directly answered by analyzing the projected images of galaxies, as this analysis only reveals the degree of compatibility between observed velocity and density fields and a given model. Instead, there are two types of observations that, in principle, can probe the degree of flatness in the overall mass distribution of certain disc galaxies. Firstly, the study of off-plane motions in our own Galaxy or neighboring galaxies can provide insights. Secondly, the analysis of high-precision gravitational lensing measurements of distant objects by nearby disc galaxies represents a powerful means to investigate the shape of a galaxy mass distribution. Forthcoming works will be devoted to the study of such issues.


\begin{acknowledgements}
{  
We thank Michael Joyce for valuable discussions. We also acknowledge an anonymous referee for their detailed comments and suggestions, which have significantly improved our presentation.
}
\end{acknowledgements}

%
 \bibliographystyle{aa} 
 

%
\section*{Appendix}


\subsection*{DDO43}

\begin{figure*}
 \quad
 \begin{subfigure}[a]{0.3\textwidth}
 \centering
\includegraphics[width=5.0cm,angle=0]{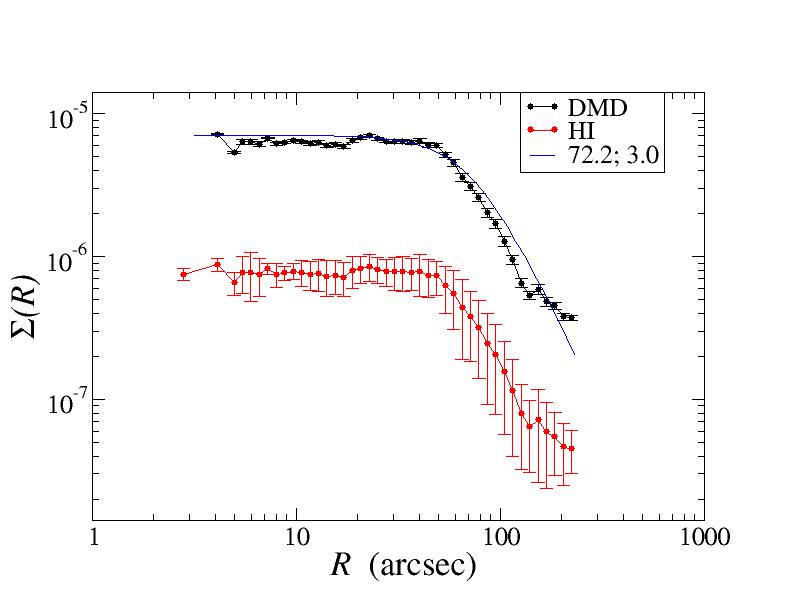}
\caption{}
\end{subfigure}
\quad
\begin{subfigure}[a]{0.3\textwidth}
\centering
\includegraphics[width=5.0cm,angle=0]{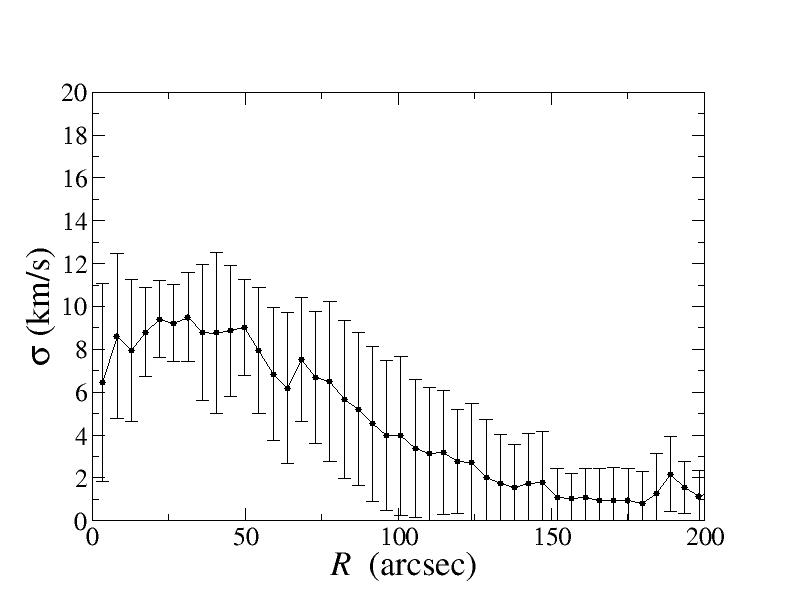}
\caption{}
\end{subfigure}
\quad
\begin{subfigure}[a]{0.3\textwidth}
\centering
\includegraphics[width=5.0cm,angle=0]{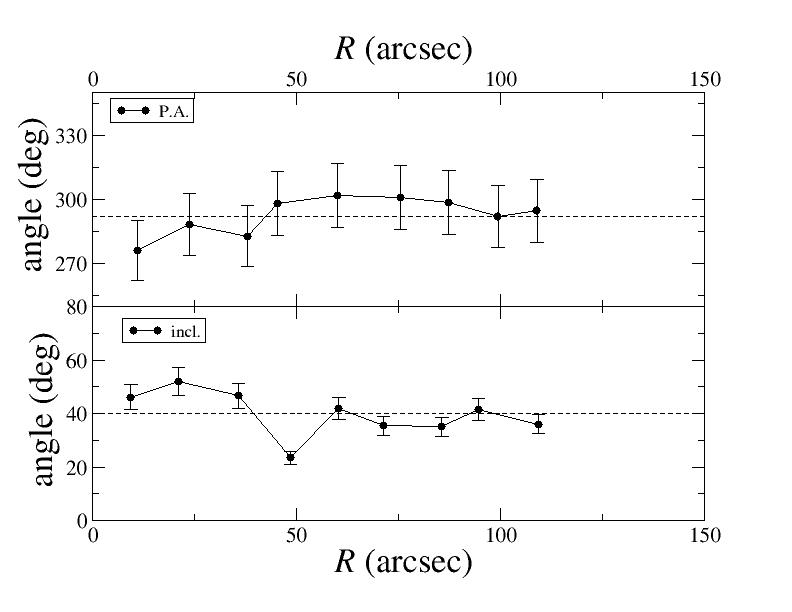}
\caption{}
\end{subfigure}
\quad
\caption{As Fig.\ref{fig:CVnIdwA2} but for DDO43} 
\label{fig:DDO43_2}
\end{figure*}

\begin{figure*}
\quad
\begin{subfigure}[a]{0.3\textwidth}
\centering
\includegraphics[width=5.0cm,angle=0]{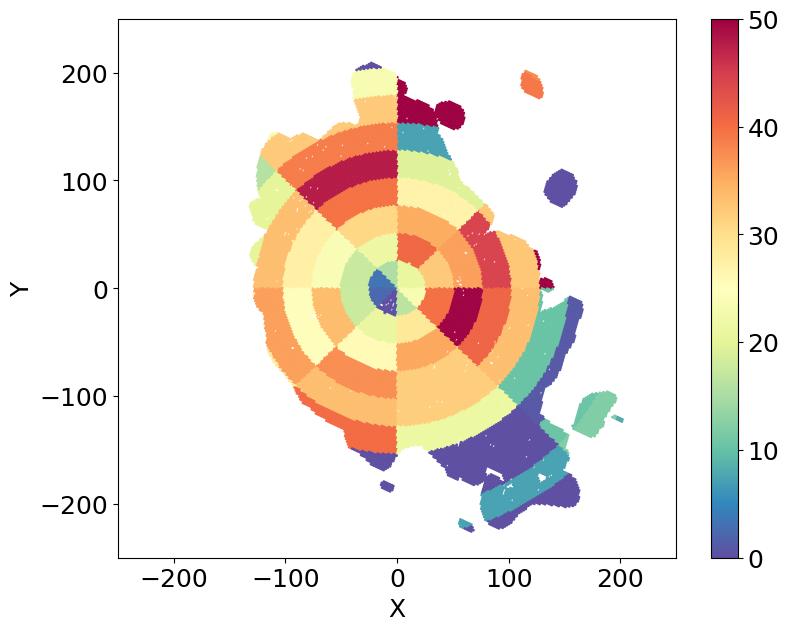}
\caption{}
\end{subfigure}
\quad
\begin{subfigure}[a]{0.3\textwidth}
\centering
\includegraphics[width=5.0cm,angle=0]{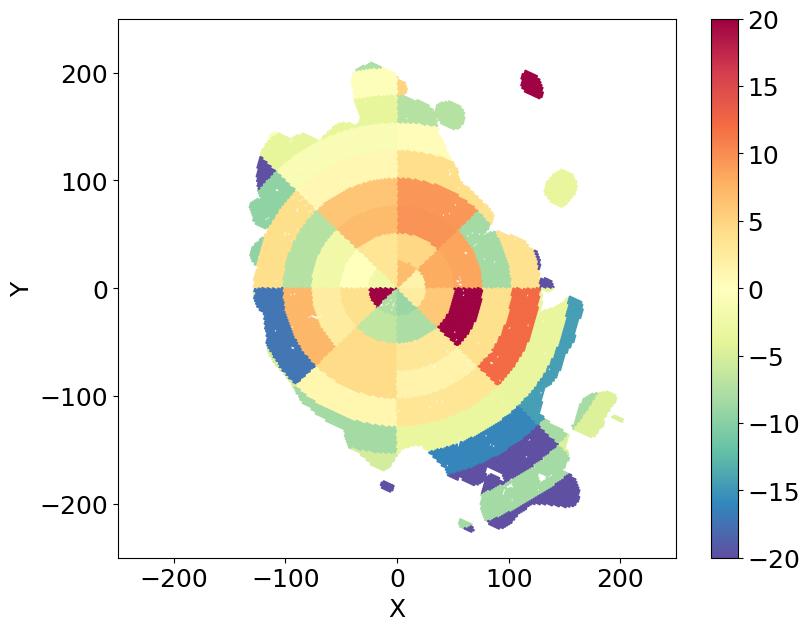}
\caption{}
\end{subfigure}
\quad
\begin{subfigure}[a]{0.3\textwidth}
\centering
\includegraphics[width=5.0cm,angle=0]{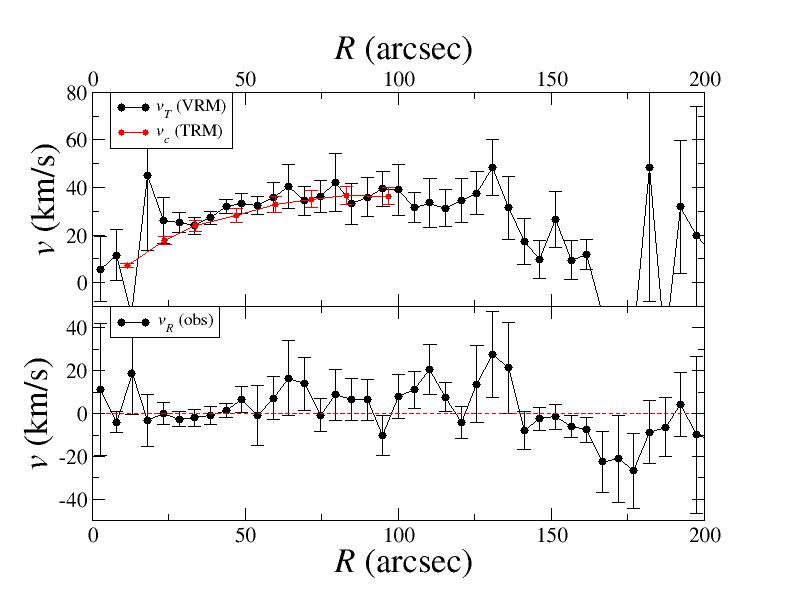}
\caption{}
\end{subfigure}
\quad
\caption{As Fig.\ref{fig:CVnIdwA3} but for DDO43} 
\label{fig:DDO43_3}
\end{figure*}

Figs.\ref{fig:DDO43_2}-\ref{fig:DDO43_3} show the results for DDO43. 
The orientation angles measured by the TRM fluctuate around a roughly constant value and thus we do not calculate the corrections to $v_\theta(R)$ and $v_R(R)$ due to a possible warp as these are negligible. 
The radial and transversal velocity component maps, reconstructed by the VRM, show relevant angular anisotropies. 
Data for the stellar components are not available and we  neglect it in the DMD fit: with only one free parameter, $\gamma_g$ (see Eq.\ref{dmd_fit}), we find a good DMD fit. 
  {However, we note that for this galaxy, as well as for others where the stellar component has negligible mass (i.e., DDO46, DDO47, and F564-V3), the DMD fit uses only one parameter, while the NFW fit uses two parameters. Thus, it is not surprising that, in the latter case, the reduced $\chi^2$ is smaller. In addition, it is worth noting that the transversal velocity component detected by the VRM is more fluctuating than the circular velocity measured by the VRM. This occurs because fluctuations in the former are perceived as variations in the orientation angles in the latter. This difference between the VRM and TRM results is clearly evident in all cases where the orientation angles vary as a function of scale.}

\subsection*{DDO46} 
\begin{figure*}
 \quad
 \begin{subfigure}[a]{0.3\textwidth}
 \centering
\includegraphics[width=5.0cm,angle=0]{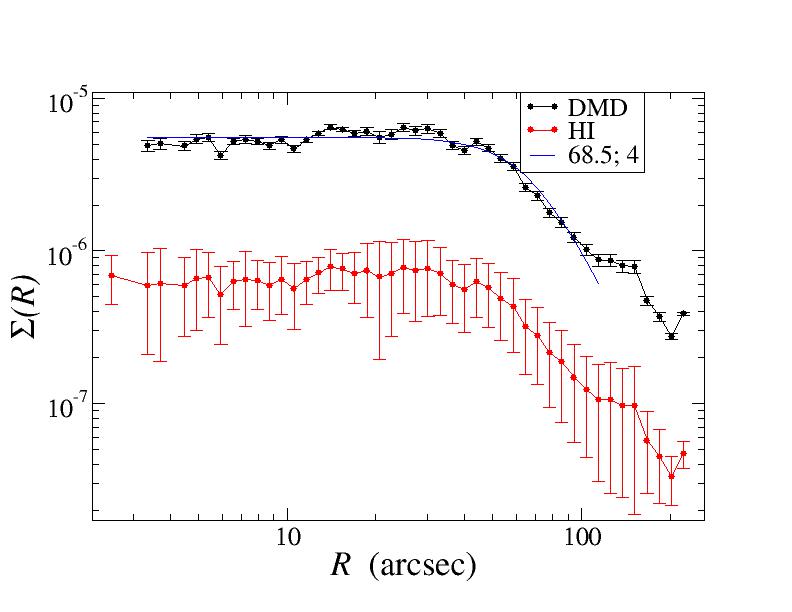}
\caption{}
\end{subfigure}
\quad
\begin{subfigure}[a]{0.3\textwidth}
\centering
\includegraphics[width=5.0cm,angle=0]{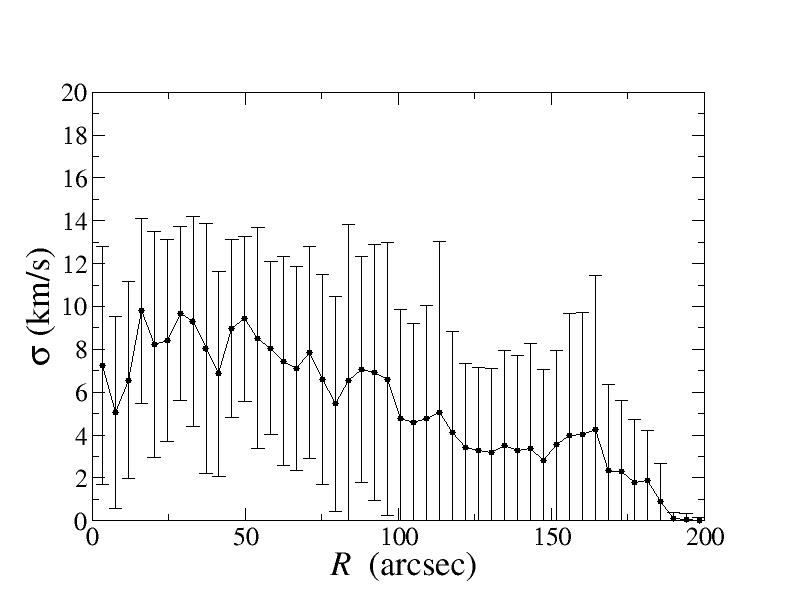}
\caption{}
\end{subfigure}
\quad
\begin{subfigure}[a]{0.3\textwidth}
\centering
\includegraphics[width=5.0cm,angle=0]{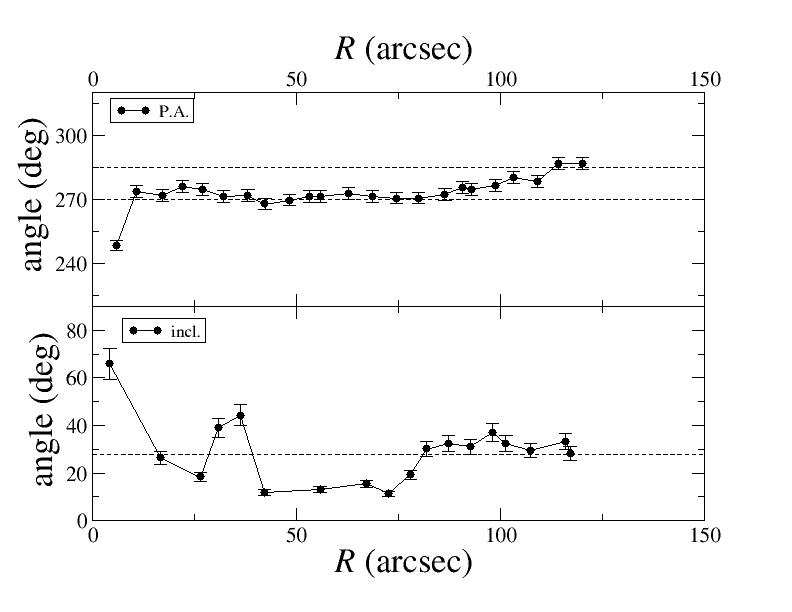}
\caption{}
\end{subfigure}
\quad
\caption{As Fig.\ref{fig:CVnIdwA2} but for DDO46.
} 
\label{fig:DDO46_2}
\end{figure*}

\begin{figure*}
\quad
\begin{subfigure}[a]{0.3\textwidth}
\centering
\includegraphics[width=5.0cm,angle=0]{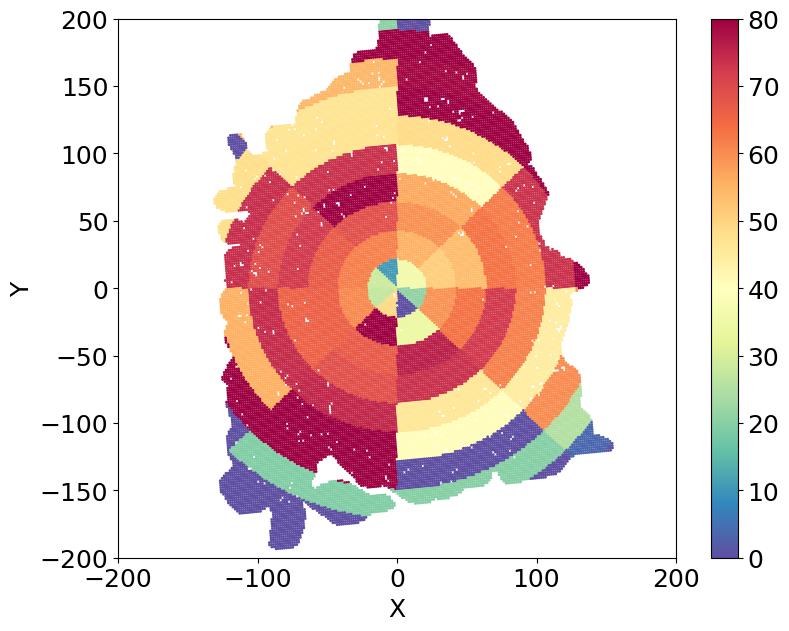}
\caption{}
\end{subfigure}
\quad
\begin{subfigure}[a]{0.3\textwidth}
\centering
\includegraphics[width=5.0cm,angle=0]{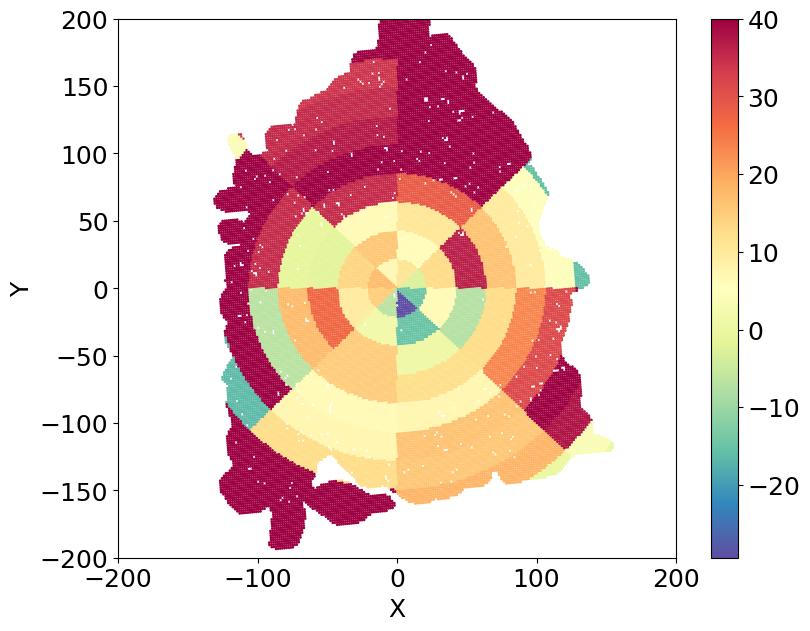}
\caption{}
\end{subfigure}
\quad
\begin{subfigure}[a]{0.3\textwidth}
\centering
\includegraphics[width=5.0cm,angle=0]{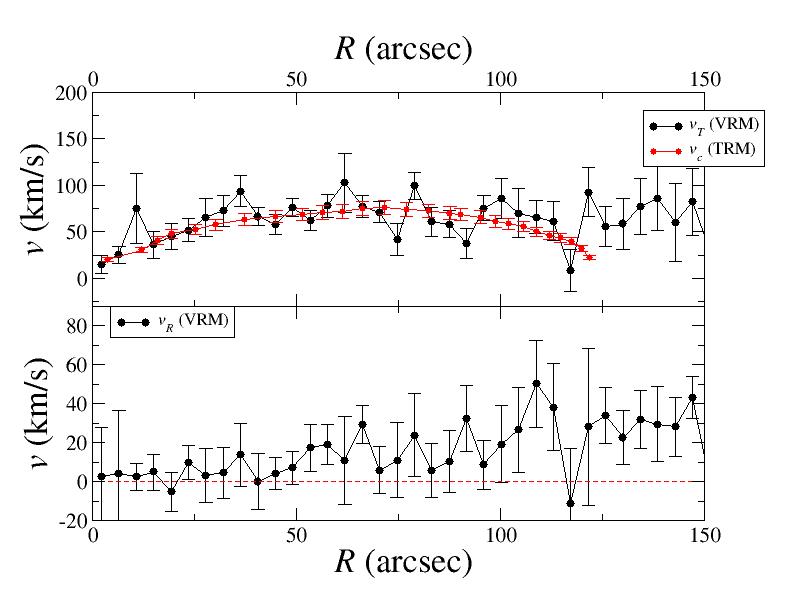}
\caption{}
\end{subfigure}
\quad
\caption{As Fig.\ref{fig:CVnIdwA3} but for DDO46.} 
\label{fig:DDO46_3}
\end{figure*}

DDO46 is similar to DDO43 (Fig.\ref{fig:DDO46_2}-\ref{fig:DDO46_3}): the orientation angles are roughly constant and the velocity field is characterized by relatively large angular anisotropies. The system is dominated by a radial outflow beyond 100" and we thus limit the analysis 
of the transversal component at smaller radii:  for $R>80"$ the effect of the warp, visible in the behavior of the PA, corresponds to the apparent growth of the mean radial velocity profile.  
Even in this case{  , as for DDO43,} there are no data for the stellar component and we make a single free parameter  best fit with the DMD. {  Note that the DMD fit is worse than the NFW fit: as in the case of DDO43, this is due to the fact that the DMD model has one fewer parameter. Additionally, it is worth noting that the agreement is poorer in the inner region of the disc, while in the outer parts, both models fit the observational data equally well.} 

\subsection*{DDO47} 
\begin{figure*}
 \quad
 \begin{subfigure}[a]{0.3\textwidth}
 \centering
\includegraphics[width=5.0cm,angle=0]{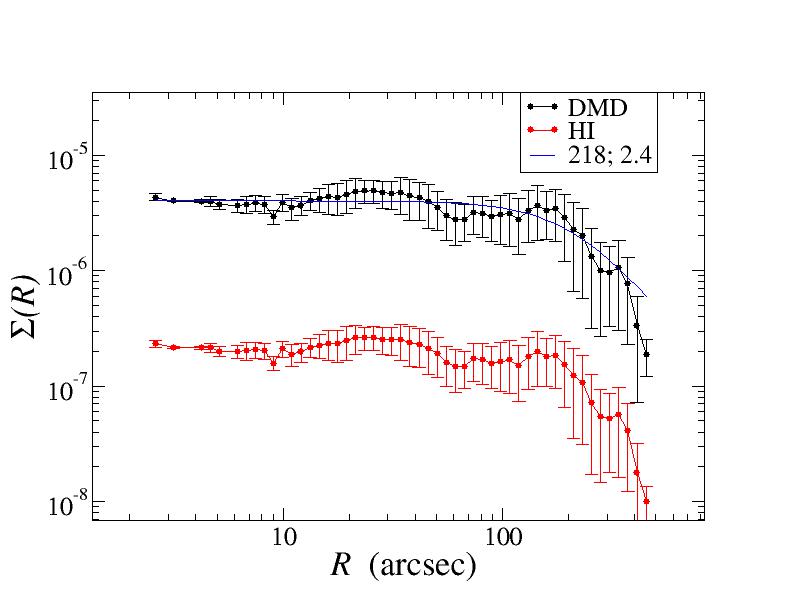}
\caption{}
\end{subfigure}
\quad
\begin{subfigure}[a]{0.3\textwidth}
\centering
\includegraphics[width=5.0cm,angle=0]{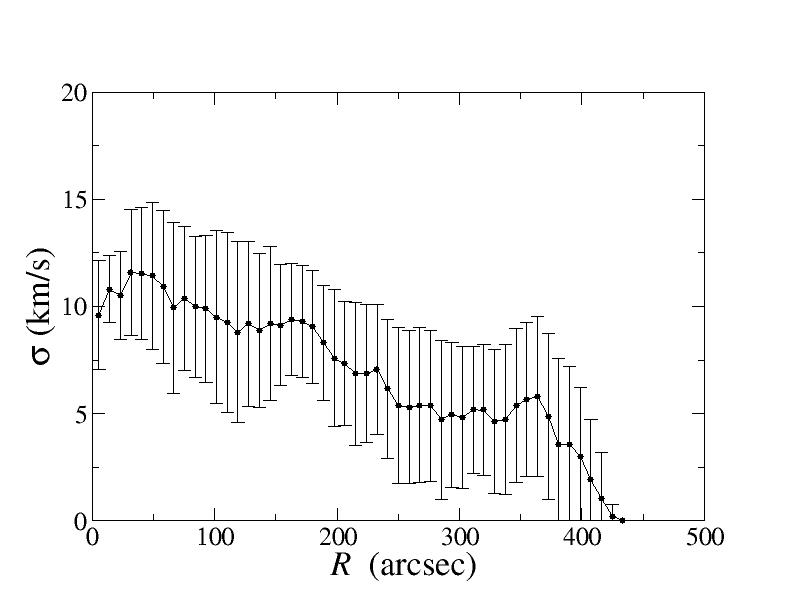}
\caption{}
\end{subfigure}
\quad
\begin{subfigure}[a]{0.3\textwidth}
\centering
\includegraphics[width=5.0cm,angle=0]{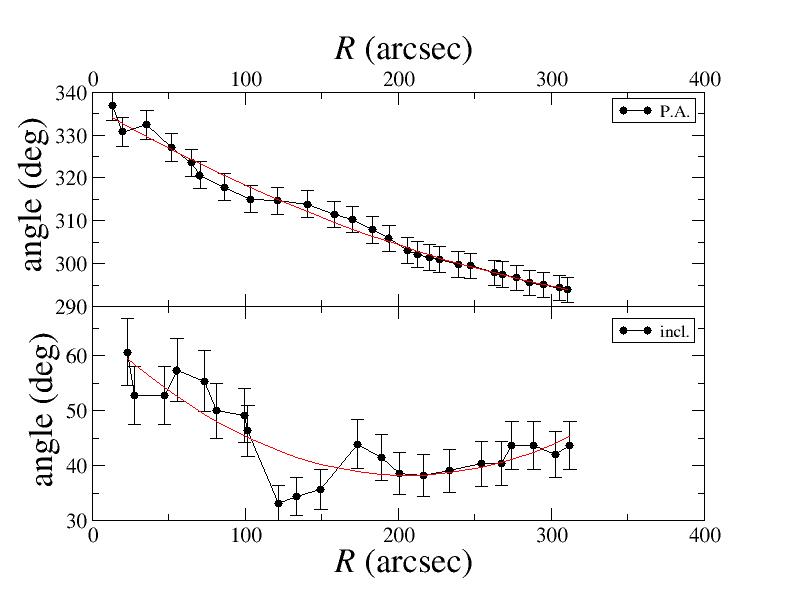}
\caption{}
\end{subfigure}
\quad
\caption{As Fig.\ref{fig:CVnIdwA2} but for DDO47.} 
\label{fig:DDO47_2}
\end{figure*}

\begin{figure*}
\quad
\begin{subfigure}[a]{0.3\textwidth}
\centering
\includegraphics[width=5.0cm,angle=0]{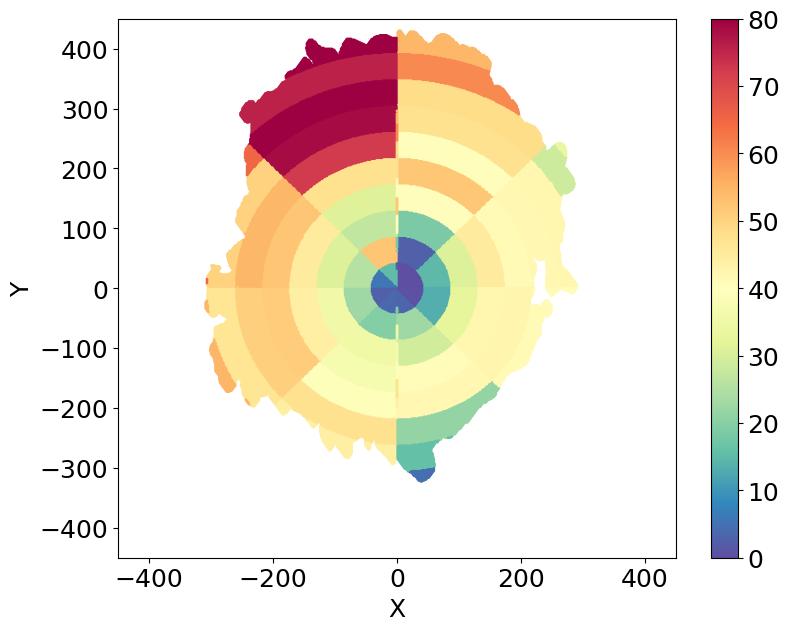}
\caption{}
\end{subfigure}
\quad
\begin{subfigure}[a]{0.3\textwidth}
\centering
\includegraphics[width=5.0cm,angle=0]{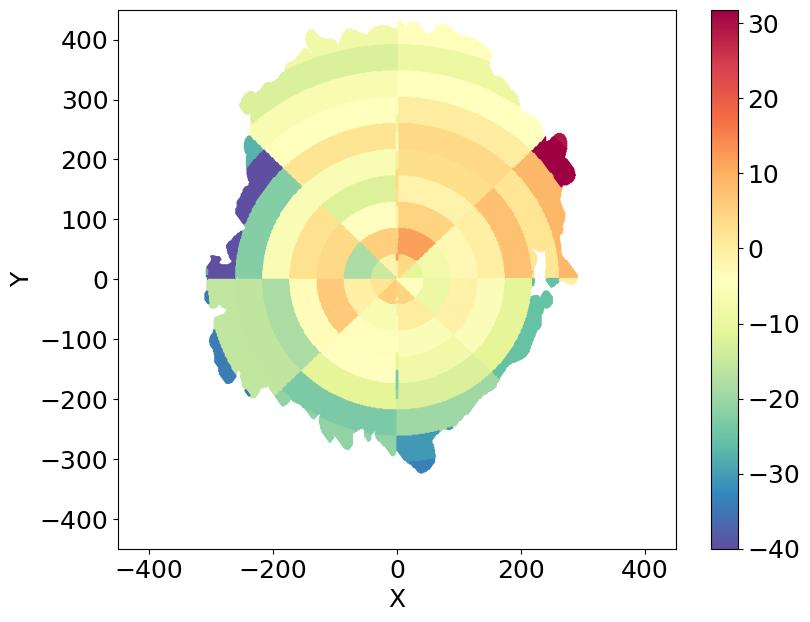}
\caption{}
\end{subfigure}
\quad
\begin{subfigure}[a]{0.3\textwidth}
\centering
\includegraphics[width=5.0cm,angle=0]{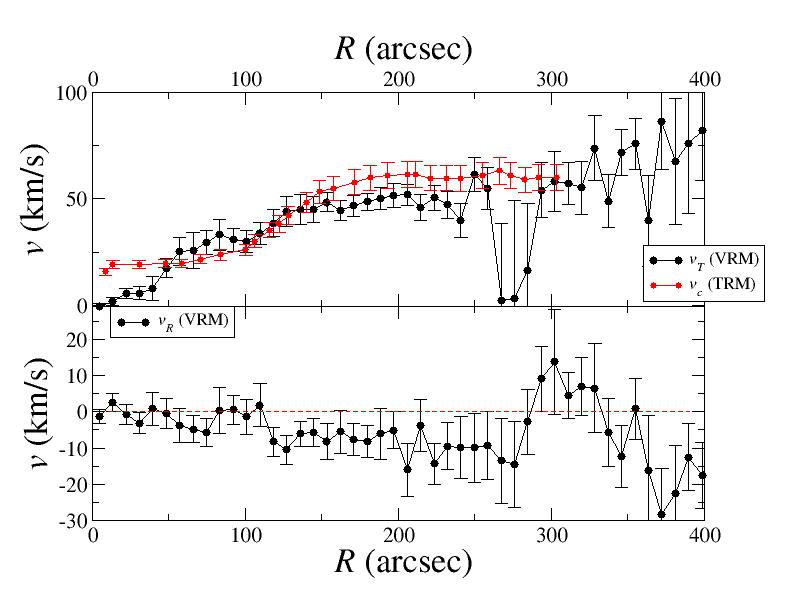}
\caption{}
\end{subfigure}
\quad
\caption{As Fig.\ref{fig:CVnIdwA3} but for DDO47.} 
\label{fig:DDO47_3}
\end{figure*}

The results for DDO47 are depicted in Figures \ref{fig:DDO47_2} to \ref{fig:DDO47_3}. In Figure \ref{fig:DDO47_2}(c), the orientation angles show a clear radial dependence, indicating the presence of a warp in the galaxy.
{  However, the results obtained with the VRM are similar to those obtained with the TRM in the inner disc where the fit with the theoretical models is done.}
Additionally, the velocity maps reveal significant perturbations in DDO47, corresponding to angular anisotropies, as shown in Figures \ref{fig:DDO47_3}(a) and \ref{fig:DDO47_3}(b). These spatial velocities increase in amplitude with distance from the galaxy's center, a behavior observed in other galaxies in our sample as well.
{  Even in this case, data for the stellar component are not available}. Consequently, we perform a fit using the DMD model with a single free parameter. Due to the limited number of adjustable parameters, the resulting fit may yield a relatively high reduced $\chi^2$ value. For the fitting process, we employ a smoothed fit to the gas velocity $v_g$ using an analytical exponential thin disc model approximation. This approach aims to regularize the behavior of the gas component. However, it is important to note that the absence of stellar component data and the simplicity of the fitting model limit the fit's ability to fully capture the system's intricacies, contributing to the relatively high reduced $\chi^2$ value.



\subsection*{DDO50} 
\begin{figure*}
 \quad
 \begin{subfigure}[a]{0.3\textwidth}
 \centering
\includegraphics[width=5.0cm,angle=0]{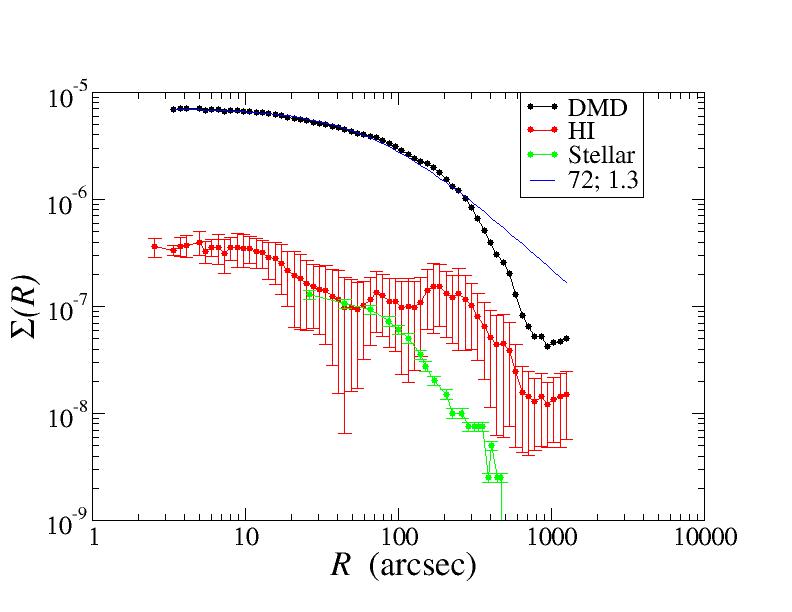}
\caption{}
\end{subfigure}
\quad
\begin{subfigure}[a]{0.3\textwidth}
\centering
\includegraphics[width=5.0cm,angle=0]{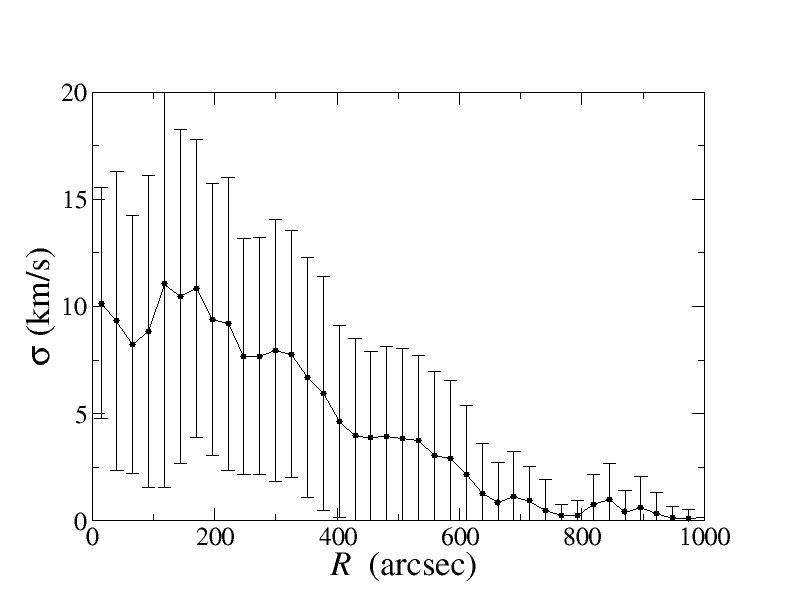}
\caption{}
\end{subfigure}
\quad
\begin{subfigure}[a]{0.3\textwidth}
\centering
\includegraphics[width=5.0cm,angle=0]{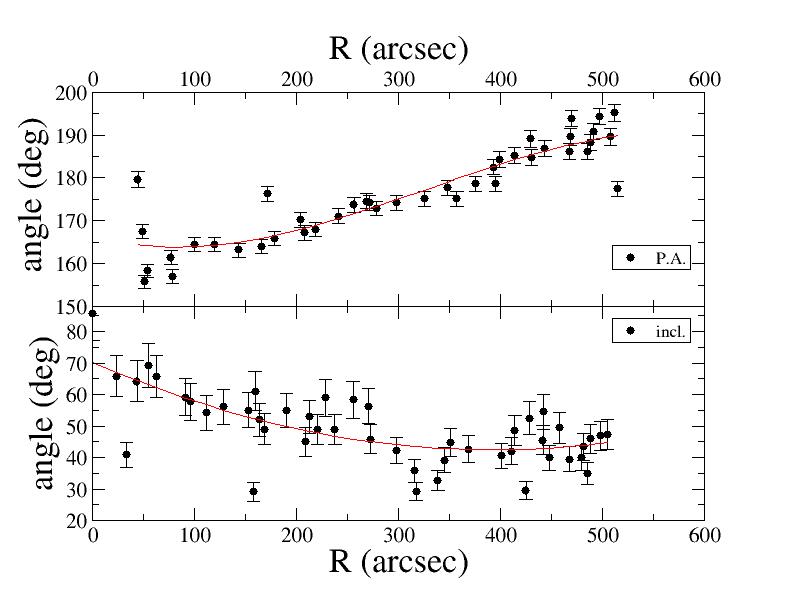}
\caption{}
\end{subfigure}
\quad
\caption{As Fig.\ref{fig:CVnIdwA2} but for DDO50
} 
\label{fig:DDO50_2}
\end{figure*}

\begin{figure*}
\quad
\begin{subfigure}[a]{0.3\textwidth}
\centering
\includegraphics[width=5.0cm,angle=0]{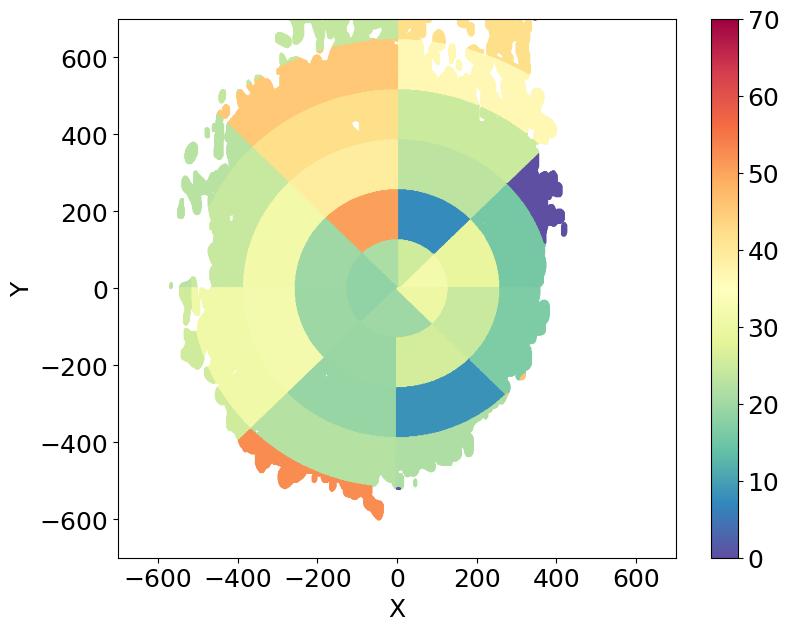}
\caption{}
\end{subfigure}
\quad
\begin{subfigure}[a]{0.3\textwidth}
\centering
\includegraphics[width=5.0cm,angle=0]{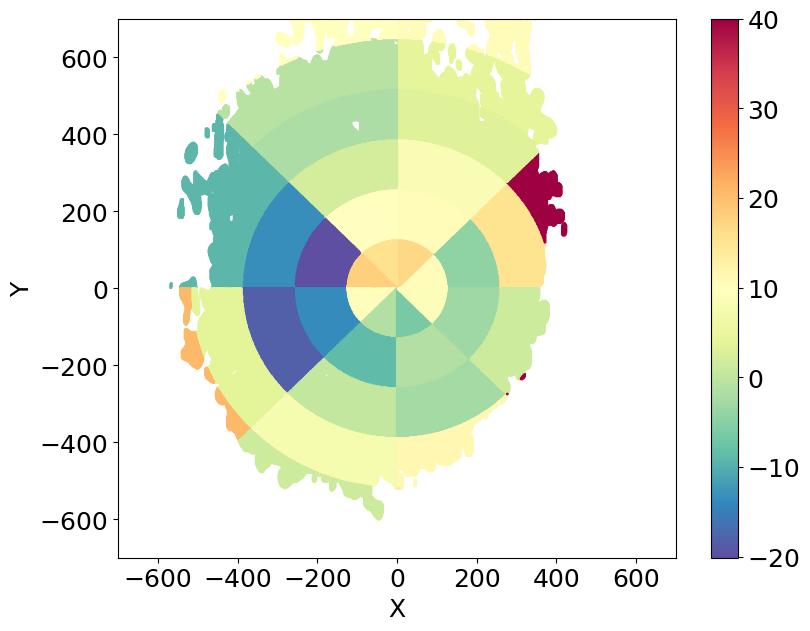}
\caption{}
\end{subfigure}
\quad
\begin{subfigure}[a]{0.3\textwidth}
\centering
\includegraphics[width=5.0cm,angle=0]{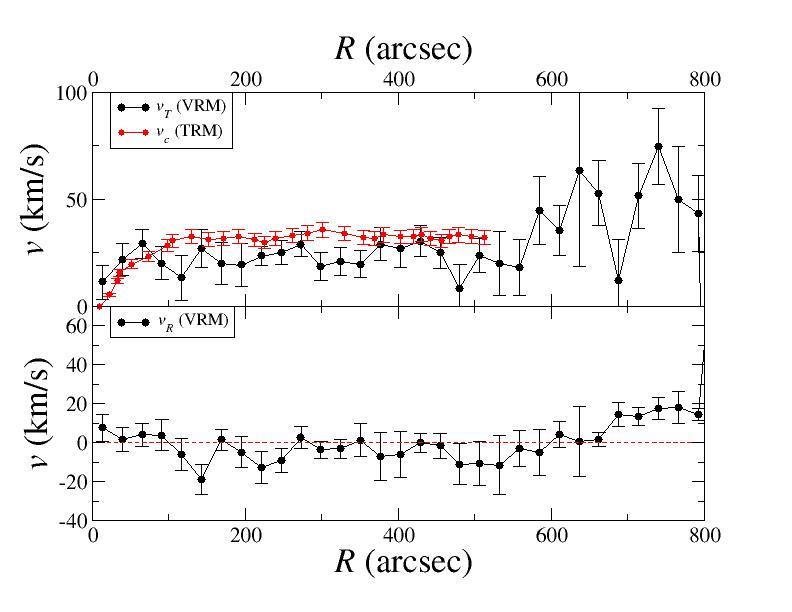}
\caption{}
\end{subfigure}
\quad
\caption{As Fig.\ref{fig:CVnIdwA3} but for DDO50.
} 
\label{fig:DDO50_3}
\end{figure*}

The orientation angles for DDO50 (Figs\ref{fig:DDO50_2}-\ref{fig:DDO50_3})  exhibit a smooth variation with radial distance, indicating the presence of a warp in the galaxy. 
{  
However, as in the case of DDO47, the results of the VRM and TRM are very close in the inner disc, indicating that the warp does not introduce significant distortions within this range of radial distances. The primary difference lies in the fact that the transversal velocity obtained by the VRM is more fluctuating. Note that we used an overall inclination angle of $i = 49.7^\circ$, whereas \citet{Iorio_etal_2017} found a smaller value, $i = 32^\circ$. This difference corresponds to a scaling factor in the rotation curve of $\sin(32^\circ)/\sin(49.7^\circ) \approx 0.7$
(in Fig.\ref{fig:fit1} the  \citet{Iorio_etal_2017}  have been rescaled by such a factor. Clearly, the total mass estimation is affected by this rescaling. However, the ratio between the NFW $M_{200}$ and $M_{\text{DMD}}$ remains unchanged.
}
The velocity maps of DDO50 reveal the presence of anisotropic patterns, characterized by localized velocity gradients. These patterns provide further insights into the dynamics of the galaxy, highlighting variations in the velocities across different regions. 
{  
The goodness of fit for the two mass models, NFW and DMD, is comparable, with the relatively high $\chi^2$ value ($\sim 2$) primarily attributable to the significant fluctuations characterizing the circular velocity.
}

\subsection*{DDO52} 
\begin{figure*}
 \quad
 \begin{subfigure}[a]{0.3\textwidth}
 \centering
\includegraphics[width=5.0cm,angle=0]{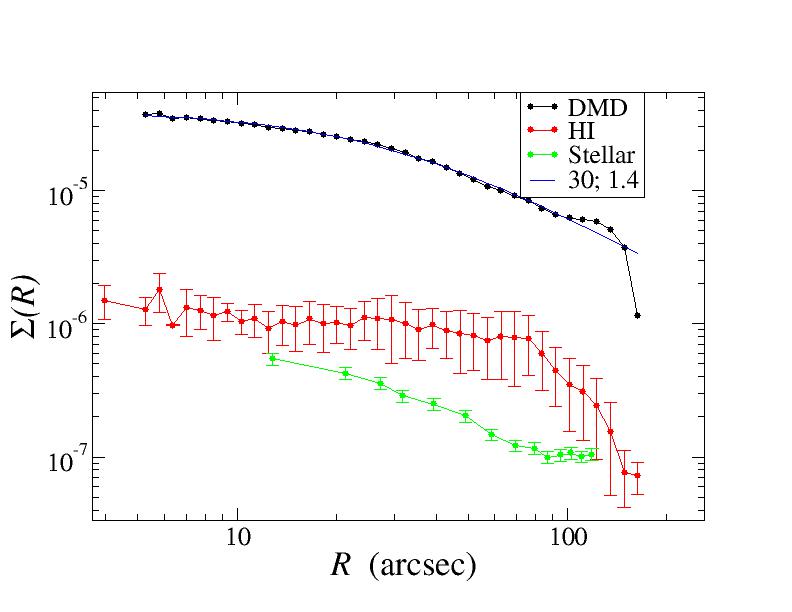}
\caption{}
\end{subfigure}
\quad
\begin{subfigure}[a]{0.3\textwidth}
\centering
\includegraphics[width=5.0cm,angle=0]{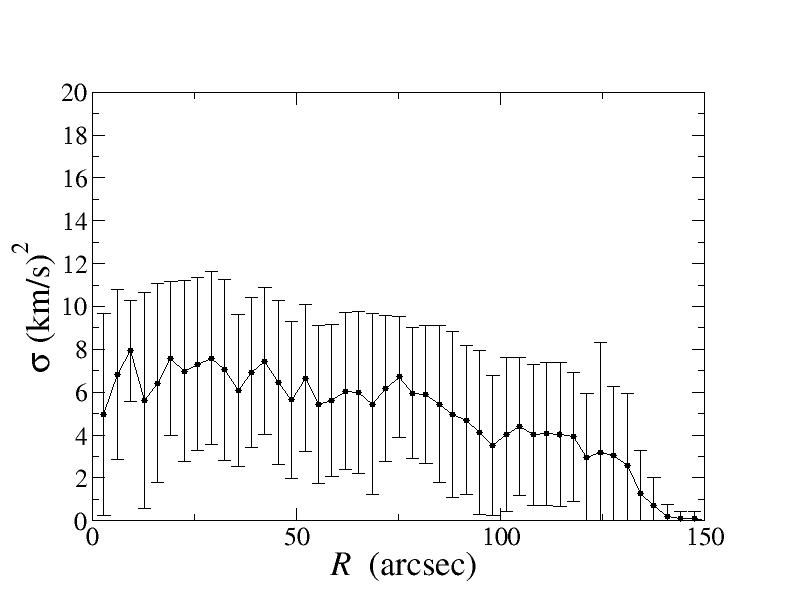}
\caption{}
\end{subfigure}
\quad
\begin{subfigure}[a]{0.3\textwidth}
\centering
\includegraphics[width=5.0cm,angle=0]{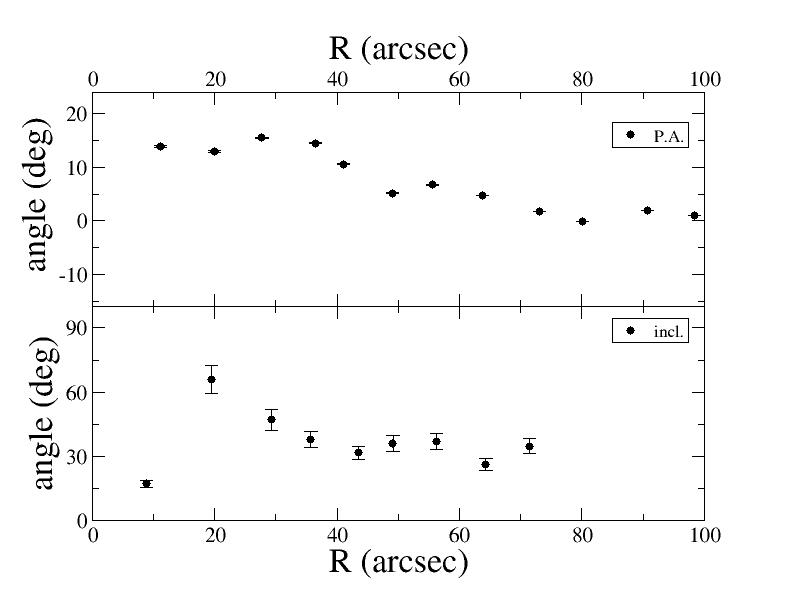}
\caption{}
\end{subfigure}
\quad
\caption{As Fig.\ref{fig:CVnIdwA2} but for DDO52.
} 
\label{fig:DDO52_2}
\end{figure*}

\begin{figure*}
\quad
\begin{subfigure}[a]{0.3\textwidth}
\centering
\includegraphics[width=5.0cm,angle=0]{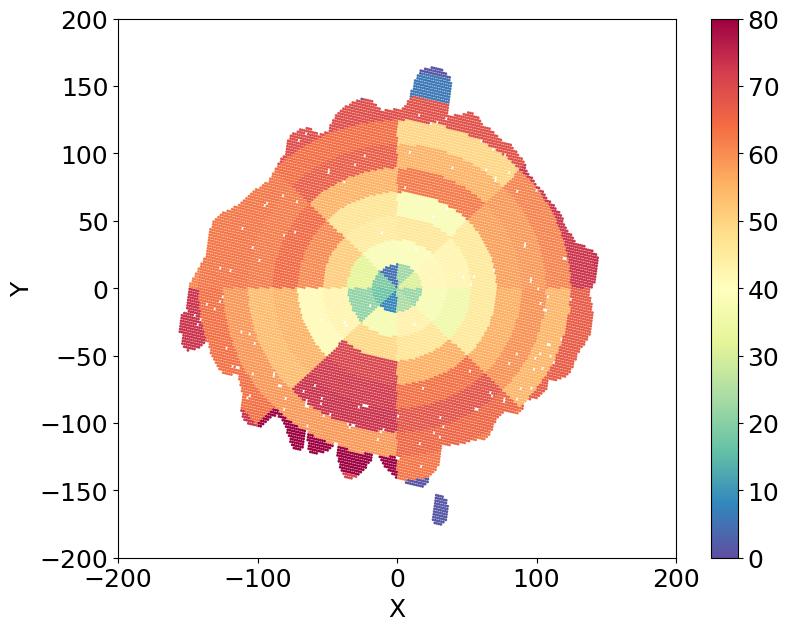}
\caption{}
\end{subfigure}
\quad
\begin{subfigure}[a]{0.3\textwidth}
\centering
\includegraphics[width=5.0cm,angle=0]{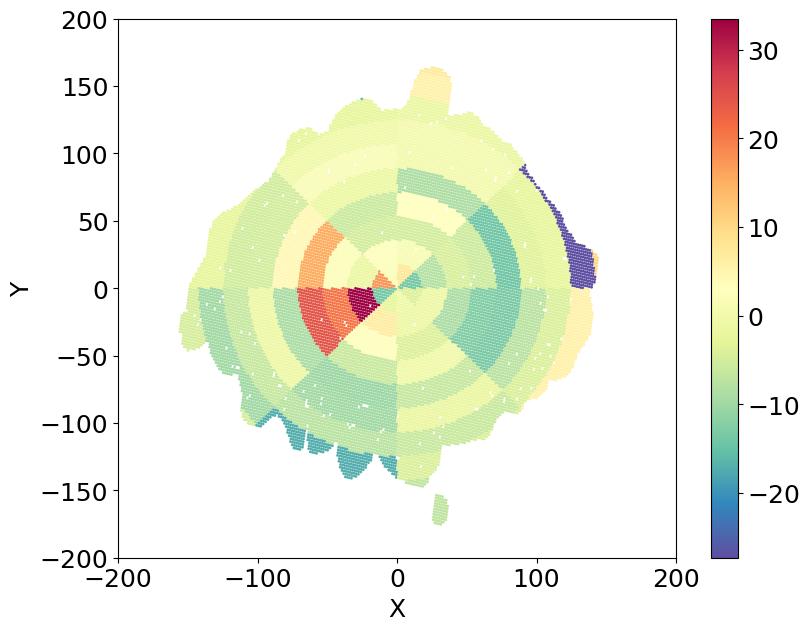}
\caption{}
\end{subfigure}
\quad
\begin{subfigure}[a]{0.3\textwidth}
\centering
\includegraphics[width=5.0cm,angle=0]{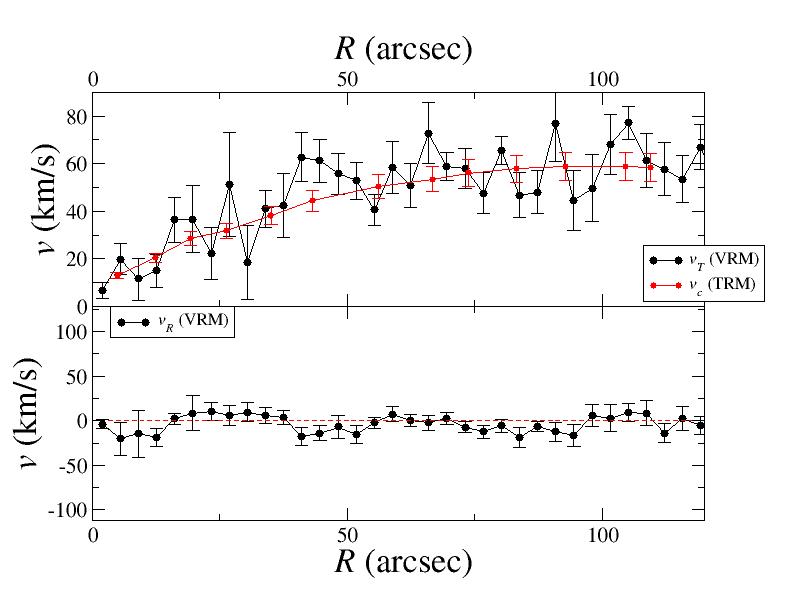}
\caption{}
\end{subfigure}
\quad
\caption{As Fig.\ref{fig:CVnIdwA3} but for DDO52.
} 
\label{fig:DDO52_3}
\end{figure*}

Both orientation angles of DDO52 (see Fig.\ref{fig:DDO52_2}-\ref{fig:DDO52_3}) exhibit limited variations in the outer regions of the system, while displaying more pronounced changes in the inner regions. This implies that the inner region of the galaxy is likely dominated by velocity gradients rather than a substantial change in the disc's geometry. 
{  Indeed, the velocity profile obtained by the VRM is more fluctuating than that measured by the TRM, but both exhibit similar behaviors: the transverse velocity profile detected by the VRM method closely aligns with the rotational velocity measured by the TRM method. This indicates a good agreement between the two methods in capturing the velocity behavior of the system.  On the other hand, the rotational velocity that we have found differs from that found by \citet{Iorio_etal_2017} because of the different inclination angle used, i.e.  $i = 43^\circ$ isetad of $i = 55^\circ$
(in Fig.\ref{fig:fit1} the  \citet{Iorio_etal_2017}  have been rescaled by a factor $\sin(55^\circ)/\sin(43^\circ)$.}
{  Furthermore, no significant net radial motions are observed; instead, large fluctuations are present, particularly in the outermost regions. Fluctuations in the velocity component profiles correspond to localized velocity anisotropies, which are evident across the entire disc, as shown in the 2D maps. These fluctuations may be attributed to various factors influencing the system's dynamics.}
{  Finally, we note that the fits with both models yield a similar reduced $\chi^2 \sim 1$, which remains relatively high due to the fluctuations in the velocity profile.} 

\subsection*{DDO53} 
\begin{figure*}
 \quad
 \begin{subfigure}[a]{0.3\textwidth}
 \centering
\includegraphics[width=5.0cm,angle=0]{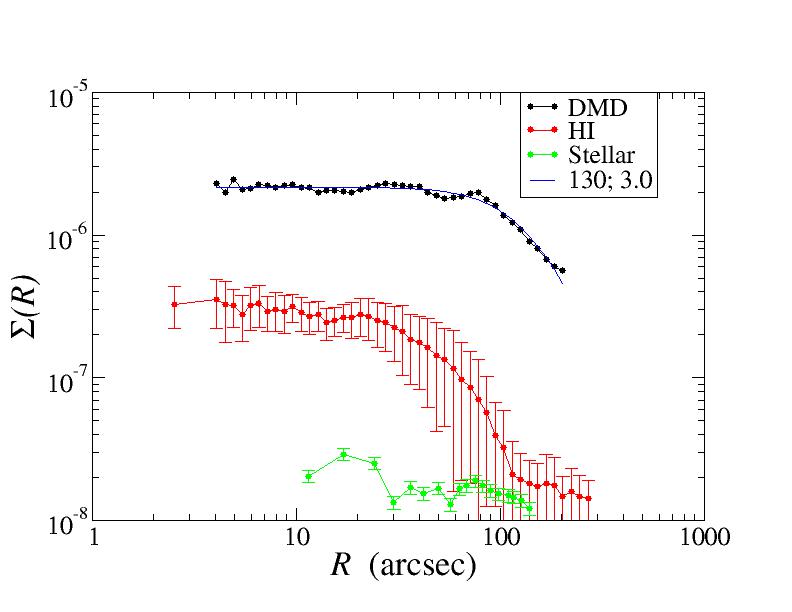}
\caption{}
\end{subfigure}
\quad
\begin{subfigure}[a]{0.3\textwidth}
\centering
\includegraphics[width=5.0cm,angle=0]{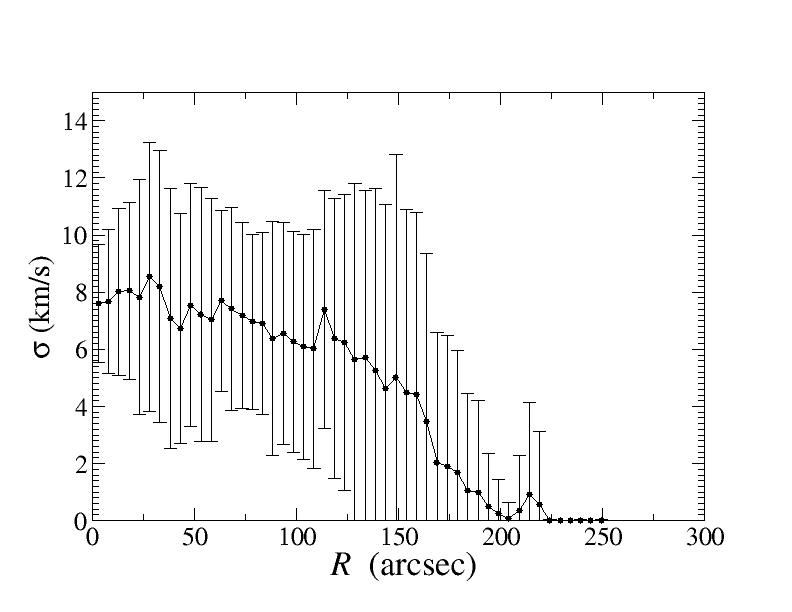}
\caption{}
\end{subfigure}
\quad
\begin{subfigure}[a]{0.3\textwidth}
\centering
\includegraphics[width=5.0cm,angle=0]{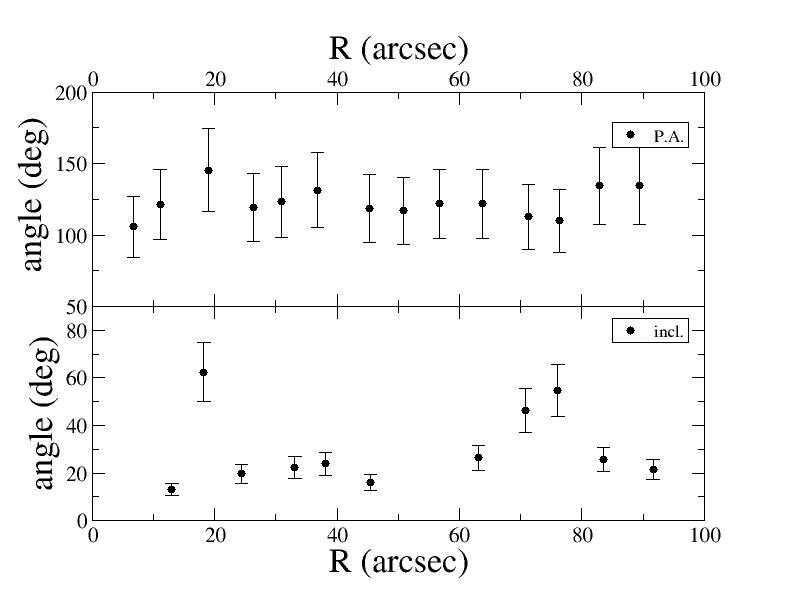}
\caption{}
\end{subfigure}
\quad
\caption{As Fig.\ref{fig:CVnIdwA2} but for  DDO53
} 
\label{fig:DDO53_2}
\end{figure*}

\begin{figure*}
\quad
\begin{subfigure}[a]{0.3\textwidth}
\centering
\includegraphics[width=5.0cm,angle=0]{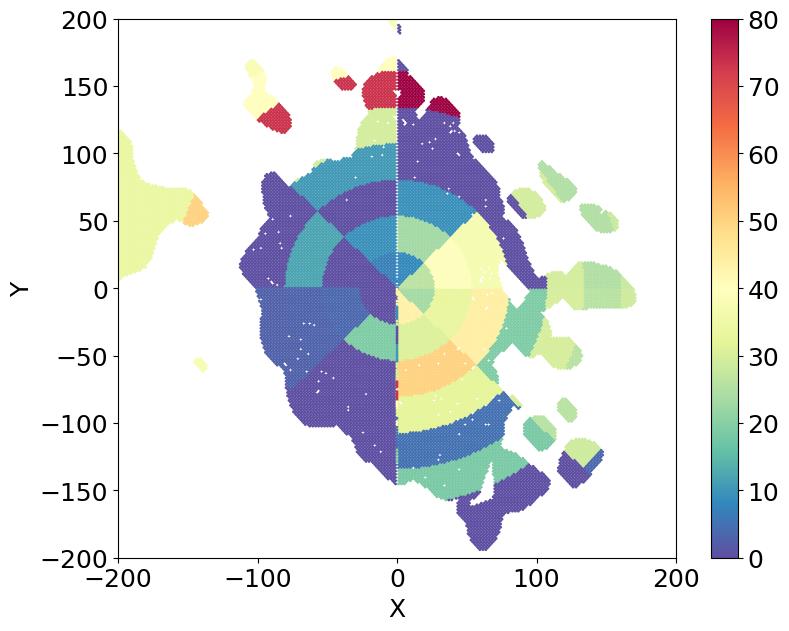}
\caption{}
\end{subfigure}
\quad
\begin{subfigure}[a]{0.3\textwidth}
\centering
\includegraphics[width=5.0cm,angle=0]{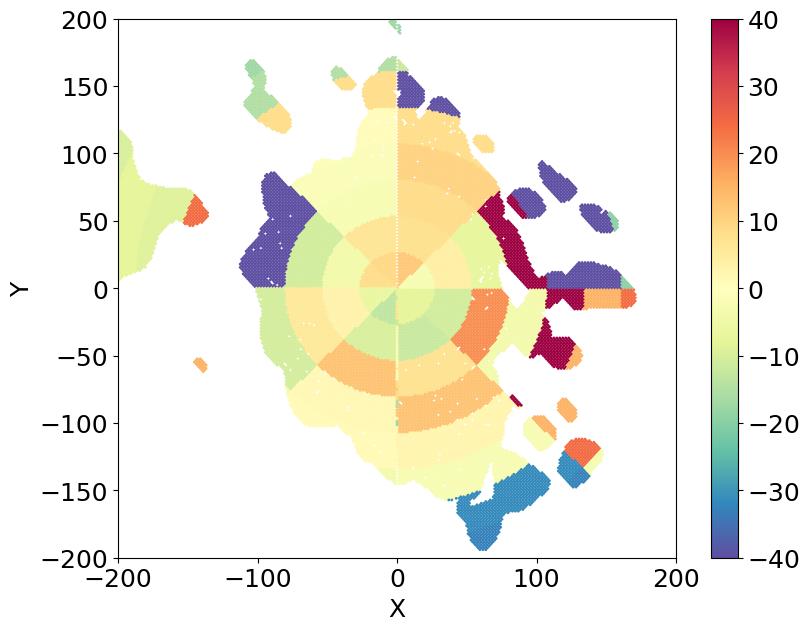}
\caption{}
\end{subfigure}
\quad
\begin{subfigure}[a]{0.3\textwidth}
\centering
\includegraphics[width=5.0cm,angle=0]{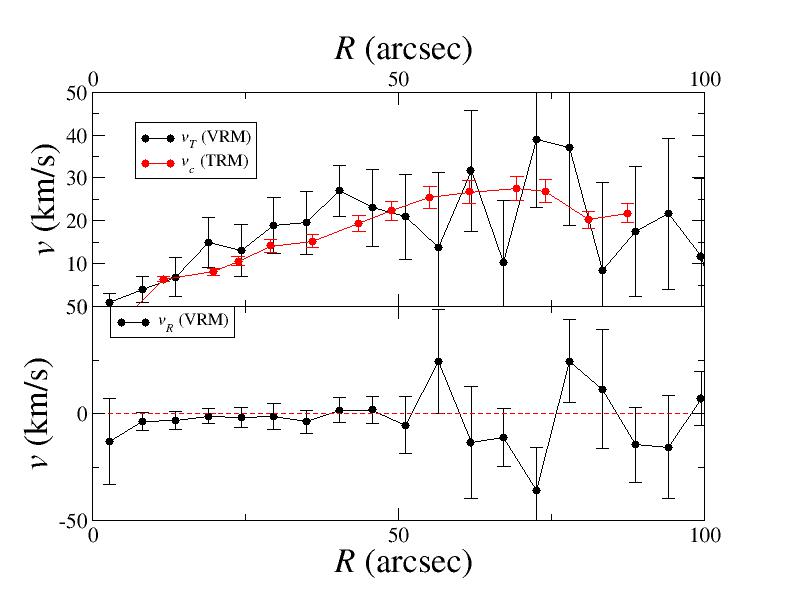}
\caption{}
\end{subfigure}
\quad
\caption{As Fig.\ref{fig:CVnIdwA3} but for  DDO53
} 
\label{fig:DDO53_3}
\end{figure*}

The orientation angles of DDO53 exhibit a roughly constant trend with respect to the radius, although they display a fluctuating behavior. The analysis conducted using the VRM interprets that these fluctuations correspond to velocity anisotropies within the galaxy. {  Indeed, the velocity profile obtained using the VRM is noisier than that measured with the TRM but showing the same trend:  this suggests a reasonable agreement between the two methods in capturing the overall velocity behavior of the system.
Note that in this case we used the inclination angle $i=27^\circ$  from \cite{Oh_etal_2015} which is smaller than that  determined by \citet{Iorio_etal_2017}  (i.e., $i=39^\circ$) and thus Fig.\ref{fig:fit1} the  \citet{Iorio_etal_2017}  result have been rescaled. 
.} 
DDO53 demonstrates an irregular LOS velocity field, as well as an \HI\ intensity map that exhibits local fluctuations (refer to Figs.\ref{fig:DDO53_2}-\ref{fig:DDO53_3}). Additionally, the LOS velocity dispersion map displays several localized fluctuations as well. These observations indicate complex dynamics within the galaxy, with variations in velocities and intensity across different regions.
{   Regarding the mass model fits, both models show good agreement, with a reduced $\chi^2 \sim 0.5$.}

\subsection*{DDO70} 

\begin{figure*}
 \quad
 \begin{subfigure}[a]{0.3\textwidth}
 \centering
\includegraphics[width=5.0cm,angle=0]{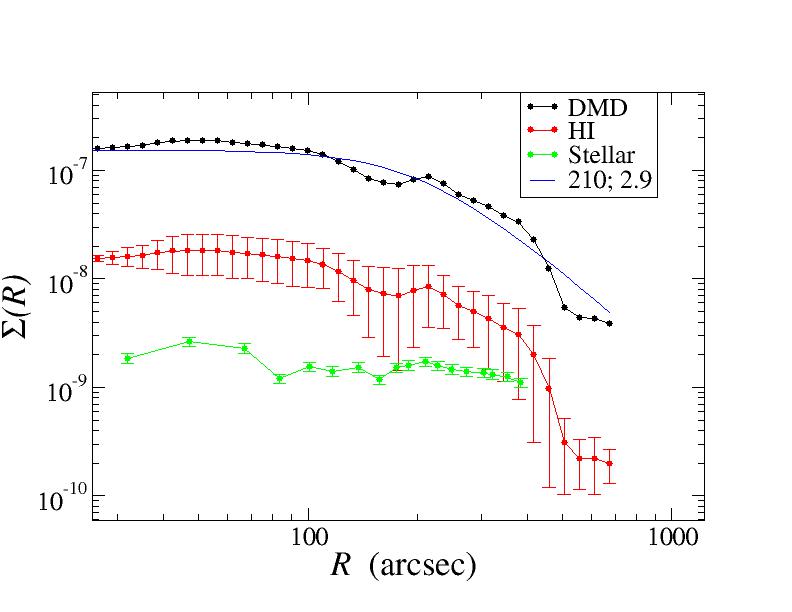}
\caption{}
\end{subfigure}
\quad
\begin{subfigure}[a]{0.3\textwidth}
\centering
\includegraphics[width=5.0cm,angle=0]{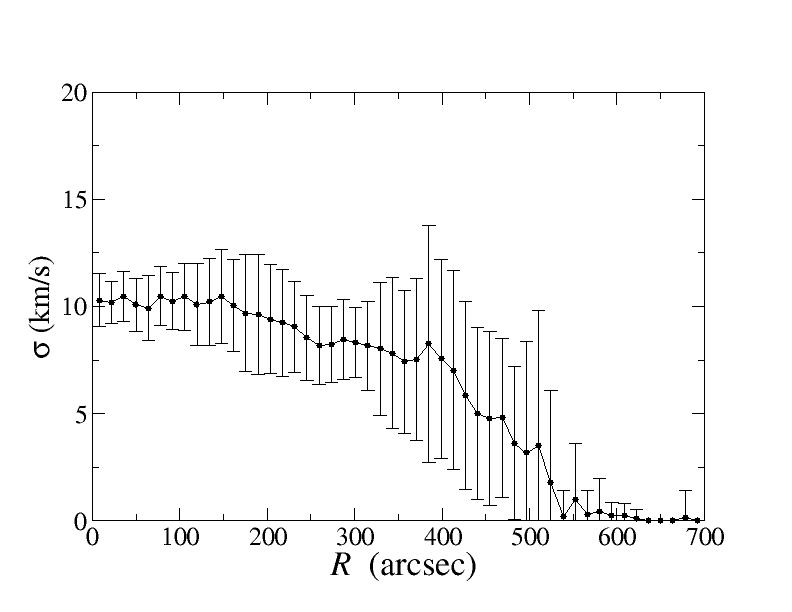}
\caption{}
\end{subfigure}
\quad
\begin{subfigure}[a]{0.3\textwidth}
\centering
\includegraphics[width=5.0cm,angle=0]{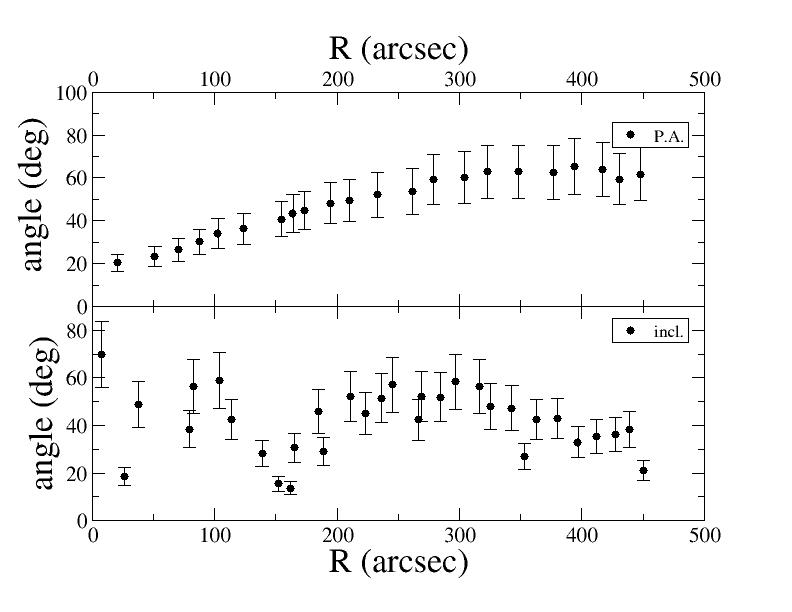}
\caption{}
\end{subfigure}
\quad
\caption{As Fig.\ref{fig:CVnIdwA2} but for  DDO70
} 
\label{fig:DDO70_2}
\end{figure*}

\begin{figure*}
\quad
\begin{subfigure}[a]{0.3\textwidth}
\centering
\includegraphics[width=5.0cm,angle=0]{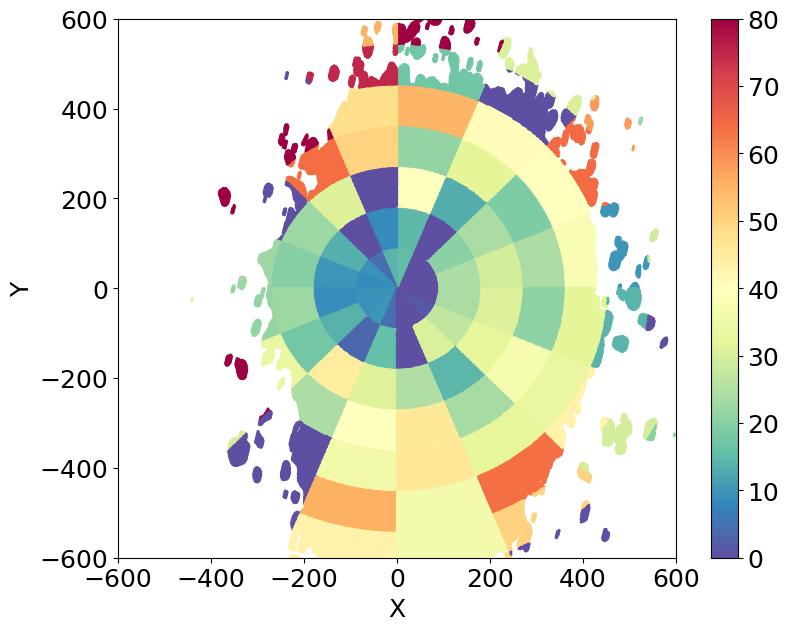}
\caption{}
\end{subfigure}
\quad
\begin{subfigure}[a]{0.3\textwidth}
\centering
\includegraphics[width=5.0cm,angle=0]{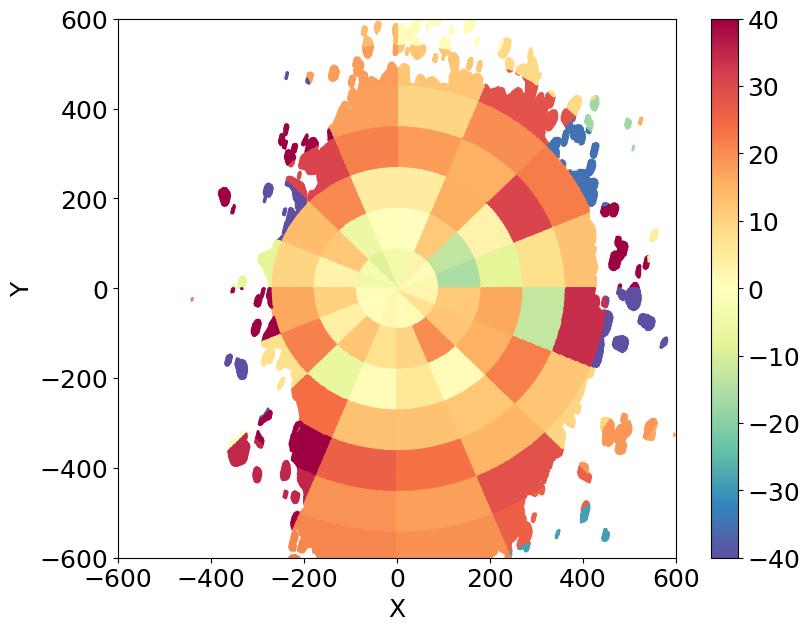}
\caption{}
\end{subfigure}
\quad
\begin{subfigure}[a]{0.3\textwidth}
\centering
\includegraphics[width=5.0cm,angle=0]{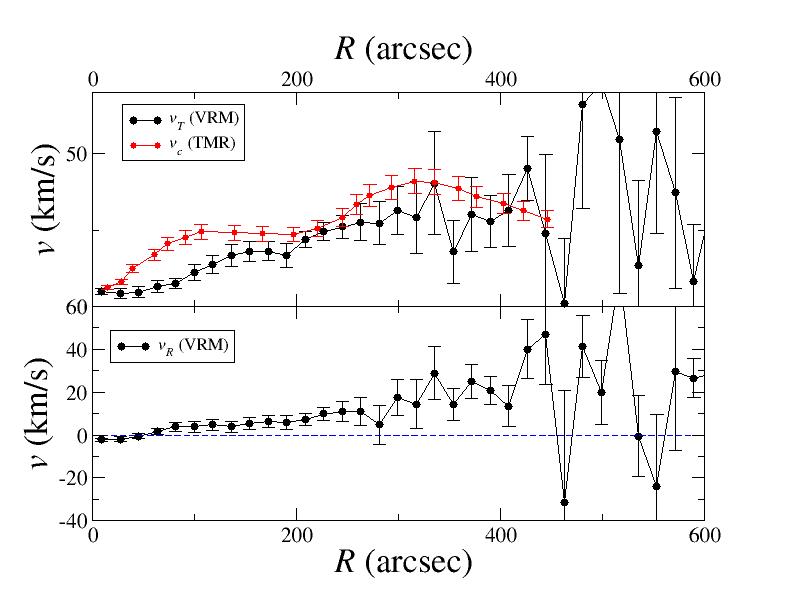}
\caption{}
\end{subfigure}
\quad
\caption{As Fig.\ref{fig:CVnIdwA3} but for  DDO70
} 
\label{fig:DDO70_3}
\end{figure*}

The presence of a warp in DDO70 is evident from the smooth variation of the orientation angles from the inner to the outer regions (refer to Figures \ref{fig:DDO70_2} to \ref{fig:DDO70_3}). %
In addition to the anisotropies induced by the warp in the reconstructed velocity maps obtained through the VRM method, we observe large anisotropies in the external regions that exhibit a greater amplitude compared to those in the inner regions. This observation is consistent with the structures visible in the \HI\ intensity map and the LOS velocity dispersion map.
{  The two mass models, i.e., the DMD and the NFW, exhibit similar $\chi^2 \sim 2$ values, with the relatively high value attributed to large fluctuations affecting the profile of the circular velocity component.} 

\subsection*{DDO87} 
\begin{figure*}
 \quad
 \begin{subfigure}[a]{0.3\textwidth}
 \centering
\includegraphics[width=5.0cm,angle=0]{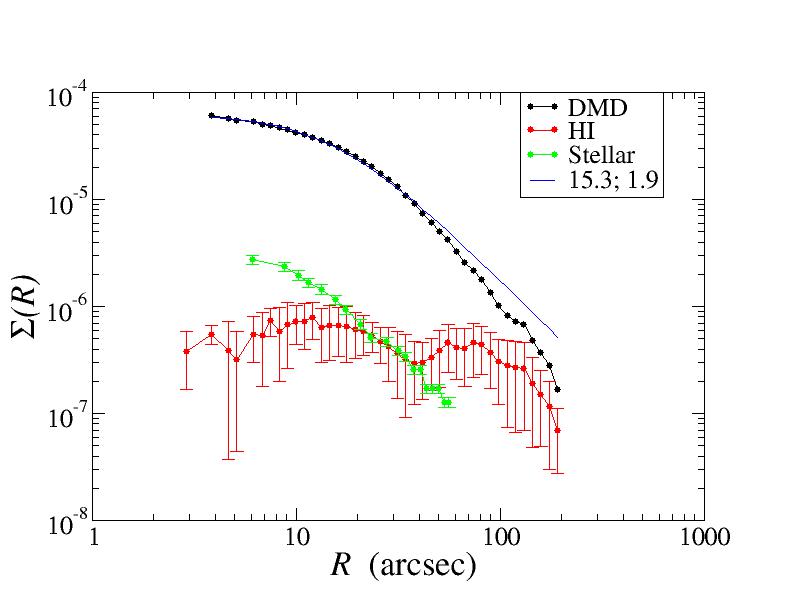}
\caption{}
\end{subfigure}
\quad
\begin{subfigure}[a]{0.3\textwidth}
\centering
\includegraphics[width=5.0cm,angle=0]{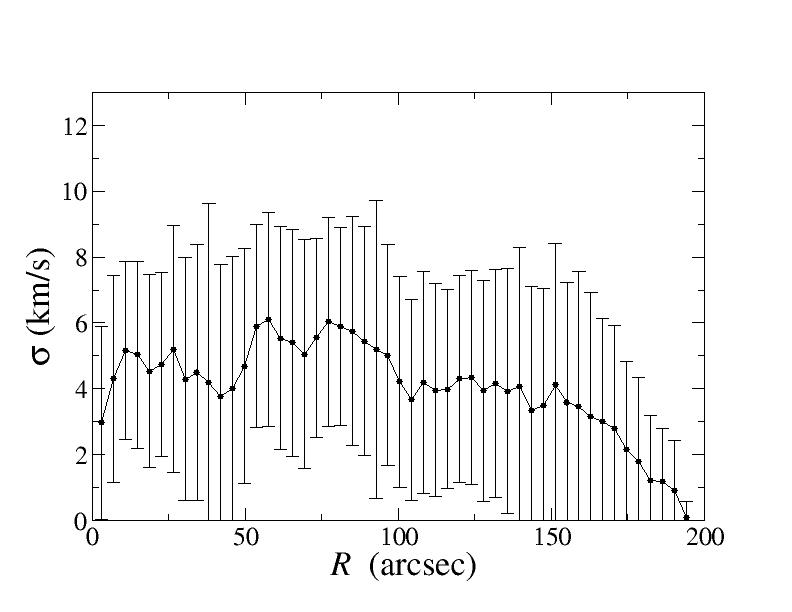}
\caption{}
\end{subfigure}
\quad
\begin{subfigure}[a]{0.3\textwidth}
\centering
\includegraphics[width=5.0cm,angle=0]{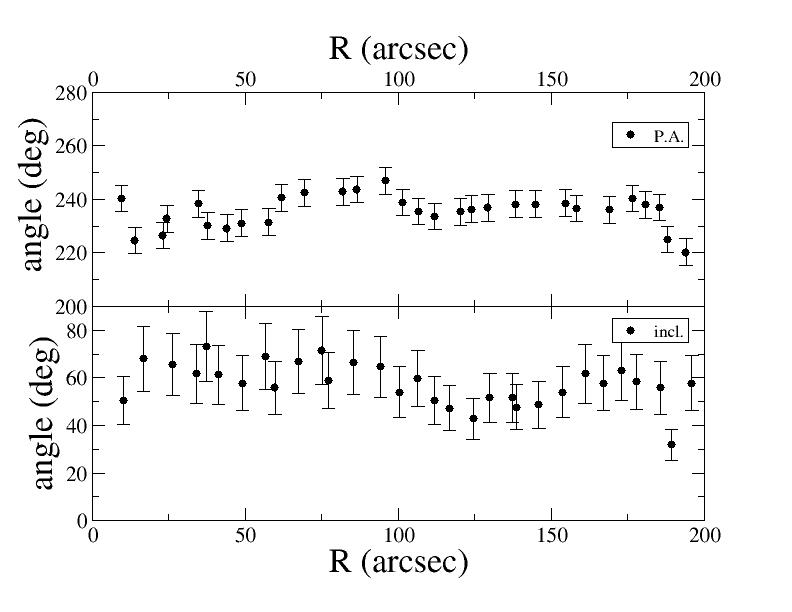}
\caption{}
\end{subfigure}
\quad
\caption{As Fig.\ref{fig:CVnIdwA2} but for DDO87}
\label{fig:DDO87_2}
\end{figure*}

\begin{figure*}
\quad
\begin{subfigure}[a]{0.3\textwidth}
\centering
\includegraphics[width=5.0cm,angle=0]{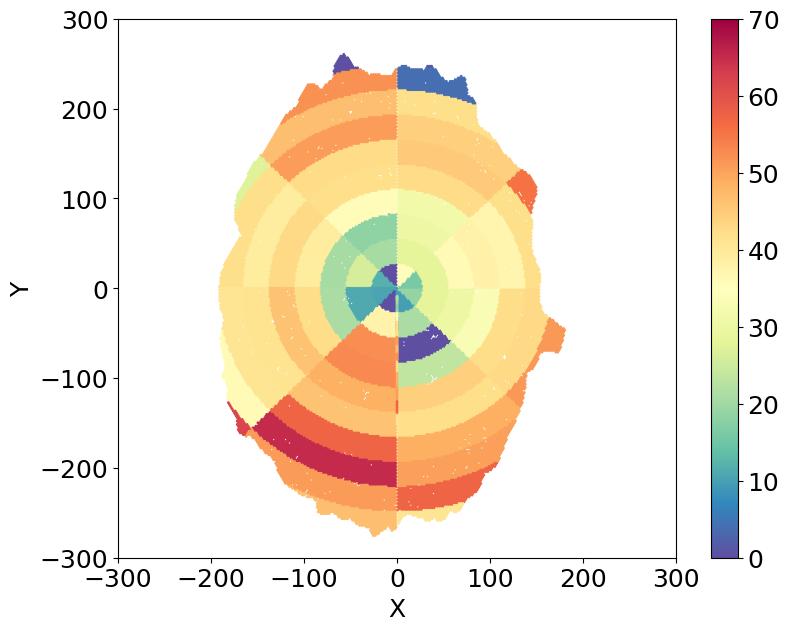}
\caption{}
\end{subfigure}
\quad
\begin{subfigure}[a]{0.3\textwidth}
\centering
\includegraphics[width=5.0cm,angle=0]{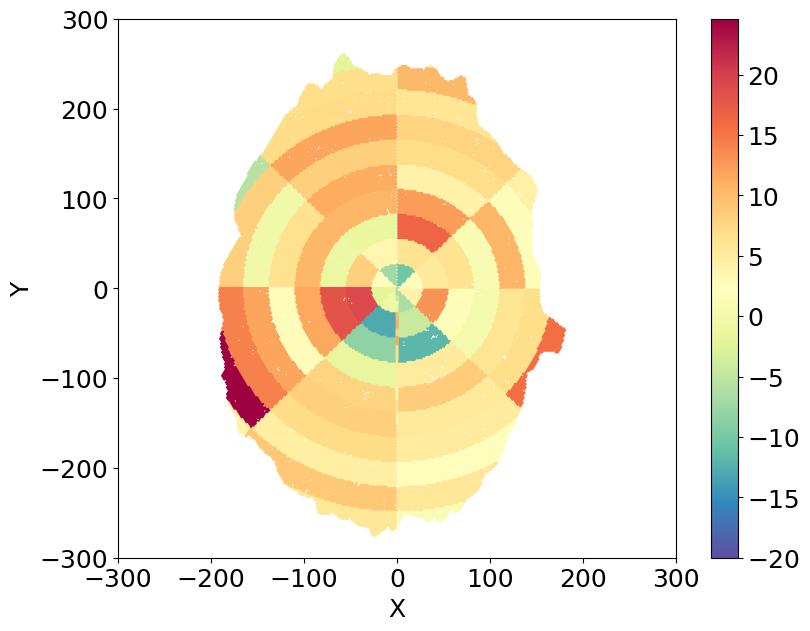}
\caption{}
\end{subfigure}
\quad
\begin{subfigure}[a]{0.3\textwidth}
\centering
\includegraphics[width=5.0cm,angle=0]{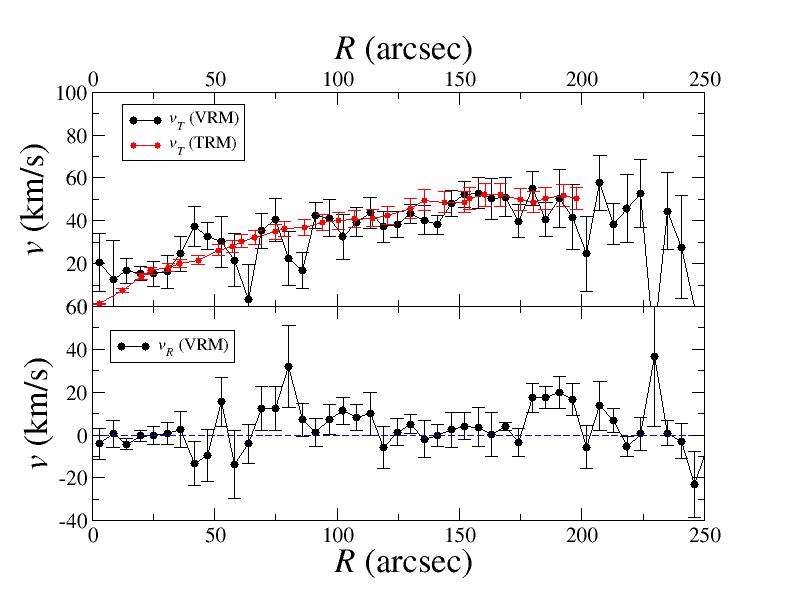}
\caption{}
\end{subfigure}
\quad
\caption{As Fig.\ref{fig:CVnIdwA3} but for  DDO87.
} 
\label{fig:DDO87_3}
\end{figure*}

The orientation angles do not exhibit any significant radial trend, indicating that a significant warp is likely not present in this galaxy
(see Figures \ref{fig:DDO87_2}-\ref{fig:DDO87_3}). However, the velocity maps reveal large angular anisotropies, which align with the fluctuating nature observed in both the \HI\; intensity map and the LOS velocity dispersion map. 
{  Even for this galaxy, the velocity profile obtained through the VRM method is similar to, but much more fluctuating than, that resulting from the TRM analysis, as, in the former case, these fluctuations are absorbed into those of the orientation angles. 
The inclination angle $i=55.5^\circ$  from \cite{Oh_etal_2015}  is larger than that  determined by \citet{Iorio_etal_2017}  (i.e., $i=45^\circ$).  
The fits using the DMD and NFW models are very similar in terms of reduced $\chi^2$ values.}

\subsection*{DDO101} 
\begin{figure*}
 \quad
 \begin{subfigure}[a]{0.3\textwidth}
 \centering
\includegraphics[width=5.0cm,angle=0]{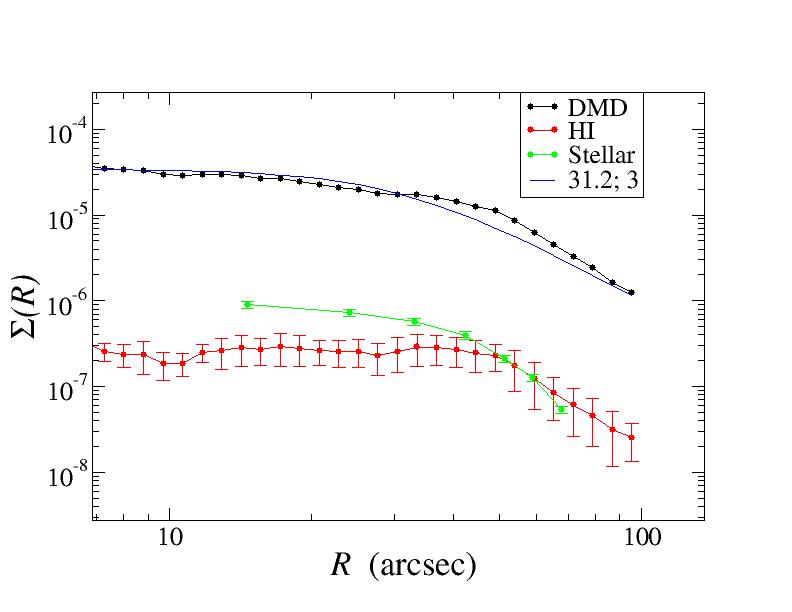}
\caption{}
\end{subfigure}
\quad
\begin{subfigure}[a]{0.3\textwidth}
\centering
\includegraphics[width=5.0cm,angle=0]{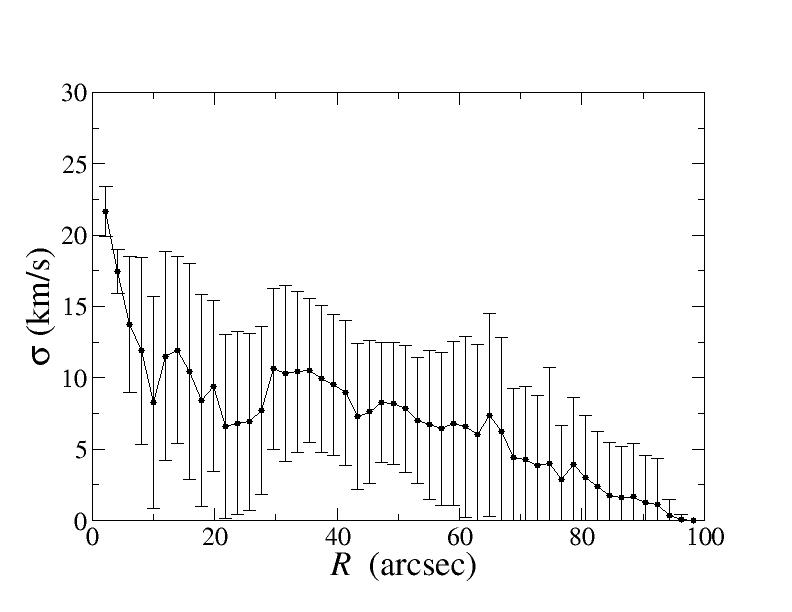}
\caption{}
\end{subfigure}
\quad
\begin{subfigure}[a]{0.3\textwidth}
\centering
\includegraphics[width=5.0cm,angle=0]{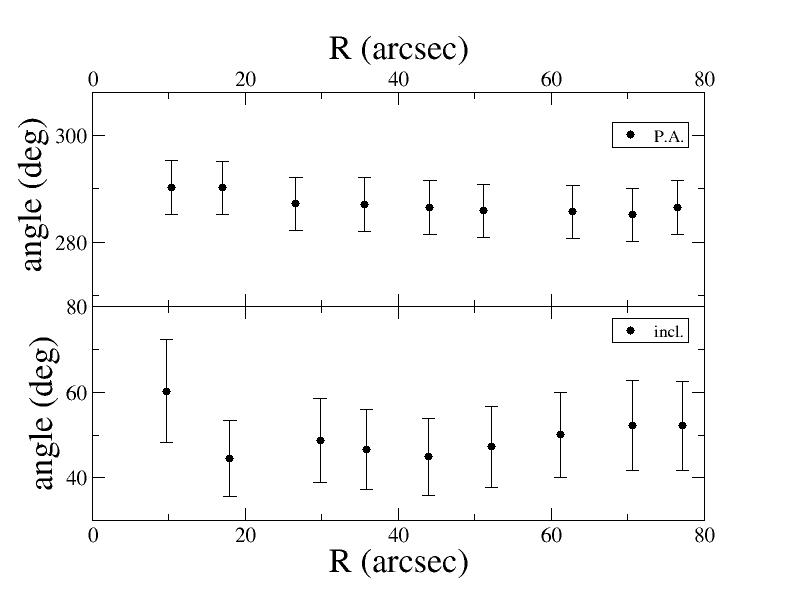}
\caption{}
\end{subfigure}
\quad
\caption{As Fig.\ref{fig:CVnIdwA2} but for  DDO101.
} 
\label{fig:DDO101_2}
\end{figure*}

\begin{figure*}
\quad
\begin{subfigure}[a]{0.3\textwidth}
\centering
\includegraphics[width=5.0cm,angle=0]{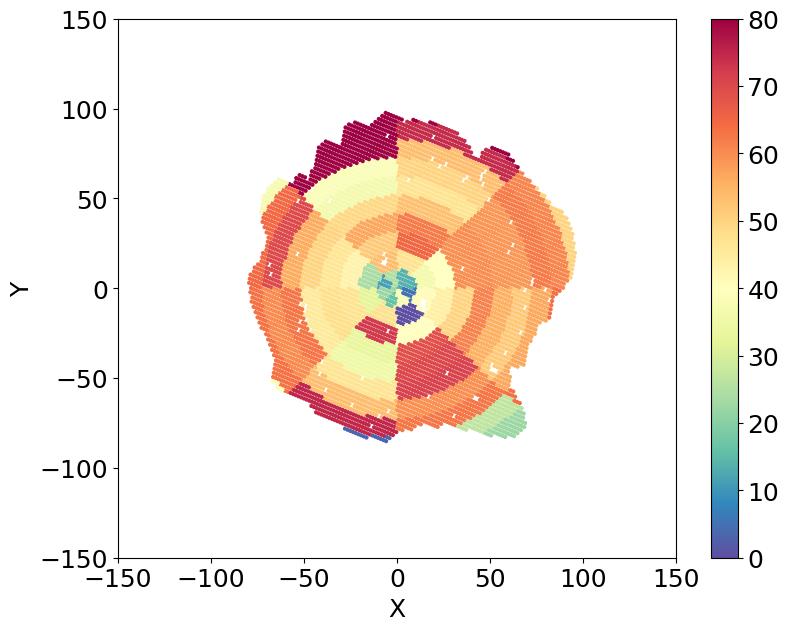}
\caption{}
\end{subfigure}
\quad
\begin{subfigure}[a]{0.3\textwidth}
\centering
\includegraphics[width=5.0cm,angle=0]{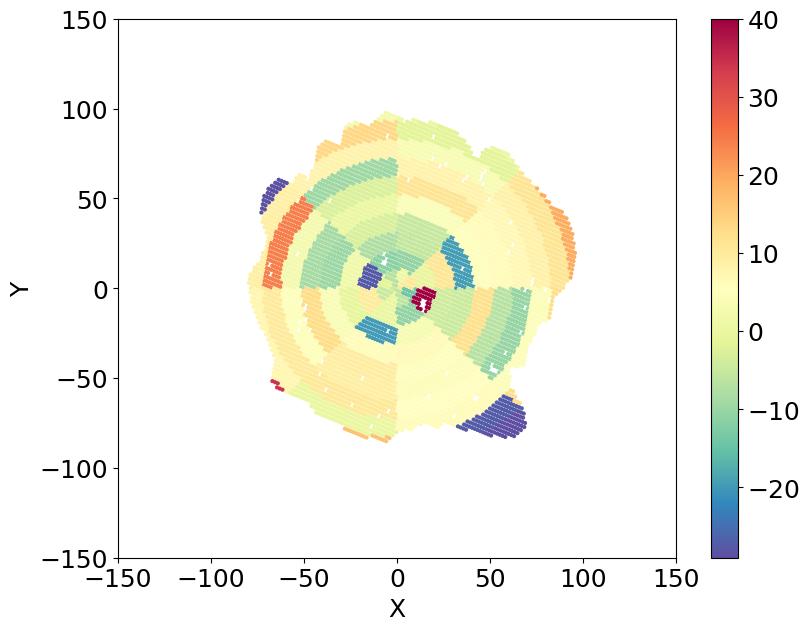}
\caption{}
\end{subfigure}
\quad
\begin{subfigure}[a]{0.3\textwidth}
\centering
\includegraphics[width=5.0cm,angle=0]{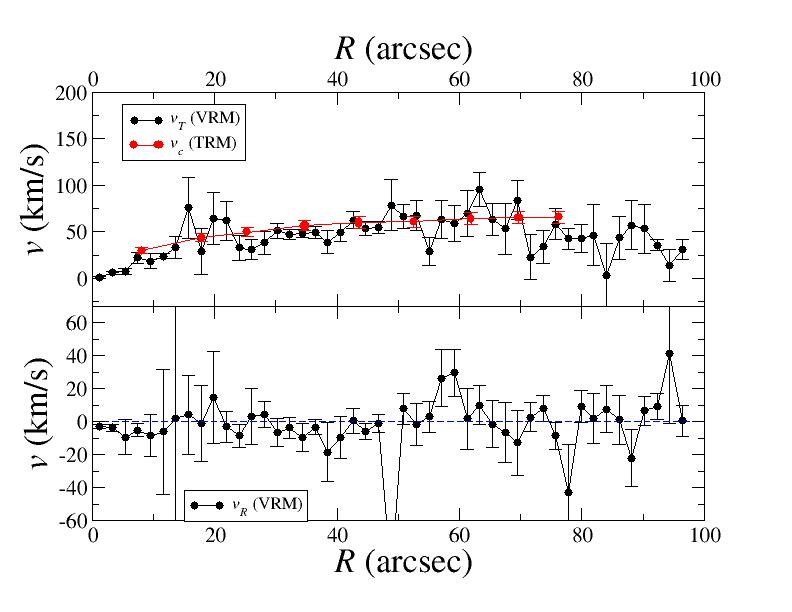}
\caption{}
\end{subfigure}
\quad
\caption{As Fig.\ref{fig:CVnIdwA3} but for  DDO101.
} 
\label{fig:DDO101_3}
\end{figure*}

DDO101 exhibits a relatively regular  LOS velocity field, indicating a coherent velocity pattern within the galaxy. However, both the intensity map and the LOS velocity dispersion maps show significant fluctuations, as depicted in Figures \ref{fig:DDO101_2}-\ref{fig:DDO101_3}.
The orientation angles in DDO101 demonstrate a roughly constant behavior as a function of radial distance. This suggests a consistent orientation pattern throughout the galaxy, with minimal variations across different regions.
The transverse velocity component obtained through the VRM aligns well with the rotational velocity measured by the TRM. In this case the inclination angle  from \cite{Oh_etal_2015}  is 
very close that  determined by \citet{Iorio_etal_2017}  (i.e., $i=52^\circ$) and the our determination well agree with both.  
{  In this case, because the orientation angles determined by the TRM are nearly constant and exhibit minimal fluctuations, the two profiles are very similar. Both the NFW and DMD models fit the data equally well, with reduced $\chi^2$ values of 2.5 and 1.4, respectively.}

\subsection*{DDO126} 
\begin{figure*}
 \quad
 \begin{subfigure}[a]{0.3\textwidth}
 \centering
\includegraphics[width=5.0cm,angle=0]{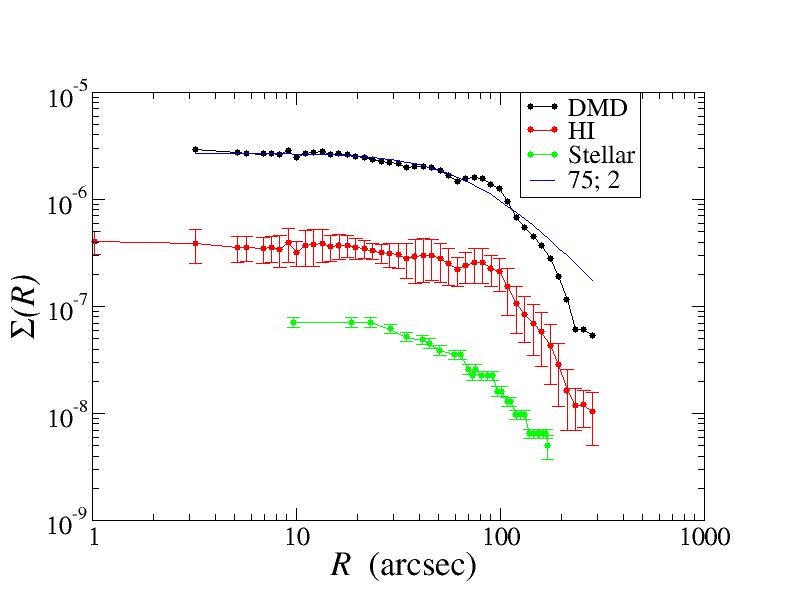}
\caption{}
\end{subfigure}
\quad
\begin{subfigure}[a]{0.3\textwidth}
\centering
\includegraphics[width=5.0cm,angle=0]{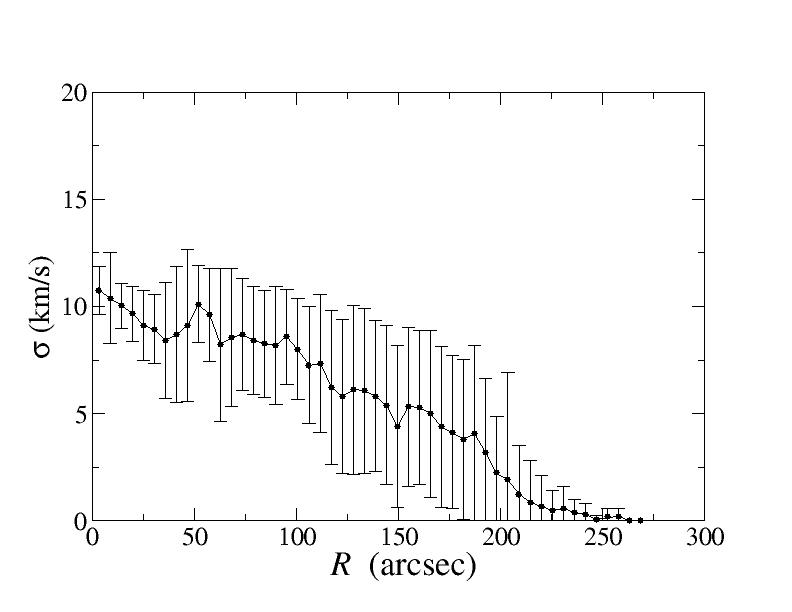}
\caption{}
\end{subfigure}
\quad
\begin{subfigure}[a]{0.3\textwidth}
\centering
\includegraphics[width=5.0cm,angle=0]{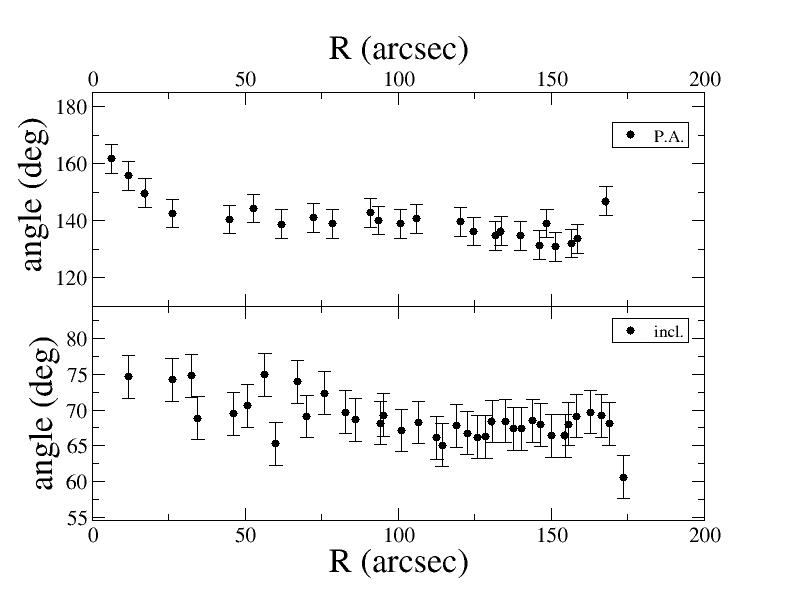}
\caption{}
\end{subfigure}
\quad
\caption{As Fig.\ref{fig:CVnIdwA2} but for  DDO126.	
} 
\label{fig:DDO126_2}
\end{figure*}

\begin{figure*}
\quad
\begin{subfigure}[a]{0.3\textwidth}
\centering
\includegraphics[width=5.0cm,angle=0]{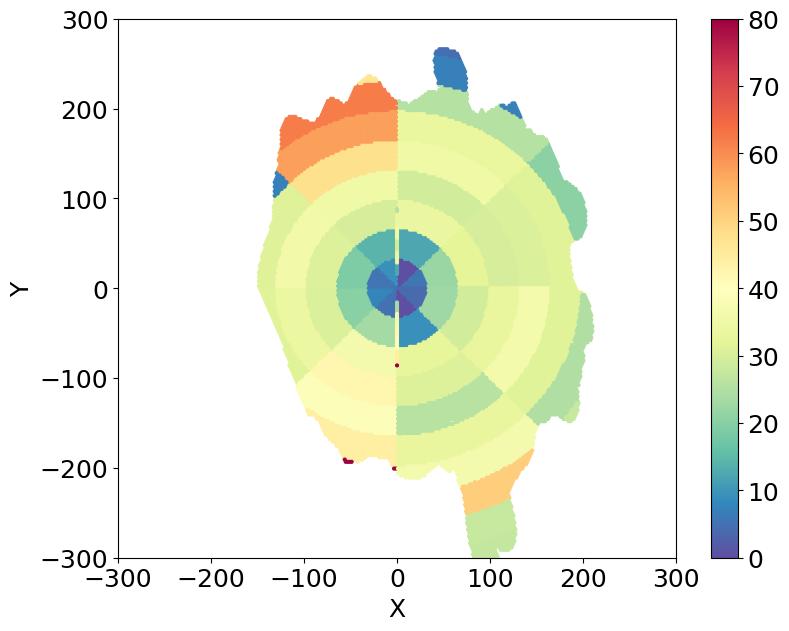}
\caption{}
\end{subfigure}
\quad
\begin{subfigure}[a]{0.3\textwidth}
\centering
\includegraphics[width=5.0cm,angle=0]{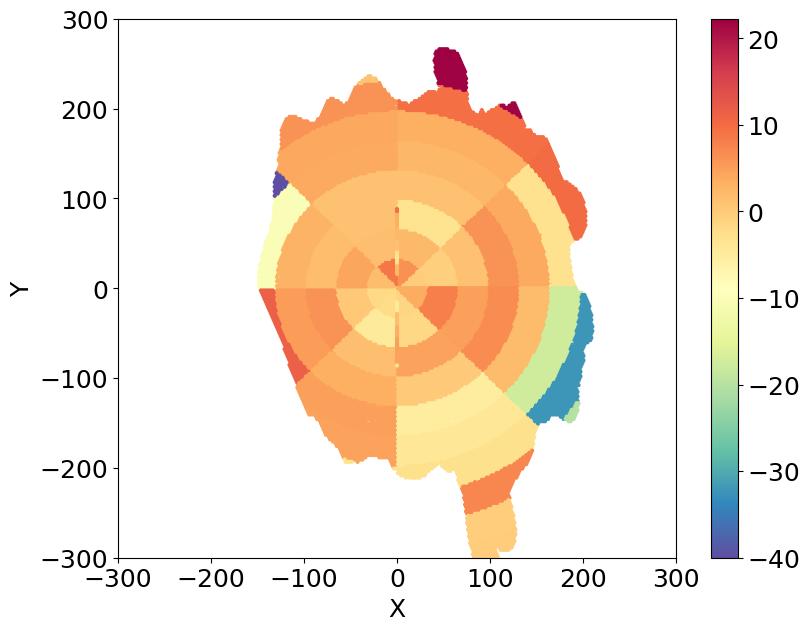}
\caption{}
\end{subfigure}
\quad
\begin{subfigure}[a]{0.3\textwidth}
\centering
\includegraphics[width=5.0cm,angle=0]{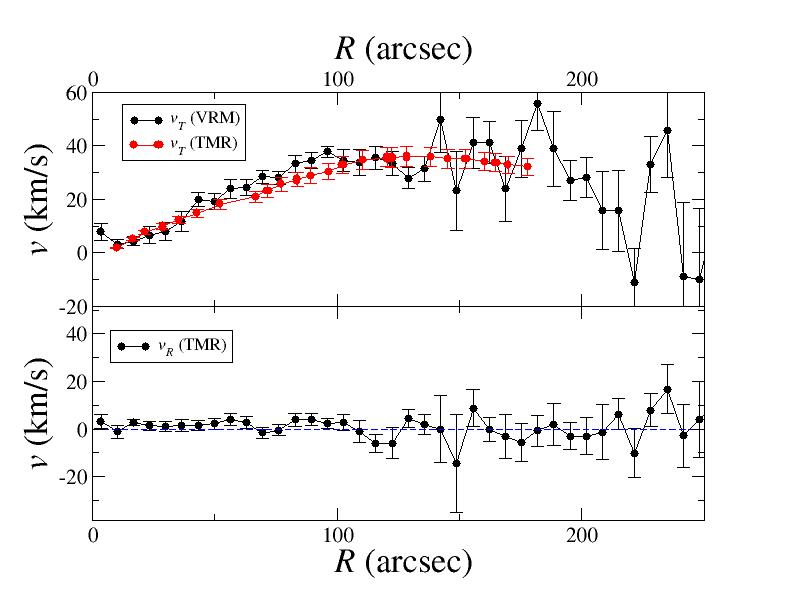}
\caption{}
\end{subfigure}
\quad
\caption{As Fig.\ref{fig:CVnIdwA3} but for  DDO126.
} 
\label{fig:DDO126_3}
\end{figure*}

Similar to DDO101, DDO126 also exhibits a regular LOS velocity map. 
{  
The P.A. shows a variation in the inner disc and remains constant as the radial distance from the galactic center increases, as shown in Figures \ref{fig:DDO126_2}-\ref{fig:DDO126_3}.
In contrast, the inclination angle exhibits a fluctuating behavior with variations of up to ten degrees. When analyzed using the VRM, these variations correspond to significant anisotropies in both velocity components in the outermost regions of the galactic disc.
The inclination angle from \citet{Oh_etal_2015} is similar to that measured by \citet{Iorio_etal_2017} (i.e., $i = 65^\circ$), and thus the different determinations of the circular velocity overlap fairly well.

These anisotropies indicate directional variations in the velocity components, suggesting complex dynamics in those regions.
Both mass model fits are affected by the relatively large fluctuations in the circular velocity component, resulting in a high reduced $\chi^2$ value of 1.7 for the DMD model and 3.0 for the NFW model. 
}

\subsection*{DDO133} 
\begin{figure*}
 \quad
 \begin{subfigure}[a]{0.3\textwidth}
 \centering
\includegraphics[width=5.0cm,angle=0]{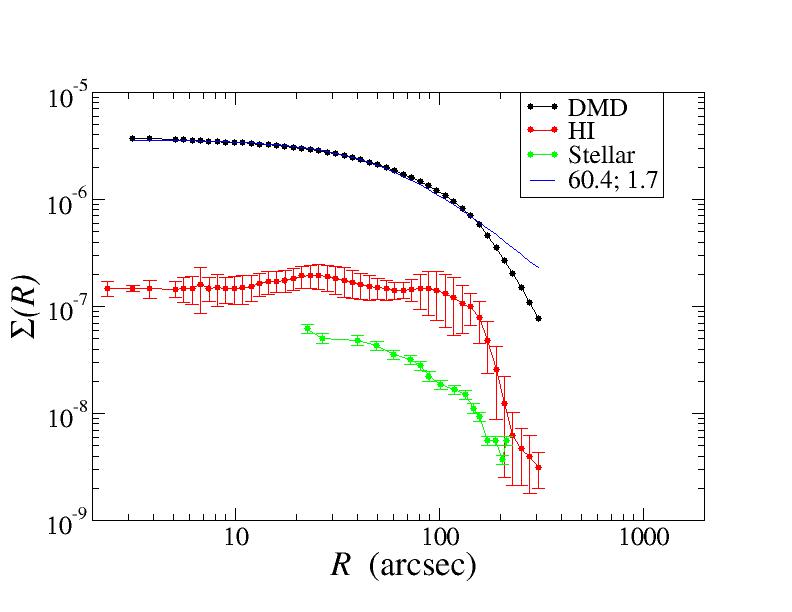}
\caption{}
\end{subfigure}
\quad
\begin{subfigure}[a]{0.3\textwidth}
\centering
\includegraphics[width=5.0cm,angle=0]{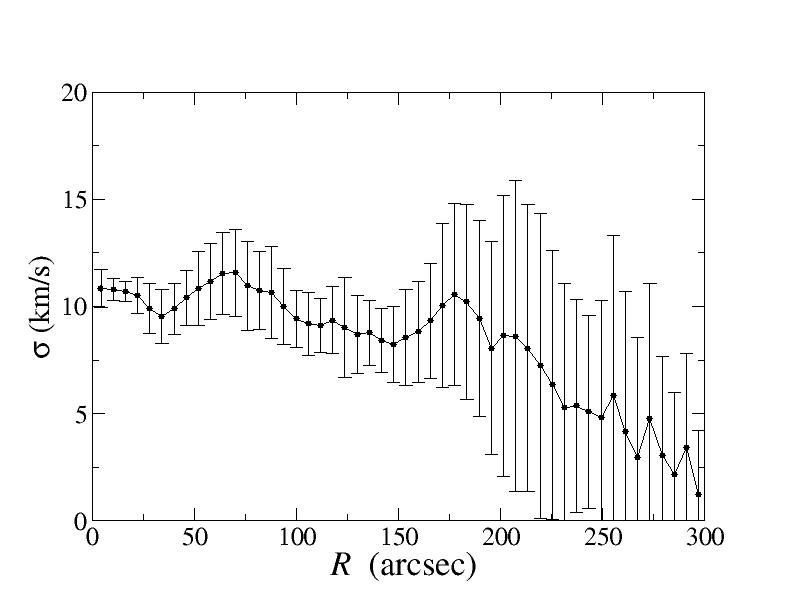}
\caption{}
\end{subfigure}
\quad
\begin{subfigure}[a]{0.3\textwidth}
\centering
\includegraphics[width=5.0cm,angle=0]{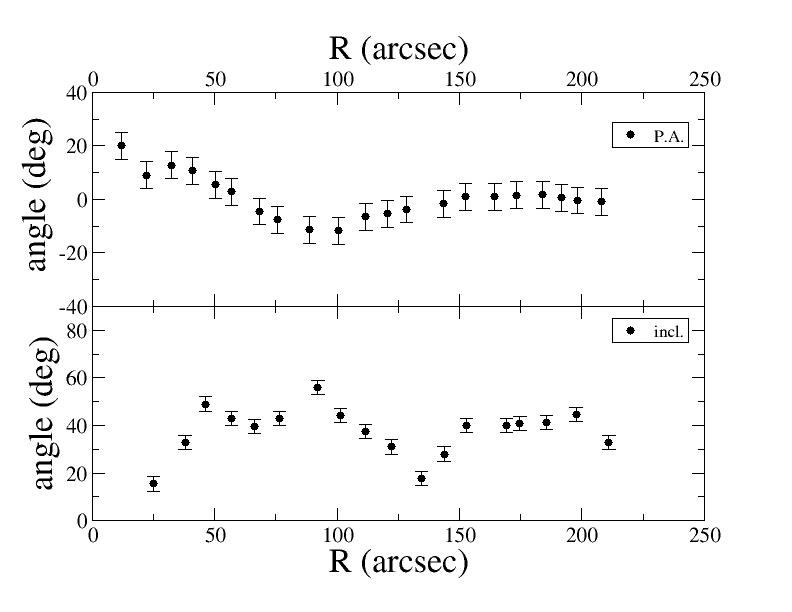}
\caption{}
\end{subfigure}
\quad
\caption{As Fig.\ref{fig:CVnIdwA2} but for  DDO133.
} 
\label{fig:DDO133_2}
\end{figure*}

\begin{figure*}
\quad
\begin{subfigure}[a]{0.3\textwidth}
\centering
\includegraphics[width=5.0cm,angle=0]{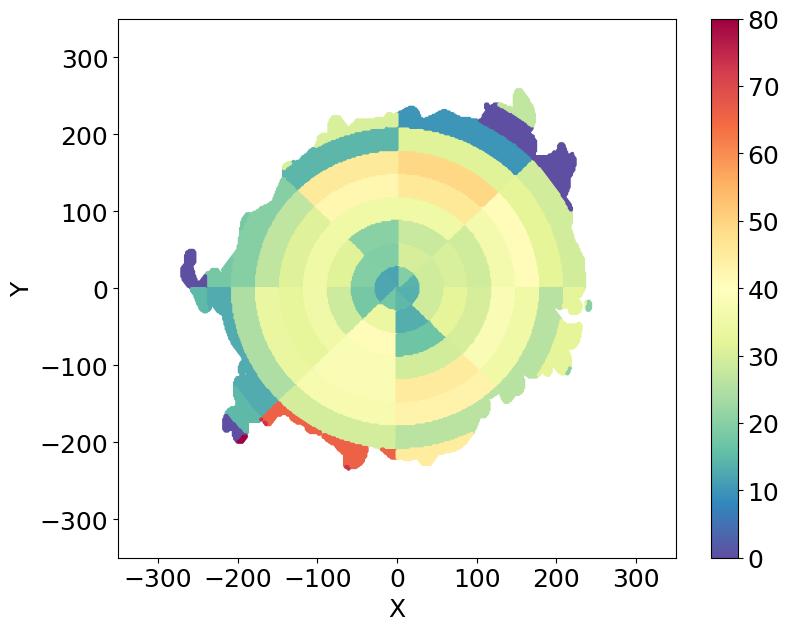}
\caption{}
\end{subfigure}
\quad
\begin{subfigure}[a]{0.3\textwidth}
\centering
\includegraphics[width=5.0cm,angle=0]{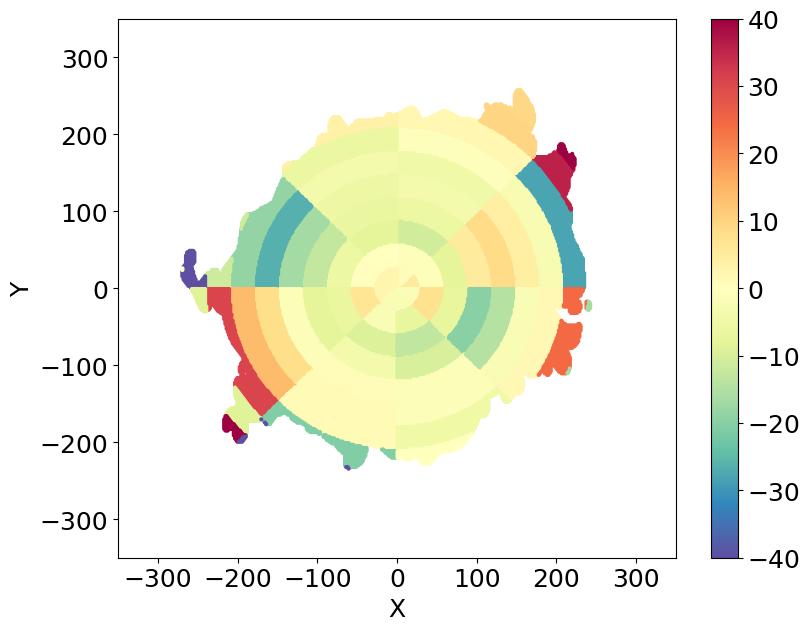}
\caption{}
\end{subfigure}
\quad
\begin{subfigure}[a]{0.3\textwidth}
\centering
\includegraphics[width=5.0cm,angle=0]{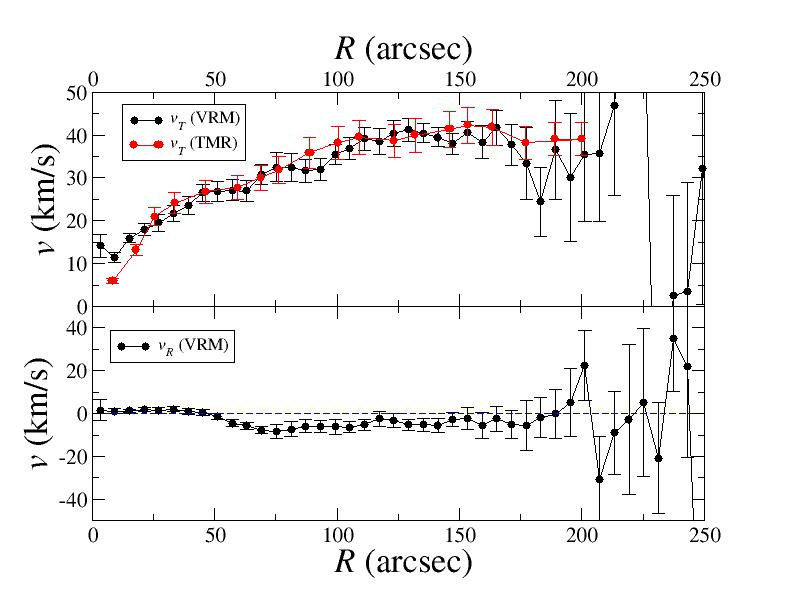}
\caption{}
\end{subfigure}
\quad
\caption{As Fig.\ref{fig:CVnIdwA3} but for  DDO133.  
} 
\label{fig:DDO133_3}
\end{figure*}

The LOS velocity field in DDO133 exhibits irregularities within the inner disc, which appear to be correlated with the structures observed in the intensity distribution, as depicted in Figures \ref{fig:DDO133_2} - \ref{fig:DDO133_3}.
Conversely, the orientation angles display large fluctuations in the inner disc but demonstrate a smooth and constant behavior in the outer regions. 
The radial and transverse velocity component maps, reconstructed using the VRM, reveal significant and localized anisotropies both in the inner and outer regions of the disc. These anisotropies indicate variations in the velocity components in different directions, occurring in specific regions of the galaxy.
{  
Even in this case the inclination angle from \citet{Oh_etal_2015} is similar to that measured by \citet{Iorio_etal_2017} (i.e., $i = 43^\circ$), and thus the different determinations of the circular velocity overlap fairly well but for a difference in the outermost 
regions of the galaxy where the P.A. shows a variation. 
Fits with both mass models show good agreement with the data (i.e., $\chi^2 = 0.65$ for the DMD and $\chi^2 = 0.36$ for the NFW), with the main source of variation attributed to the intrinsic fluctuations in the circular velocity profile.
}.

\subsection*{DDO154} 

\begin{figure*}
 \quad
 \begin{subfigure}[a]{0.3\textwidth}
 \centering
\includegraphics[width=5.0cm,angle=0]{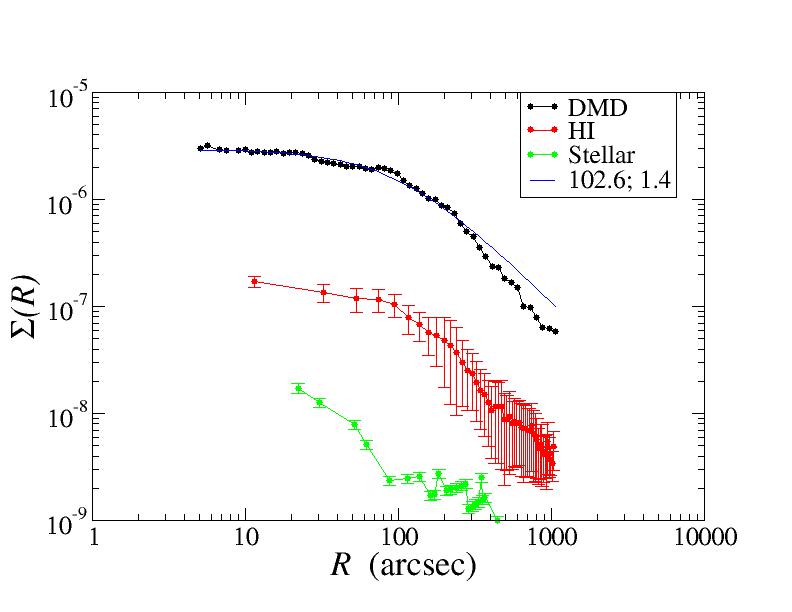}
\caption{}
\end{subfigure}
\quad
\begin{subfigure}[a]{0.3\textwidth}
\centering
\includegraphics[width=5.0cm,angle=0]{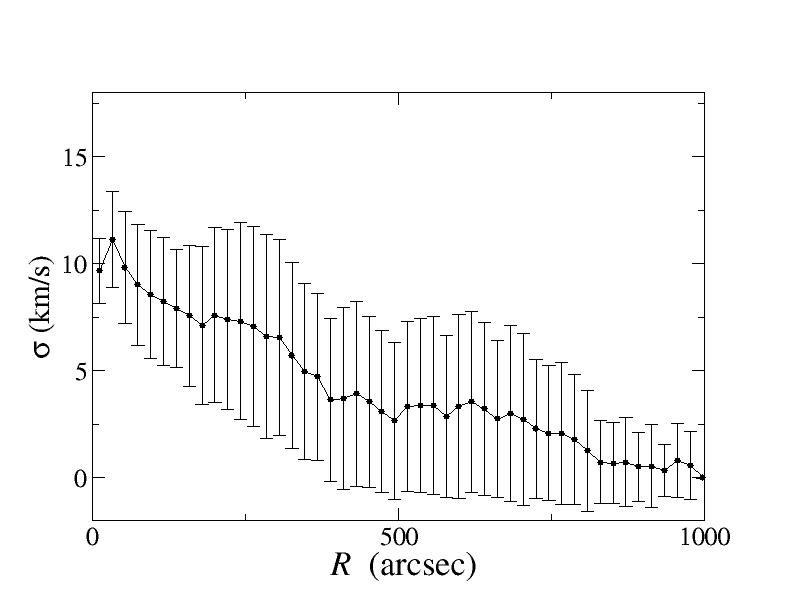}
\caption{}
\end{subfigure}
\quad
\begin{subfigure}[a]{0.3\textwidth}
\centering
\includegraphics[width=5.0cm,angle=0]{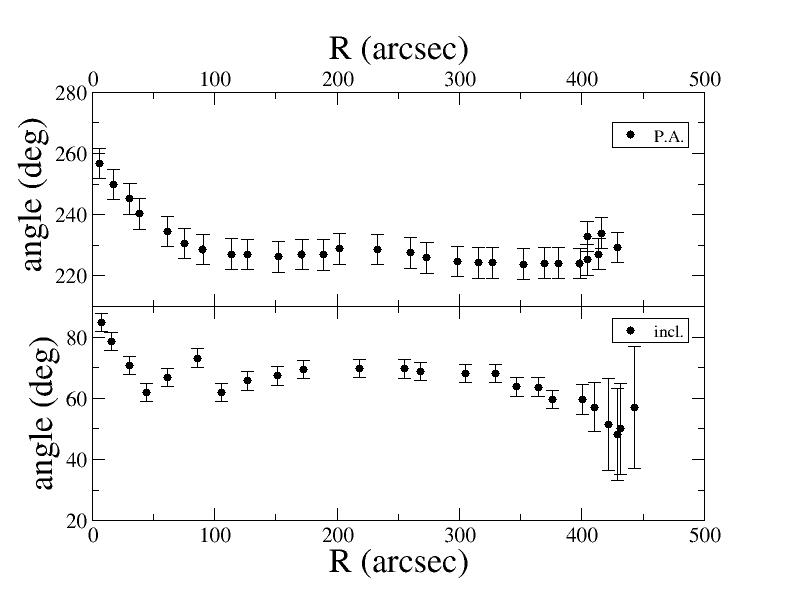}
\caption{}
\end{subfigure}
\quad
\caption{As Fig.\ref{fig:CVnIdwA2} but for  DDO154.	
} 
\label{fig:DDO154_2}
\end{figure*}

\begin{figure*}
\quad
\begin{subfigure}[a]{0.3\textwidth}
\centering
\includegraphics[width=5.0cm,angle=0]{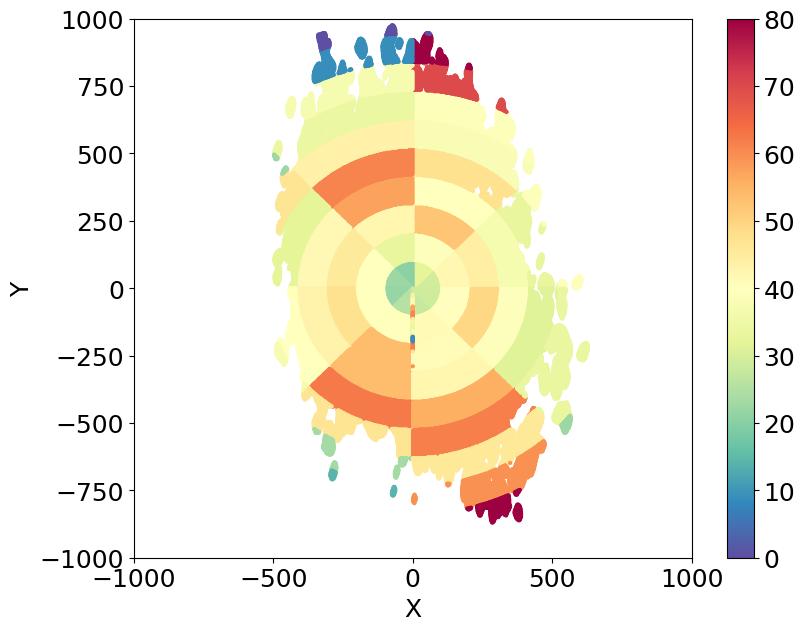}
\caption{}
\end{subfigure}
\quad
\begin{subfigure}[a]{0.3\textwidth}
\centering
\includegraphics[width=5.0cm,angle=0]{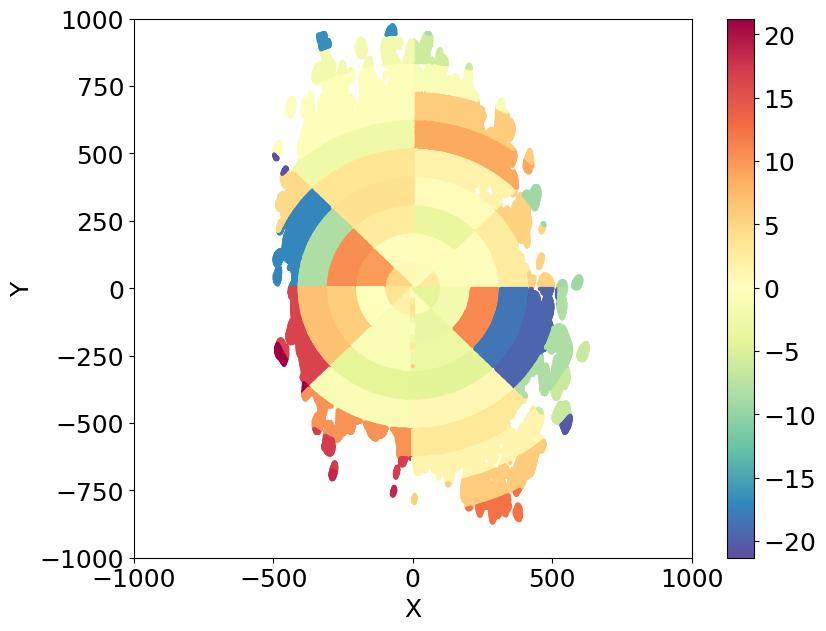}
\caption{}
\end{subfigure}
\quad
\begin{subfigure}[a]{0.3\textwidth}
\centering
\includegraphics[width=5.0cm,angle=0]{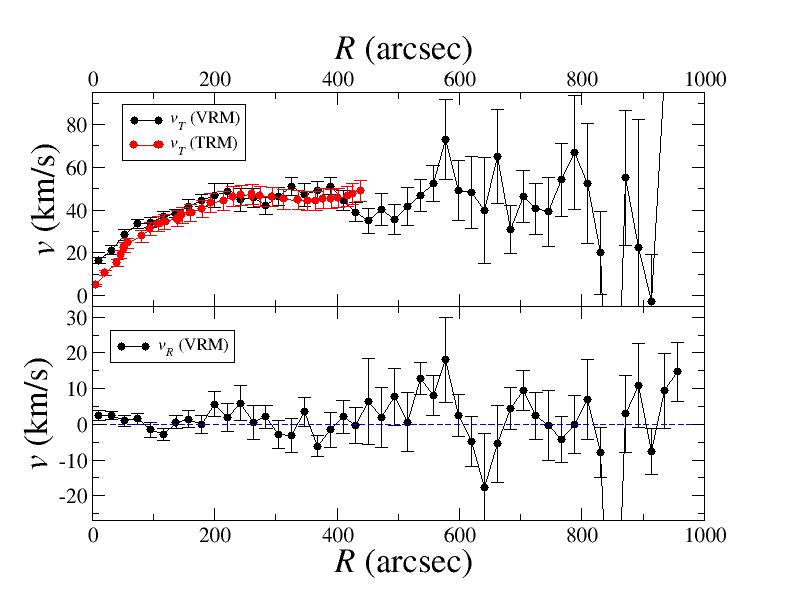}
\caption{}
\end{subfigure}
\quad
\caption{As Fig.\ref{fig:CVnIdwA3} but for  DDO154.  
} 
\label{fig:DDO154_3}
\end{figure*}

From Figures \ref{fig:DDO154_2}-\ref{fig:DDO154_3}, it can be observed that DDO154 exhibits a LOS velocity field that demonstrates the typical pattern expected for a rotating disc. 
{  
However, while the inclination angle appears nearly flat throughout the entire disc, the position angle (P.A.) shows a decrease with radius in the inner disc, which causes the difference between our determination of the rotational velocity and that measured by \citet{Iorio_etal_2017} and \citet{Oh_etal_2015}.  
On the other hand, at larger radii, the rotation curves measured by the TRM and the VRM show very similar results. In addition, fluctuations in both orientation angles are observed in the most distant radial bins, suggesting the presence of local anisotropies that become more significant in the outermost regions.}
This similarity suggests that the warp in DDO154 is not significant.
{  Fits with mass models yield similar results in terms of the reduced $\chi^2$, i.e., $\sim 1.5$, with the relatively large value being attributed to the outermost data points.  
} 

\subsection*{DDO168} 
\begin{figure*}
 \quad
 \begin{subfigure}[a]{0.3\textwidth}
 \centering
\includegraphics[width=5.0cm,angle=0]{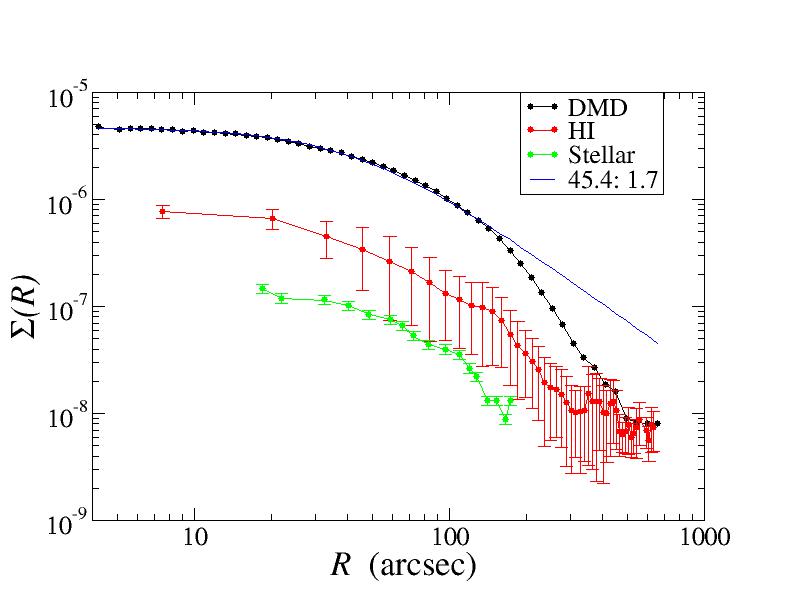}
\caption{}
\end{subfigure}
\quad
\begin{subfigure}[a]{0.3\textwidth}
\centering
\includegraphics[width=5.0cm,angle=0]{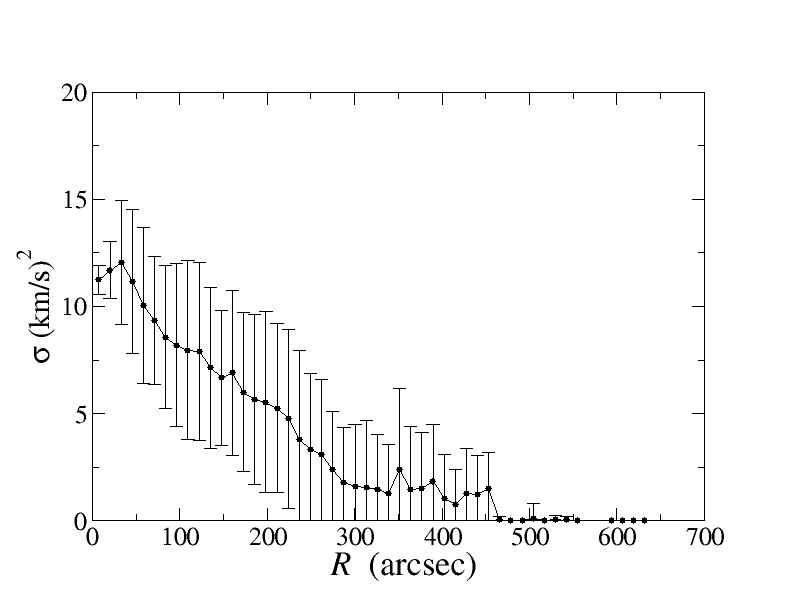}
\caption{}
\end{subfigure}
\quad
\begin{subfigure}[a]{0.3\textwidth}
\centering
\includegraphics[width=5.0cm,angle=0]{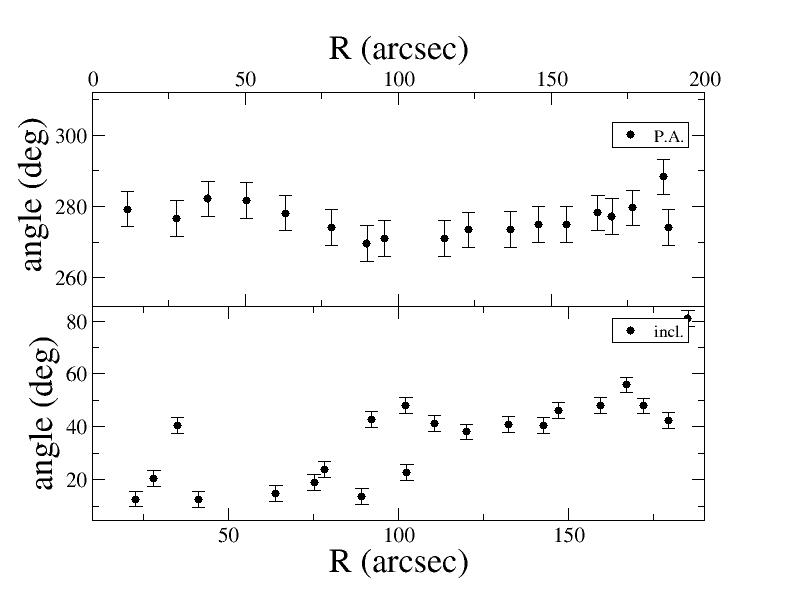}
\caption{}
\end{subfigure}
\quad
\caption{As Fig.\ref{fig:CVnIdwA2} but for DDO168.	
} 
\label{fig:DDO168_2}
\end{figure*}

\begin{figure*}
\quad
\begin{subfigure}[a]{0.3\textwidth}
\centering
\includegraphics[width=5.0cm,angle=0]{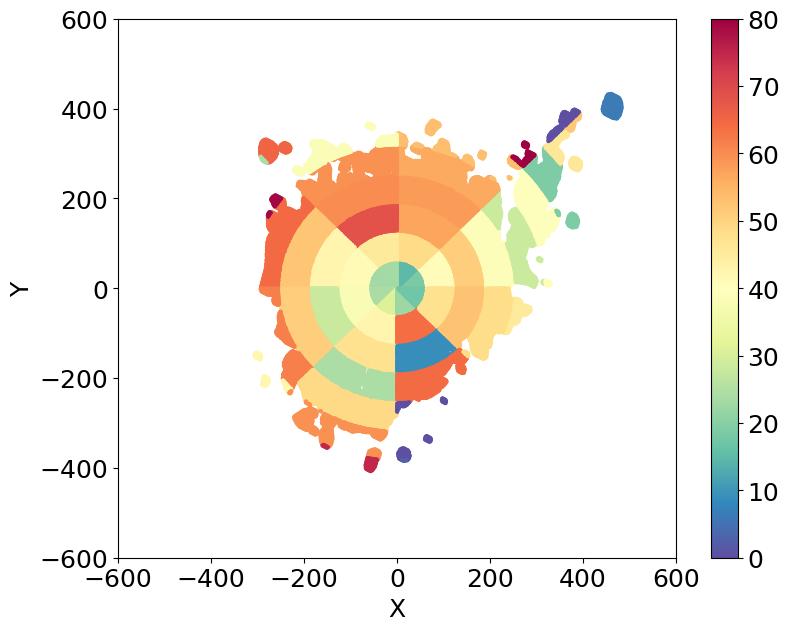}
\caption{}
\end{subfigure}
\quad
\begin{subfigure}[a]{0.3\textwidth}
\centering
\includegraphics[width=5.0cm,angle=0]{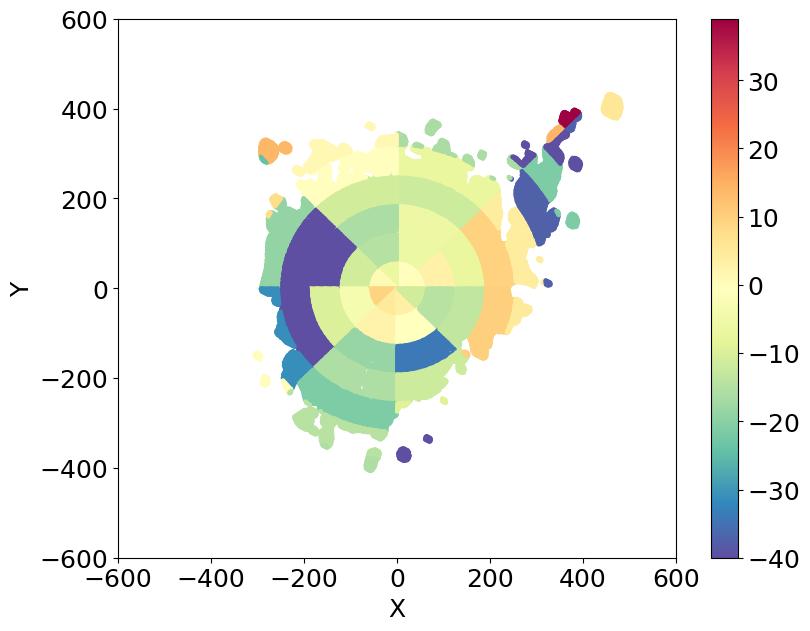}
\caption{}
\end{subfigure}
\quad
\begin{subfigure}[a]{0.3\textwidth}
\centering
\includegraphics[width=5.0cm,angle=0]{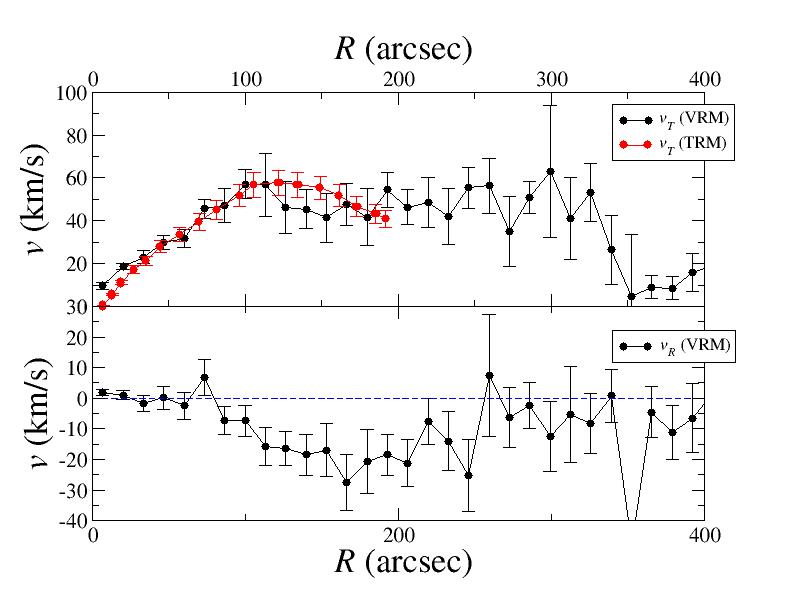}
\caption{}
\end{subfigure}
\quad
\caption{As Fig.\ref{fig:CVnIdwA3} but for DDO168.
} 
\label{fig:DDO168_3}
\end{figure*}

DDO168 exhibits a regular LOS velocity map in the inner part of its disc, with an almost aligned kinematic axis (see Figs. \ref{fig:DDO168_2} - \ref{fig:DDO168_3}). However, the outer regions of the galaxy display irregularities in the LOS velocity distribution.
The orientation angles in DDO168 experience relatively large fluctuations but do not exhibit significant radial trends. 
{ 
The inclination angle used in our analysis, i.e., $i = 46.5^\circ$ \citep{Oh_etal_2015}, is smaller than that measured by \citet{Iorio_etal_2017} (i.e., $i \approx 60^\circ$). In addition, the P.A. shows a radial variation, and these trends cause the relatively small difference between our determination of the circular velocity and theirs.
} 

The transverse and radial component maps reveal localized and substantial anisotropies, indicating variations in the velocity components in different directions within specific regions of the galaxy. Additionally, the radial velocity profile decreases in the external part of the disc, implying a declining trend in the radial velocities towards the outer regions.
{ 
The determination of the circular velocity using the VRM and TRM methods yields similar results. However, as in other cases, the VRM determination is significantly more fluctuating because the orientation angles are kept fixed.  
Fitting with mass models yields a reduced $\chi^2$ value around 1.0 in both cases within the radial range [0, 7]~kpc. At larger radial distances, the fluctuations become too large to obtain reliable results.  
}

\subsection*{DDO210} 
\begin{figure*}
 \quad
 \begin{subfigure}[a]{0.3\textwidth}
 \centering
\includegraphics[width=5.0cm,angle=0]{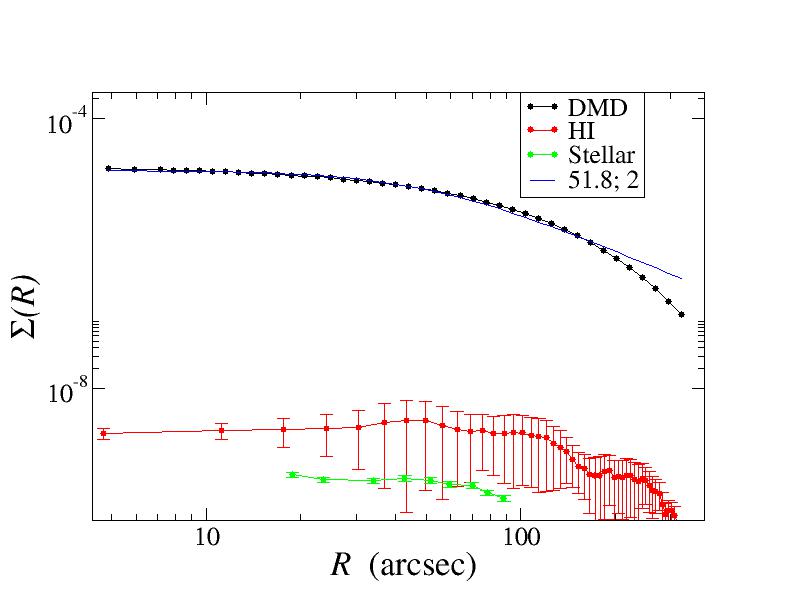}
\caption{}
\end{subfigure}
\quad
\begin{subfigure}[a]{0.3\textwidth}
\centering
\includegraphics[width=5.0cm,angle=0]{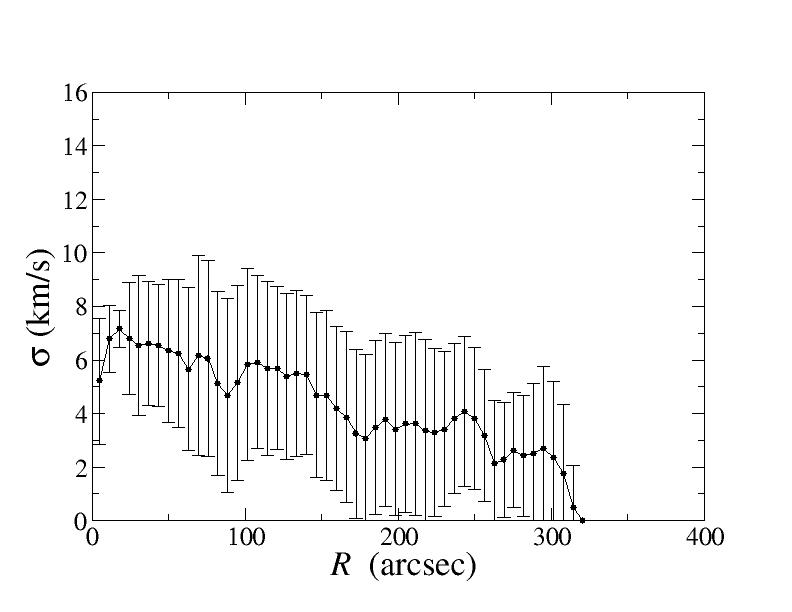}
\caption{}
\end{subfigure}
\quad
\begin{subfigure}[a]{0.3\textwidth}
\centering
\includegraphics[width=5.0cm,angle=0]{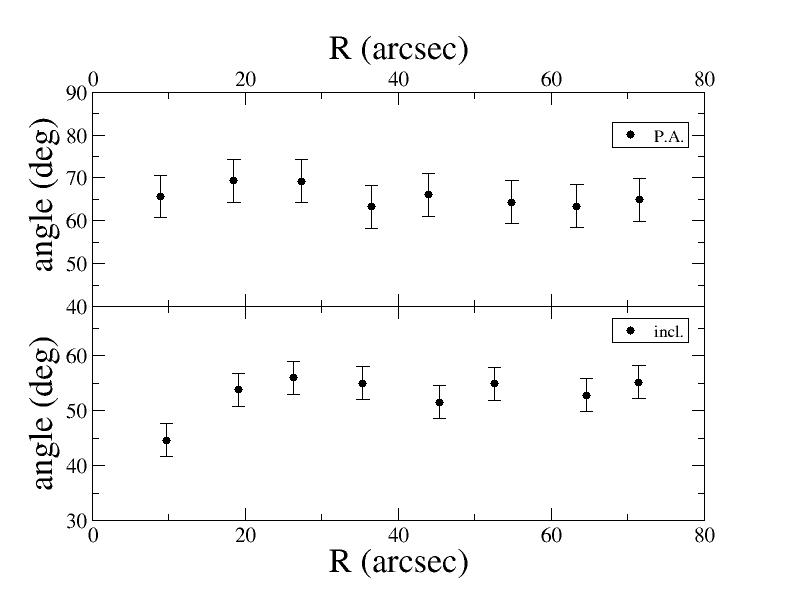}
\caption{}
\end{subfigure}
\quad
\caption{As Fig.\ref{fig:CVnIdwA2} but for DDO210.
} 
\label{fig:DDO210_2}
\end{figure*}

\begin{figure*}
\quad
\begin{subfigure}[a]{0.3\textwidth}
\centering
\includegraphics[width=5.0cm,angle=0]{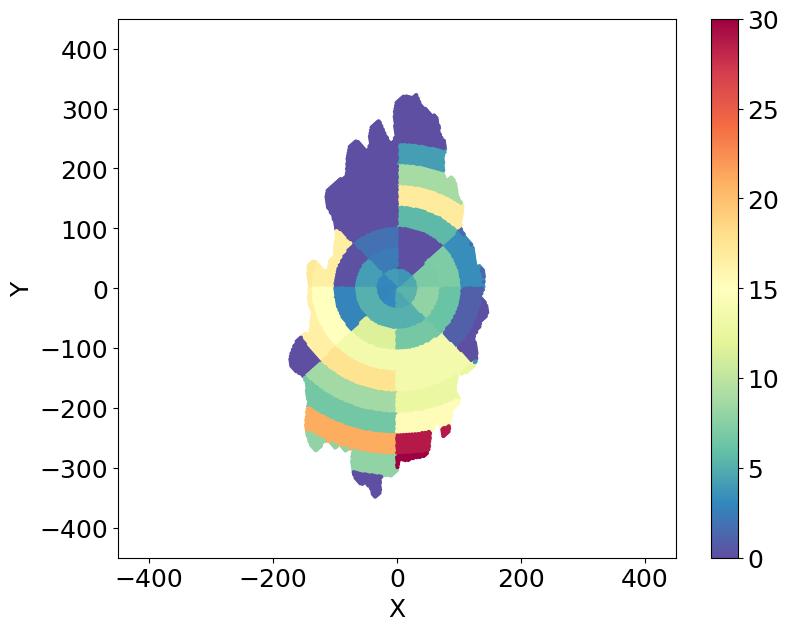}
\caption{}
\end{subfigure}
\quad
\begin{subfigure}[a]{0.3\textwidth}
\centering
\includegraphics[width=5.0cm,angle=0]{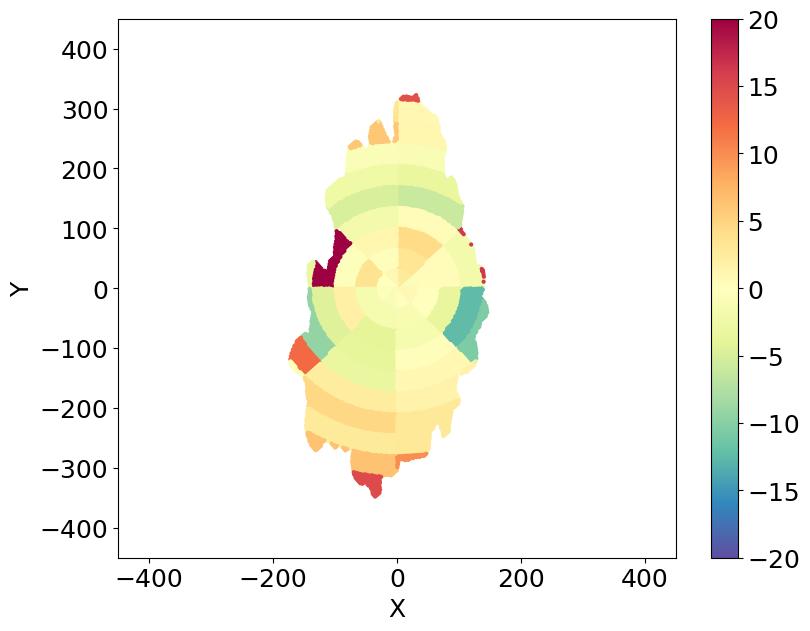}
\caption{}
\end{subfigure}
\quad
\begin{subfigure}[a]{0.3\textwidth}
\centering
\includegraphics[width=5.0cm,angle=0]{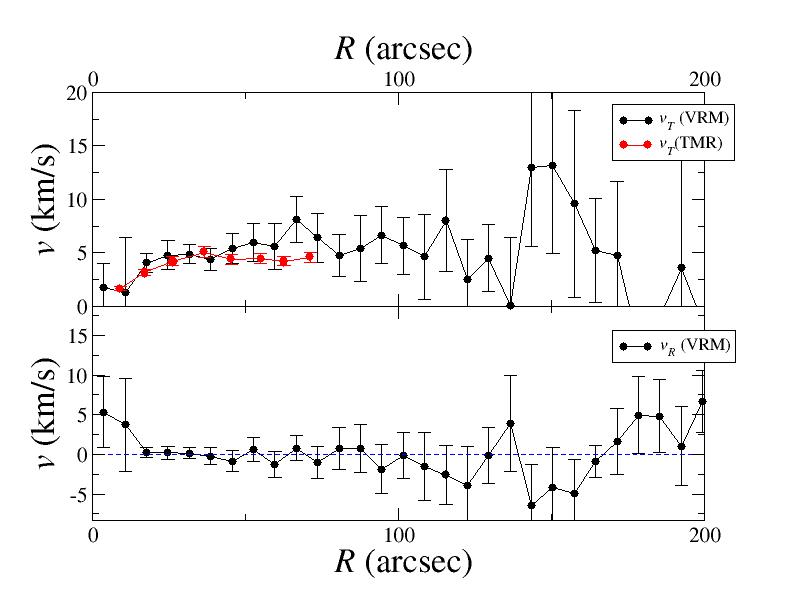}
\caption{}
\end{subfigure}
\quad
\caption{As Fig.\ref{fig:CVnIdwA3} but for DDO210.  
} 
\label{fig:DDO210_3}
\end{figure*}

{ 
In the case of DDO210, the LOS velocity anisotropy field appears irregular, as depicted in Figures~\ref{fig:DDO210_2} and \ref{fig:DDO210_3}. The orientation angles do not exhibit a radial trend, suggesting that the assumption of a flat disc made by the VRM is a good approximation for DDO210. However, they both show relevant fluctuations. For this reason both the radial and transverse velocity component fields show significant anisotropy, particularly in the outermost regions of the galactic disc. These anisotropies indicate variations in the velocity components in different directions, with pronounced effects in specific regions of the galaxy.  
However, we find a significant discrepancy between our determination of the asymmetric drift and that of \citet{Iorio_etal_2017}, which is due to the different methods used to measure it. 
The velocity profiles obtained through the VRM and TRM are very similar, indicating consistency between the two methods and aligning with the small variation in the orientation angles. This similarity suggests that the warp in DDO210 is not significant and that both methods provide reliable measurements of the rotation curve. 
There  

The NFW model provides a better fit ($\chi^2 = 0.18$) compared to the DMD model ($\chi^2 = 1.3$). This difference arises from the fact that both the stellar and gas profiles are highly fluctuating.  
}

\subsection*{DDO216} 

\begin{figure*}
 \quad
 \begin{subfigure}[a]{0.3\textwidth}
 \centering
\includegraphics[width=5.0cm,angle=0]{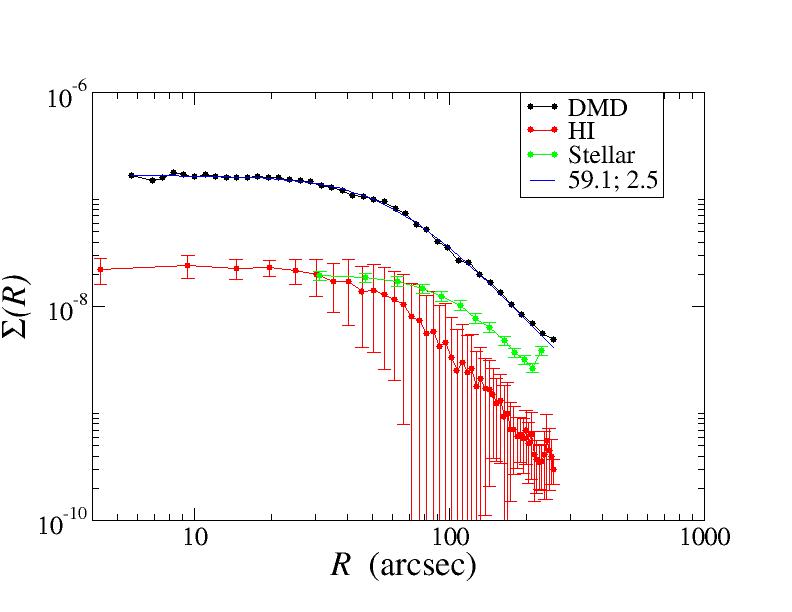}
\caption{}
\end{subfigure}
\quad
\begin{subfigure}[a]{0.3\textwidth}
\centering
\includegraphics[width=5.0cm,angle=0]{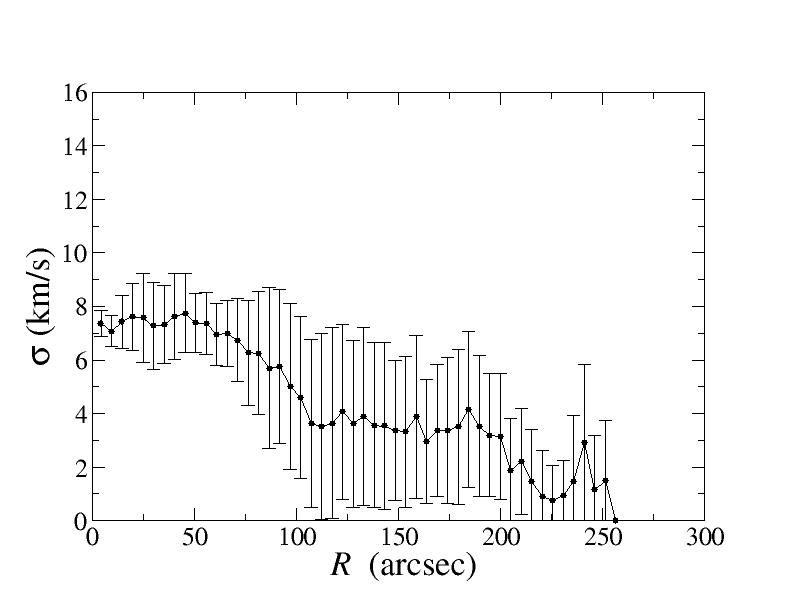}
\caption{}
\end{subfigure}
\quad
\begin{subfigure}[a]{0.3\textwidth}
\centering
\includegraphics[width=5.0cm,angle=0]{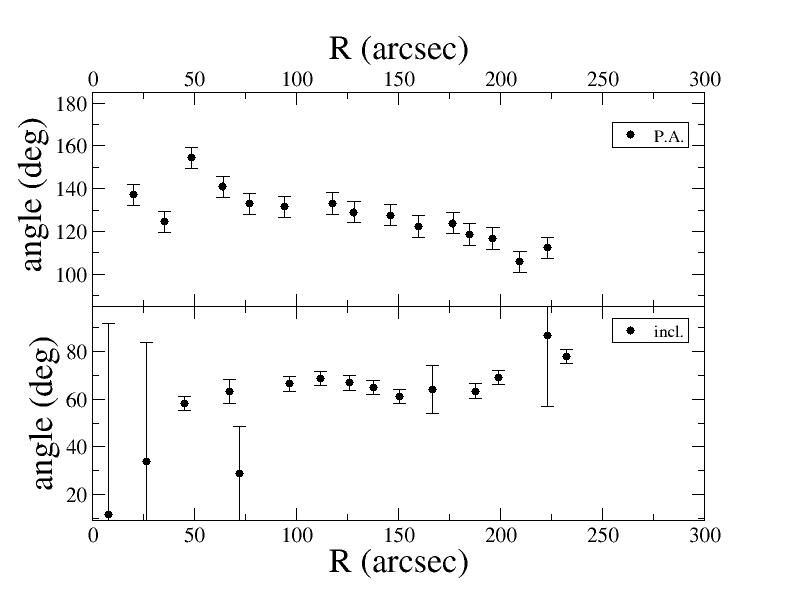}
\caption{}
\end{subfigure}
\quad
\caption{As Fig.\ref{fig:CVnIdwA2} but for DDO216.
} 
\label{fig:DDO216_2}
\end{figure*}

\begin{figure*}
\quad
\begin{subfigure}[a]{0.3\textwidth}
\centering
\includegraphics[width=5.0cm,angle=0]{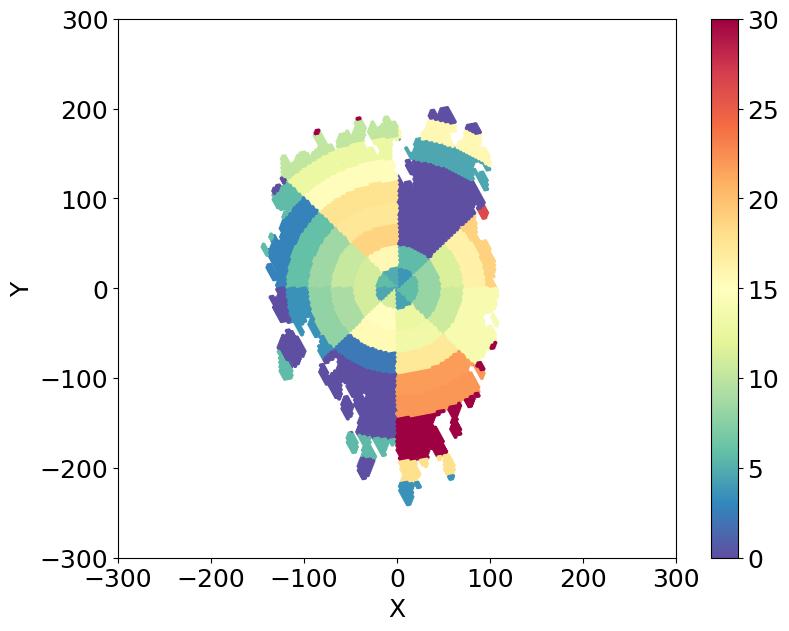}
\caption{}
\end{subfigure}
\quad
\begin{subfigure}[a]{0.3\textwidth}
\centering
\includegraphics[width=5.0cm,angle=0]{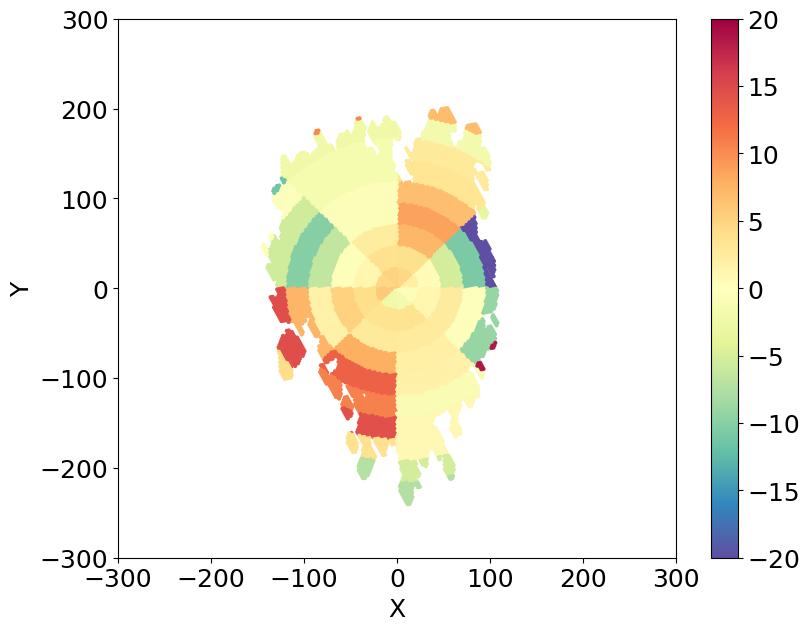}
\caption{}
\end{subfigure}
\quad
\begin{subfigure}[a]{0.3\textwidth}
\centering
\includegraphics[width=5.0cm,angle=0]{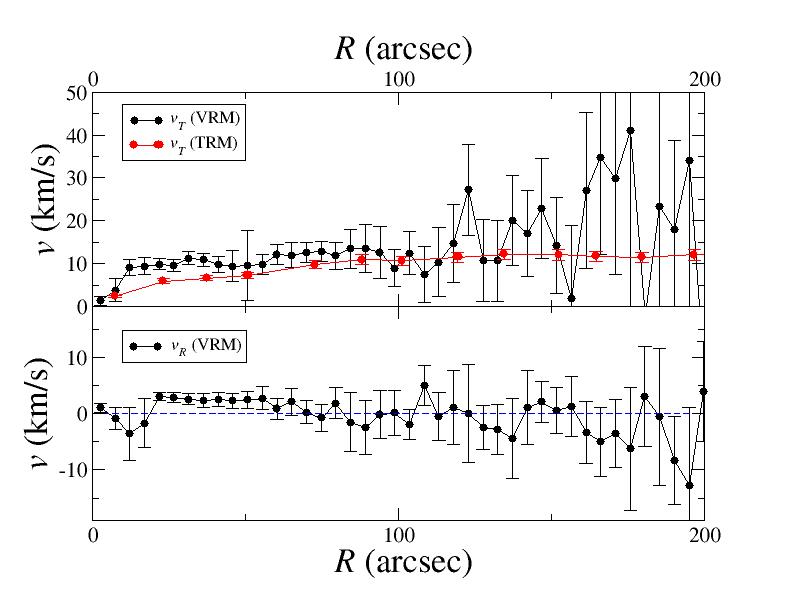}
\caption{}
\end{subfigure}
\quad
\caption{As Fig.\ref{fig:CVnIdwA3} but for DDO216.  
} 
\label{fig:DDO216_3}
\end{figure*}

{ 
Both orientation angles show noticeable fluctuations superimposed on a radial trend toward the outermost part of the disc. These radial changes originate the difference between our determination of the circular velocity and that of \citet{Iorio_etal_2017} (see Figures \ref{fig:DDO216_2}-\ref{fig:DDO216_3}).  
Thus, the entire galactic disc is characterized by velocity anisotropies in both components, which are larger toward the outskirts of the galaxy. These outer regions,
where a warp may possibile be present,  are not included in the fits for the mass models. The fits with both mass models are comparable and are clearly influenced by the highly fluctuating behavior of the circular velocity.
}

\subsection*{F564-V3} 
\begin{figure*}
 \quad
 \begin{subfigure}[a]{0.3\textwidth}
 \centering
\includegraphics[width=5.0cm,angle=0]{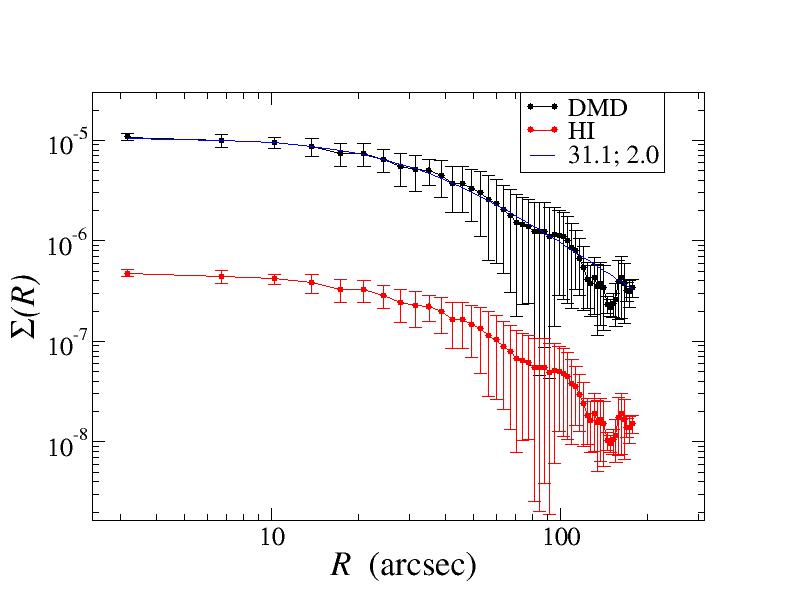}
\caption{}
\end{subfigure}
\quad
\begin{subfigure}[a]{0.3\textwidth}
\centering
\includegraphics[width=5.0cm,angle=0]{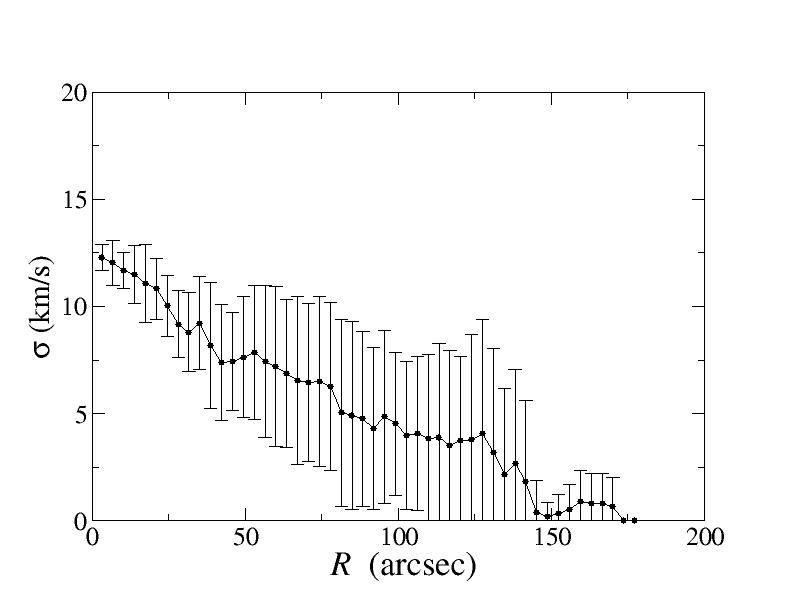}
\caption{}
\end{subfigure}
\quad
\begin{subfigure}[a]{0.3\textwidth}
\centering
\includegraphics[width=5.0cm,angle=0]{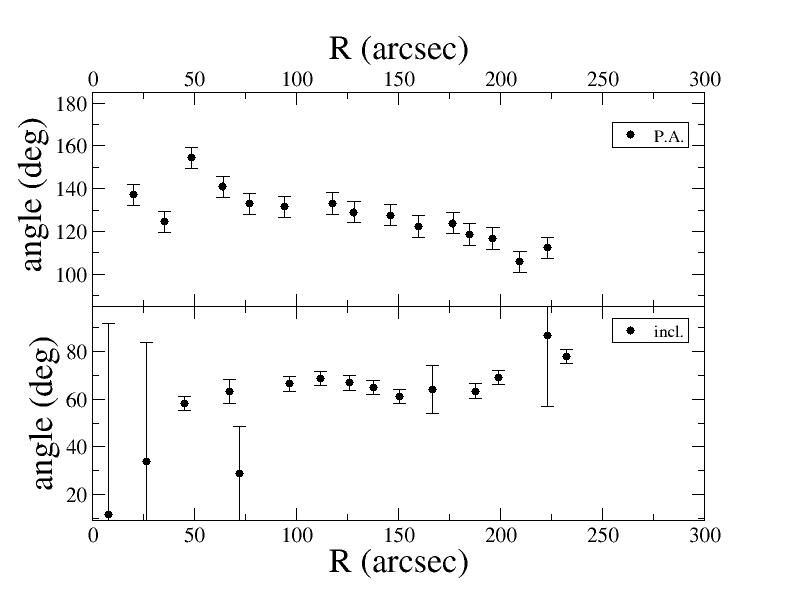}
\caption{}
\end{subfigure}
\quad
\caption{As Fig.\ref{fig:CVnIdwA2} but for F564-V3.	
} 
\label{fig:F564-V3_2}
\end{figure*}

\begin{figure*}
\quad
\begin{subfigure}[a]{0.3\textwidth}
\centering
\includegraphics[width=5.0cm,angle=0]{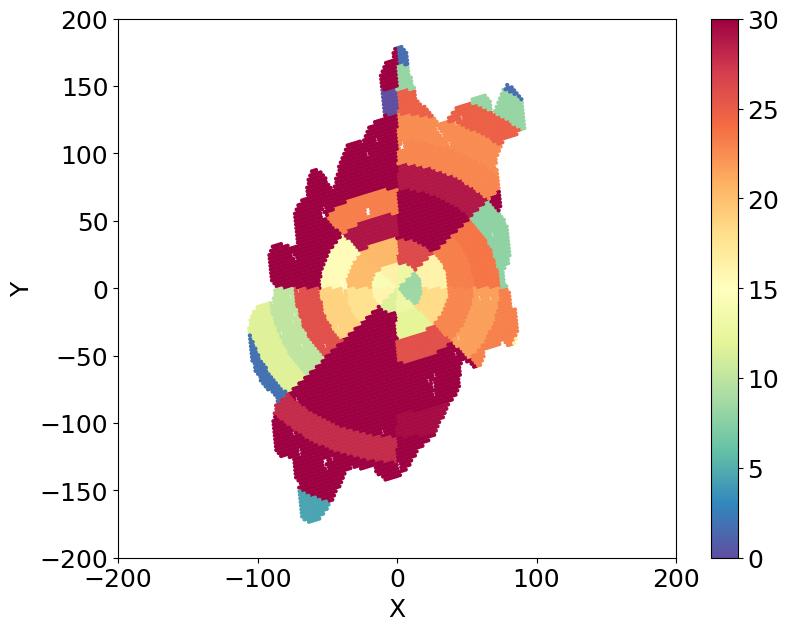}
\caption{}
\end{subfigure}
\quad
\begin{subfigure}[a]{0.3\textwidth}
\centering
\includegraphics[width=5.0cm,angle=0]{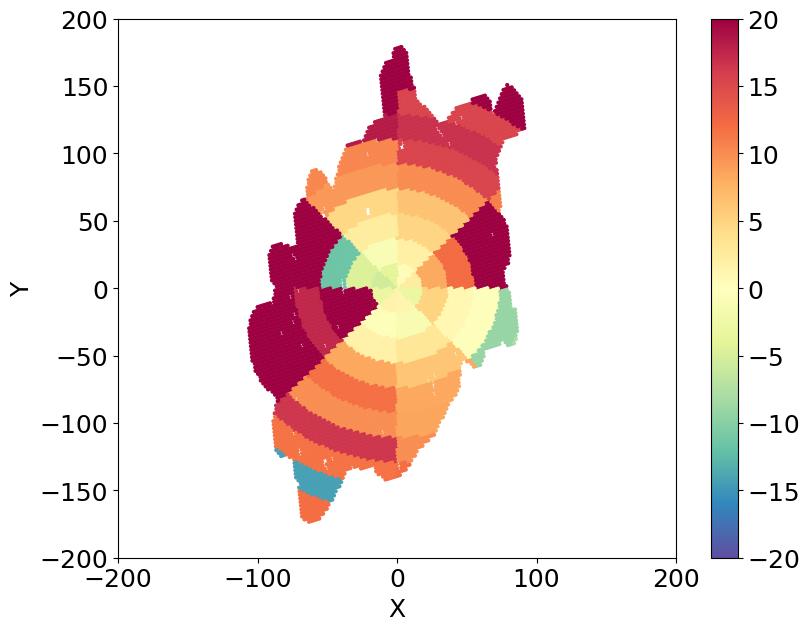}
\caption{}
\end{subfigure}
\quad
\begin{subfigure}[a]{0.3\textwidth}
\centering
\includegraphics[width=5.0cm,angle=0]{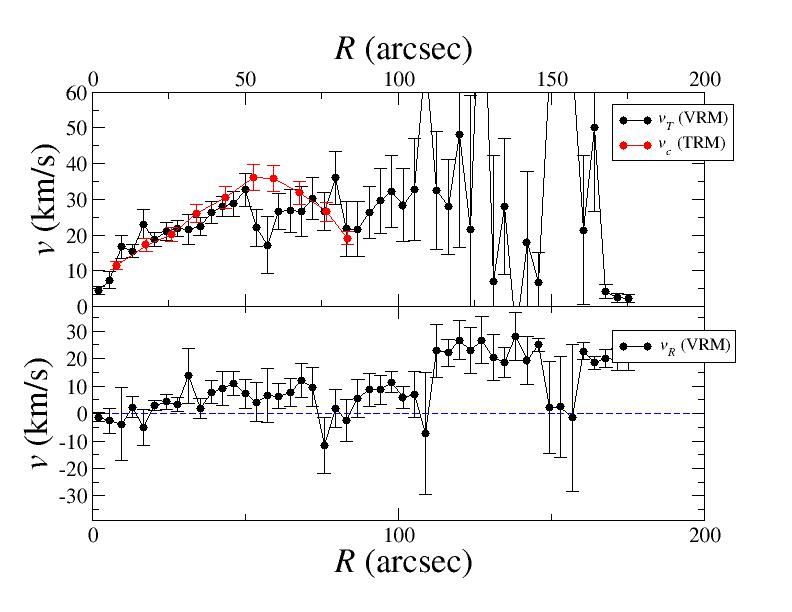}
\caption{}
\end{subfigure}
\quad
\caption{As Fig.\ref{fig:CVnIdwA3} but for F564-V3.
} 
\label{fig:F564-V3_3}
\end{figure*}

{  
For the inner galactic disc of F564-V3 (with $R < 100"$), the orientation angles remain relatively constant as a function of radial distance, as shown in Figures \ref{fig:F564-V3_2}-\ref{fig:F564-V3_3}.
However, both angles exhibit variations in the outer regions of the galaxy. In particular, the P.A. shows a declining trend, while the inclination angle exhibits large fluctuations.
These trends correspond, within the VRM framework, to significant spatial anisotropies and explain the difference between the rotational velocity detected by the VRM and that measured by the TRM at large radial distances.
Since F564-V3 is primarily gas-dominated, a one-parameter fit was performed by varying only the gas component parameter. Consequently, the NFW mass model fit yields a slightly smaller $\chi^2$.
}

%

\subsection*{IC10} 
\begin{figure*}
 \quad
 \begin{subfigure}[a]{0.3\textwidth}
 \centering
\includegraphics[width=5.0cm,angle=0]{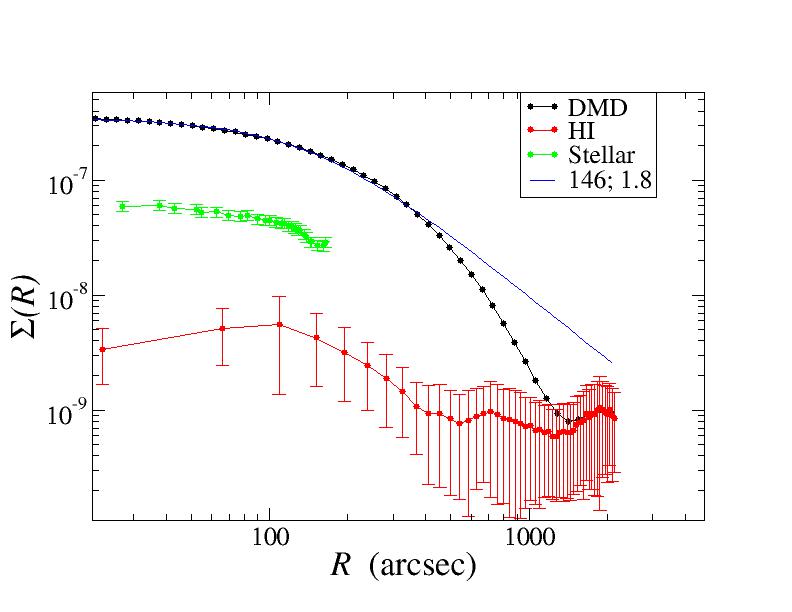}
\caption{}
\end{subfigure}
\quad
\begin{subfigure}[a]{0.3\textwidth}
\centering
\includegraphics[width=5.0cm,angle=0]{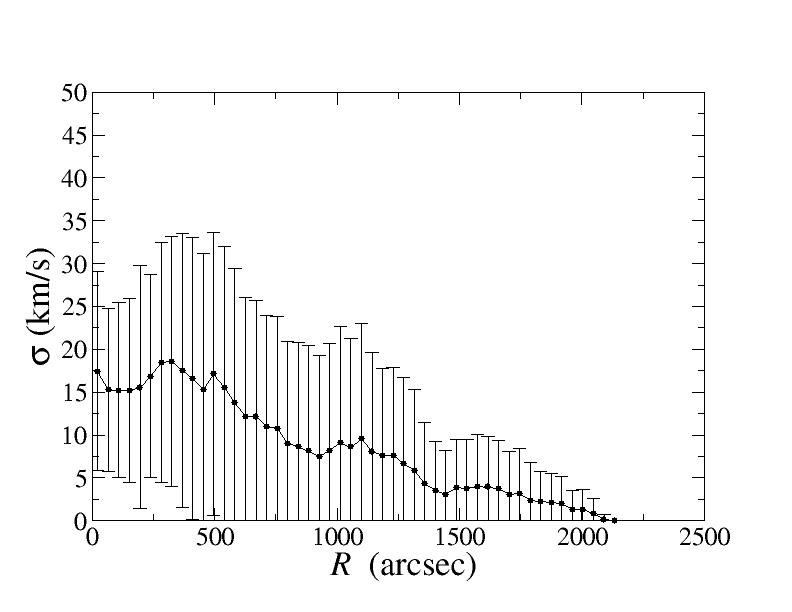}
\caption{}
\end{subfigure}
\quad
\begin{subfigure}[a]{0.3\textwidth}
\centering
\includegraphics[width=5.0cm,angle=0]{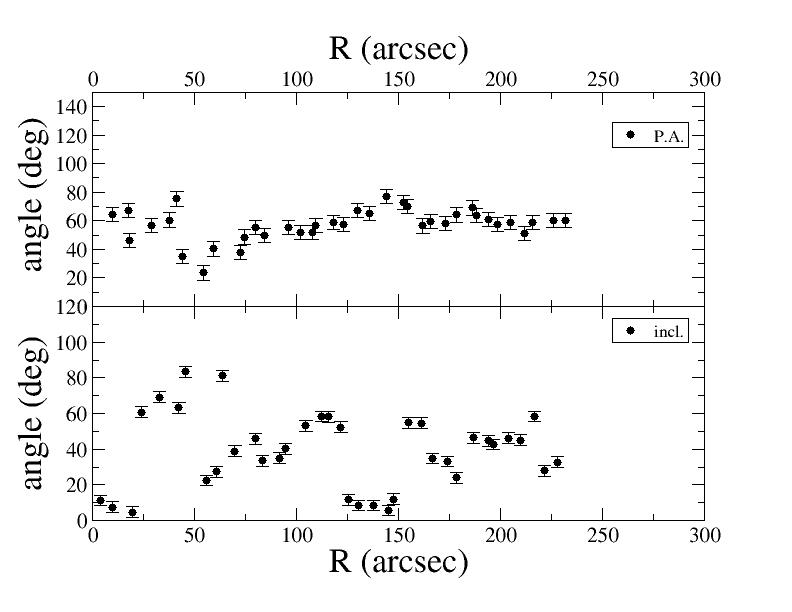}
\caption{}
\end{subfigure}
\quad
\caption{As Fig.\ref{fig:CVnIdwA2} but for IC10.
} 
\label{fig:IC10_2}
\end{figure*}

\begin{figure*}
\quad
\begin{subfigure}[a]{0.3\textwidth}
\centering
\includegraphics[width=5.0cm,angle=0]{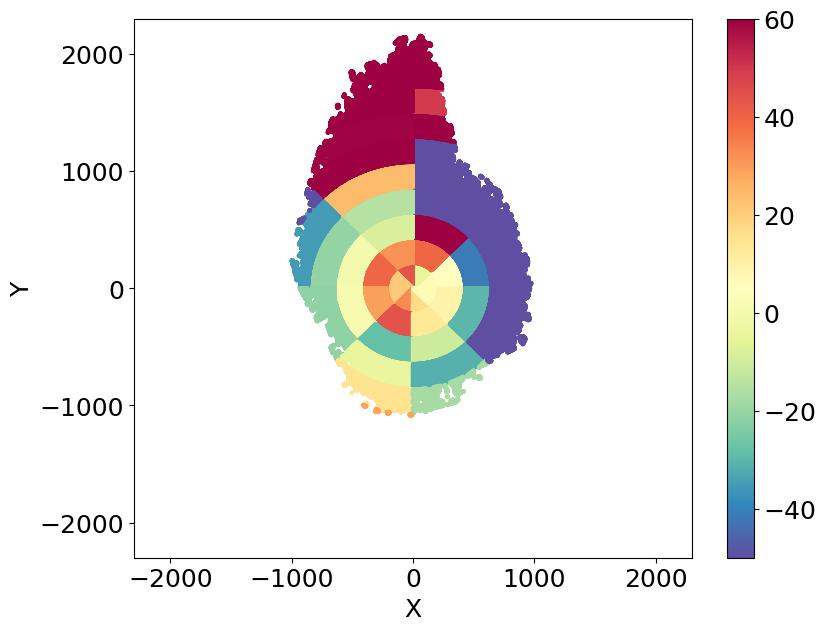}
\caption{}
\end{subfigure}
\quad
\begin{subfigure}[a]{0.3\textwidth}
\centering
\includegraphics[width=5.0cm,angle=0]{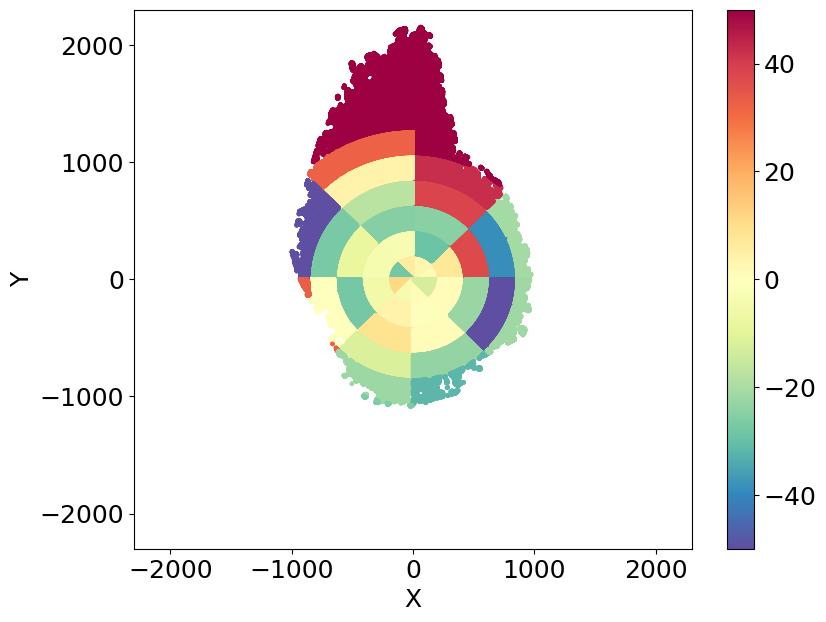}
\caption{}
\end{subfigure}
\quad
\begin{subfigure}[a]{0.3\textwidth}
\centering
\includegraphics[width=5.0cm,angle=0]{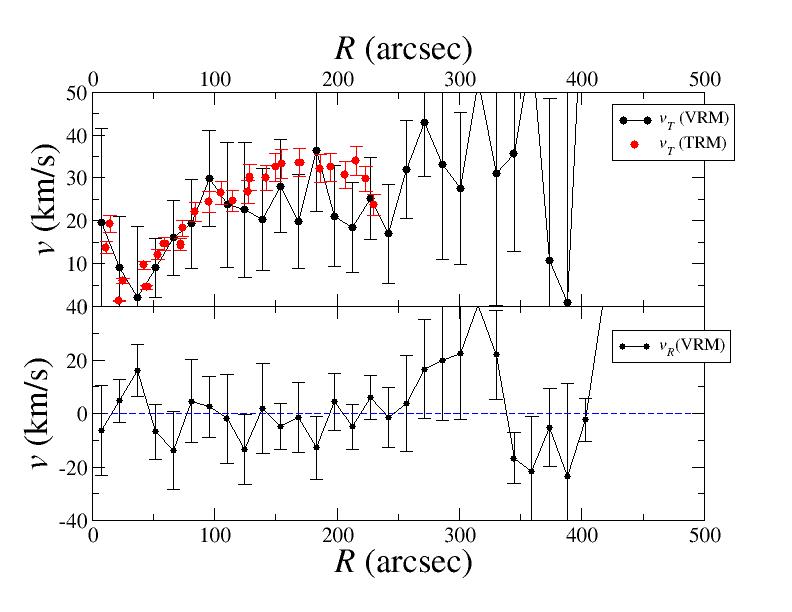}
\caption{}
\end{subfigure}
\quad
\caption{As Fig.\ref{fig:CVnIdwA3} but for IC10.
} 
\label{fig:IC10_3}
\end{figure*}

{  
The LOS velocity field of IC10 is characterized by extreme irregularities and anisotropy, as shown in Figures \ref{fig:IC10_2}-\ref{fig:IC10_3}. This velocity field exhibits highly fluctuating and anisotropic behavior, resulting in significant variations in the orientation angles. Although these angles  do not show any clear radial dependence, they are marked by substantial fluctuations, which are closely linked to the irregular and anisotropic velocity field identified by the VRM.

In the inner part of the disc, the rotation curve is comparatively more regular but remains affected by considerable fluctuations. Moreover, discrepancies arise between the measurements obtained using the TRM and the VRM, primarily due to the pronounced variations in the orientation angles.

Both mass models yield fits with very small $\chi^2$ values and display nearly identical behavior.
} 

\subsection*{IC1613} 

\begin{figure*}
 \quad
 \begin{subfigure}[a]{0.3\textwidth}
 \centering
\includegraphics[width=5.0cm,angle=0]{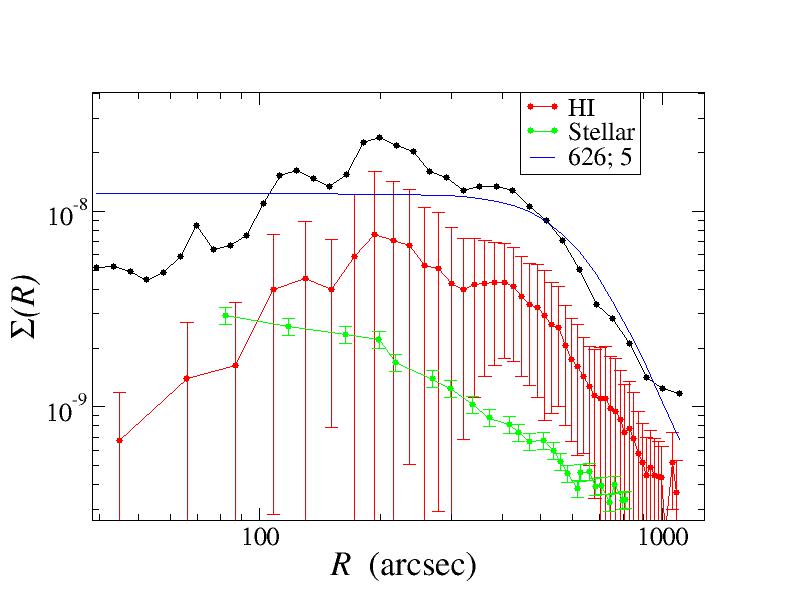}
\caption{}
\end{subfigure}
\quad
\begin{subfigure}[a]{0.3\textwidth}
\centering
\includegraphics[width=5.0cm,angle=0]{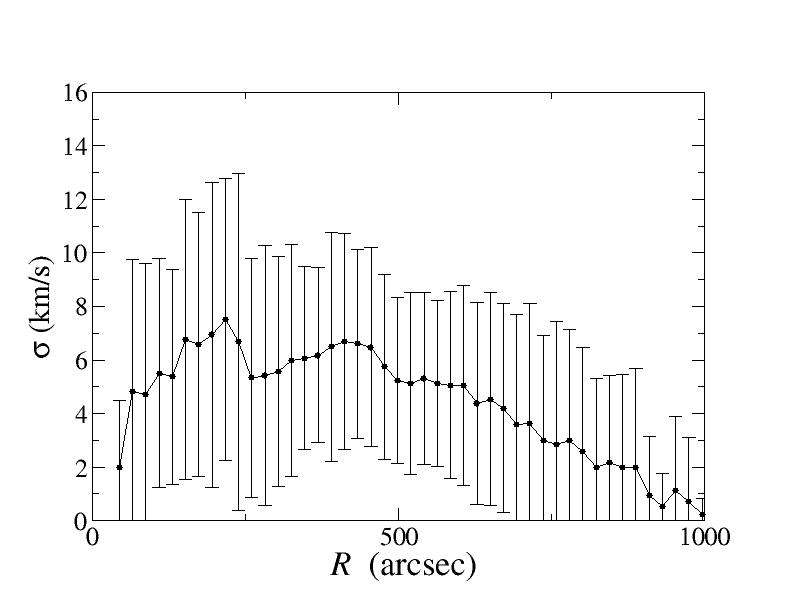}
\caption{}
\end{subfigure}
\quad
\begin{subfigure}[a]{0.3\textwidth}
\centering
\includegraphics[width=5.0cm,angle=0]{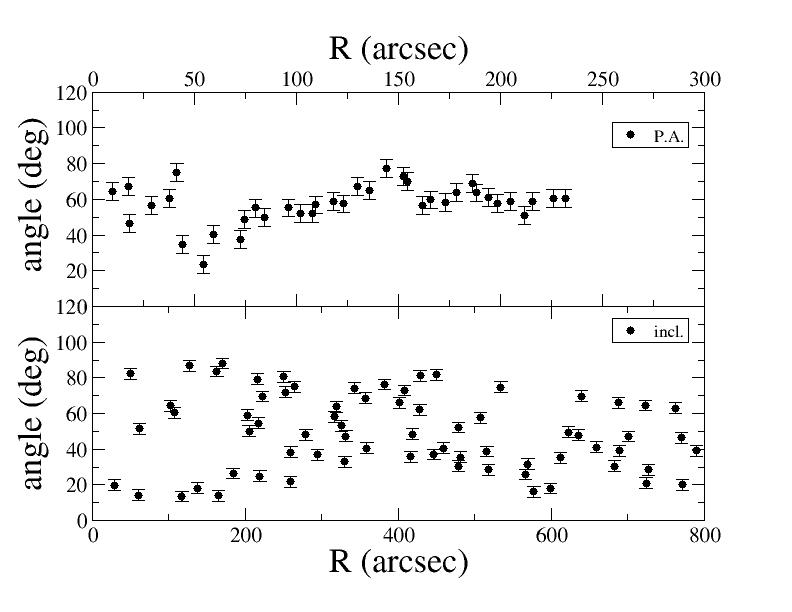}
\caption{}
\end{subfigure}
\quad
\caption{As Fig.\ref{fig:CVnIdwA2} but for IC1613.}
\label{fig:IC1613_2}
\end{figure*}

\begin{figure*}
\quad
\begin{subfigure}[a]{0.3\textwidth}
\centering
\includegraphics[width=5.0cm,angle=0]{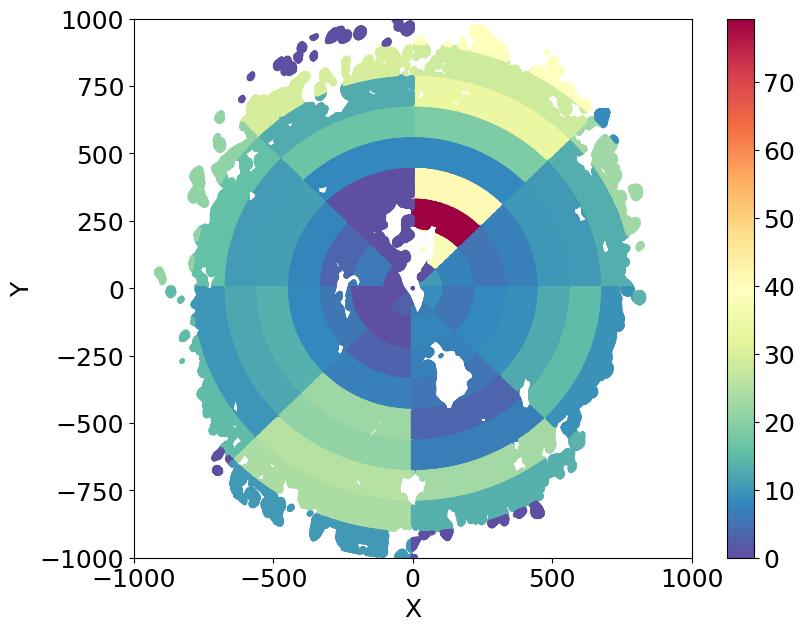}
\caption{}
\end{subfigure}
\quad
\begin{subfigure}[a]{0.3\textwidth}
\centering
\includegraphics[width=5.0cm,angle=0]{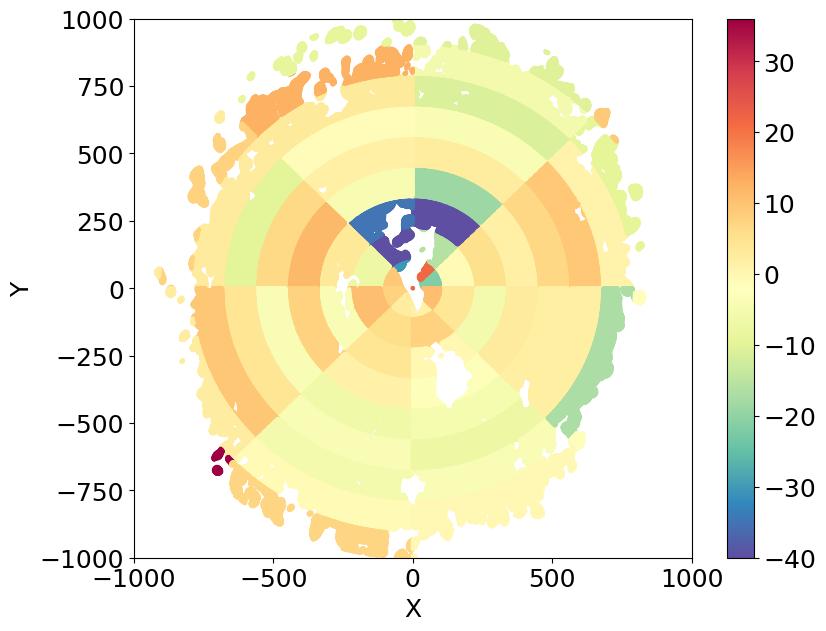}
\caption{}
\end{subfigure}
\quad
\begin{subfigure}[a]{0.3\textwidth}
\centering
\includegraphics[width=5.0cm,angle=0]{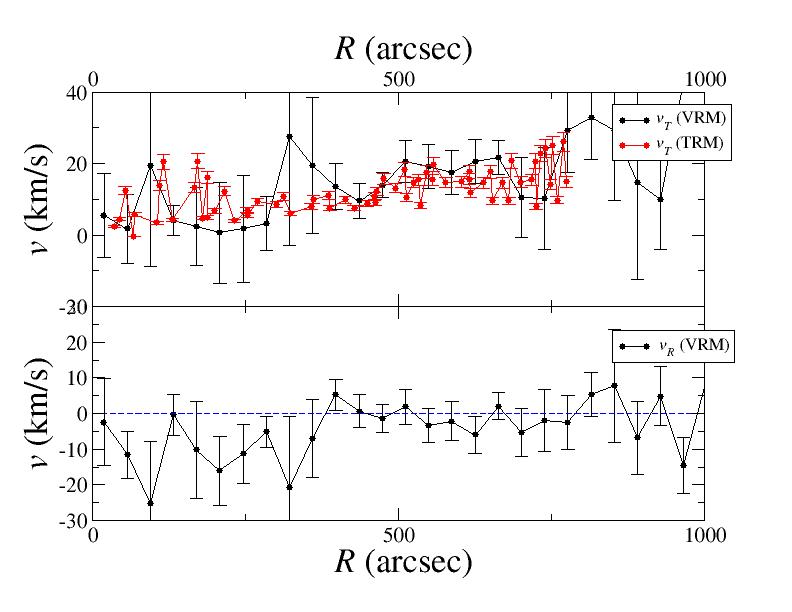}
\caption{}
\end{subfigure}
\quad
\caption{As Fig.\ref{fig:CVnIdwA3} but for IC1613.  
} 
\label{fig:IC1613_3}
\end{figure*}

IC1613 also exhibits a highly irregular LOS velocity map, as shown in Figures \ref{fig:IC1613_2} to \ref{fig:IC1613_3}. 
The orientation angles exhibit a fluctuating behavior. These fluctuations in the orientation angles can be interpreted as significant spatial anisotropies in the velocity field when observed using the VRM.
The presence of these fluctuations indicates variations in the velocity components in different directions within specific regions of the galaxy, contributing to the irregularity observed in the LOS velocity map.
{  
Both mass models exhibit excellent agreement with the data.
}

\subsection*{NGC2366} 
\begin{figure*}
 \quad
 \begin{subfigure}[a]{0.3\textwidth}
 \centering
\includegraphics[width=5.0cm,angle=0]{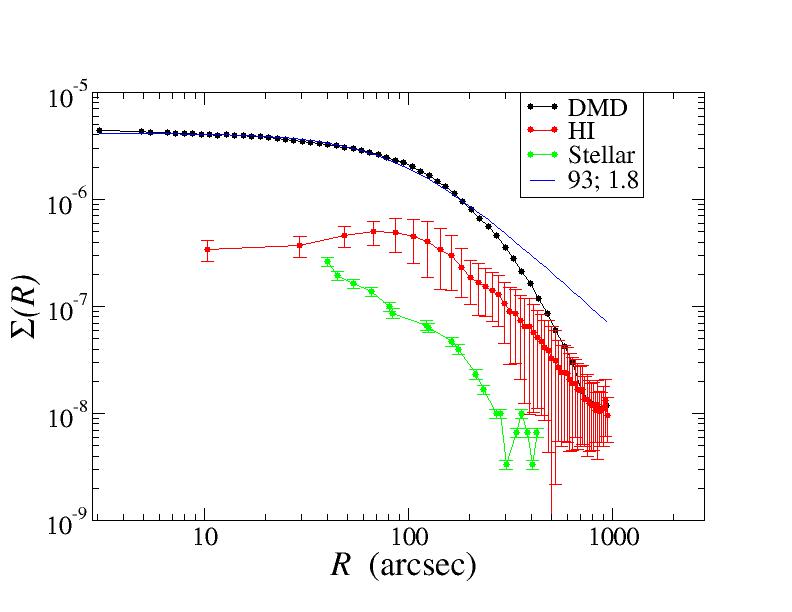}
\caption{}
\end{subfigure}
\quad
\begin{subfigure}[a]{0.3\textwidth}
\centering
\includegraphics[width=5.0cm,angle=0]{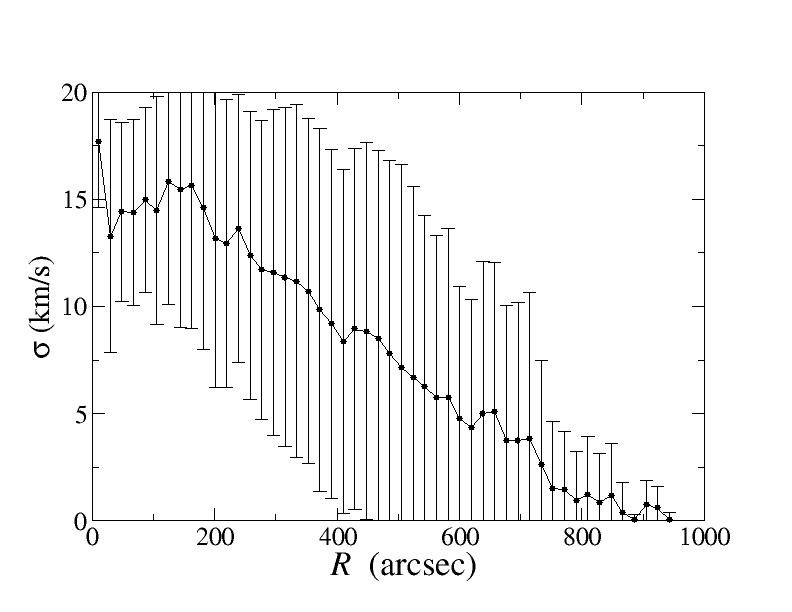}
\caption{}
\end{subfigure}
\quad
\begin{subfigure}[a]{0.3\textwidth}
\centering
\includegraphics[width=5.0cm,angle=0]{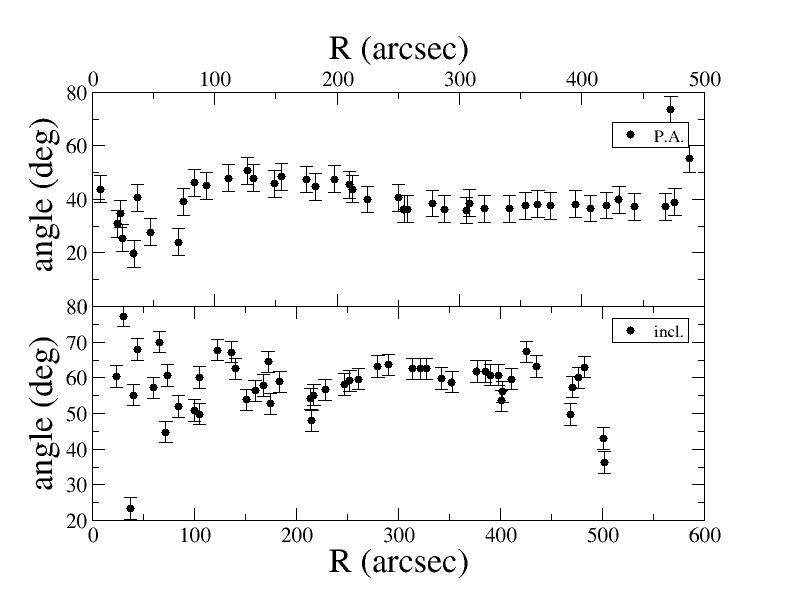}
\caption{}
\end{subfigure}
\quad
\caption{As Fig.\ref{fig:CVnIdwA2} but for NGC2366.
} 
\label{fig:NGC2366_2}
\end{figure*}

\begin{figure*}
\quad
\begin{subfigure}[a]{0.3\textwidth}
\centering
\includegraphics[width=5.0cm,angle=0]{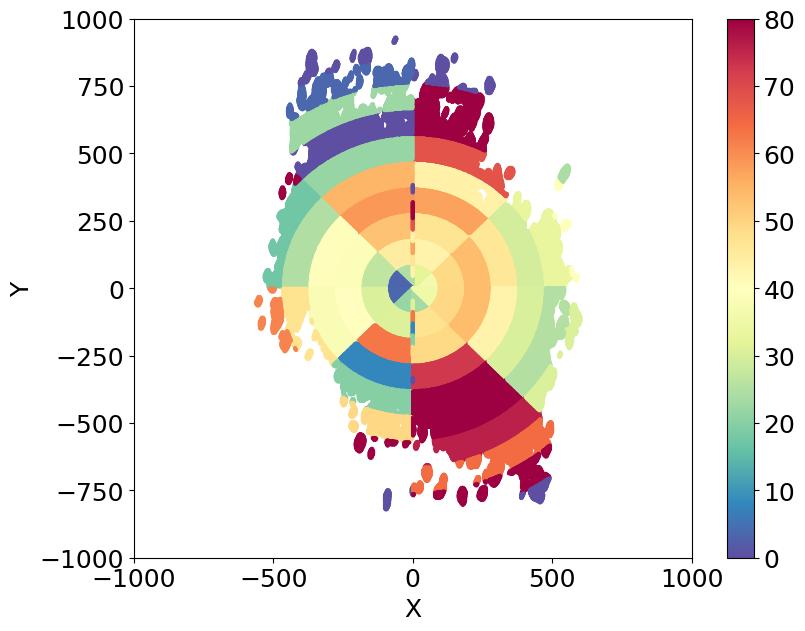}
\caption{}
\end{subfigure}
\quad
\begin{subfigure}[a]{0.3\textwidth}
\centering
\includegraphics[width=5.0cm,angle=0]{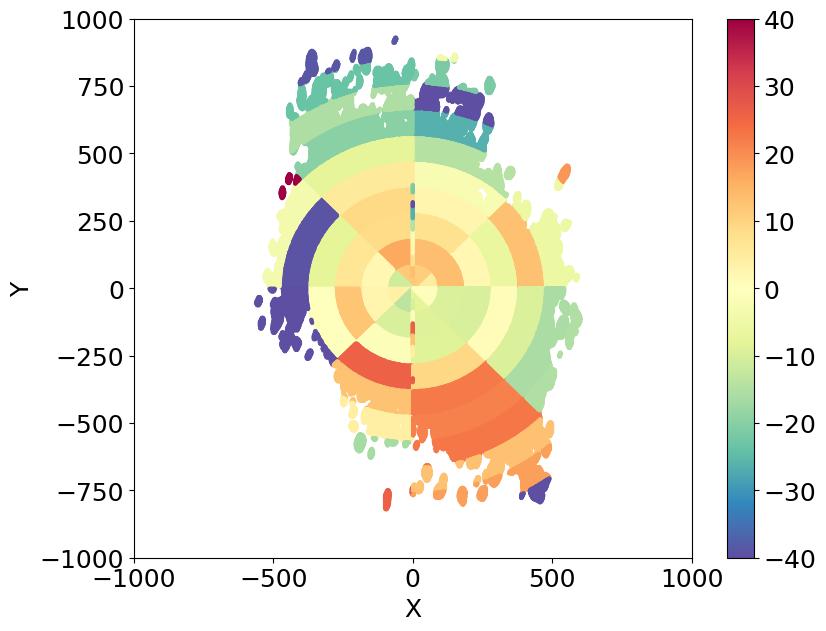}
\caption{}
\end{subfigure}
\quad
\begin{subfigure}[a]{0.3\textwidth}
\centering
\includegraphics[width=5.0cm,angle=0]{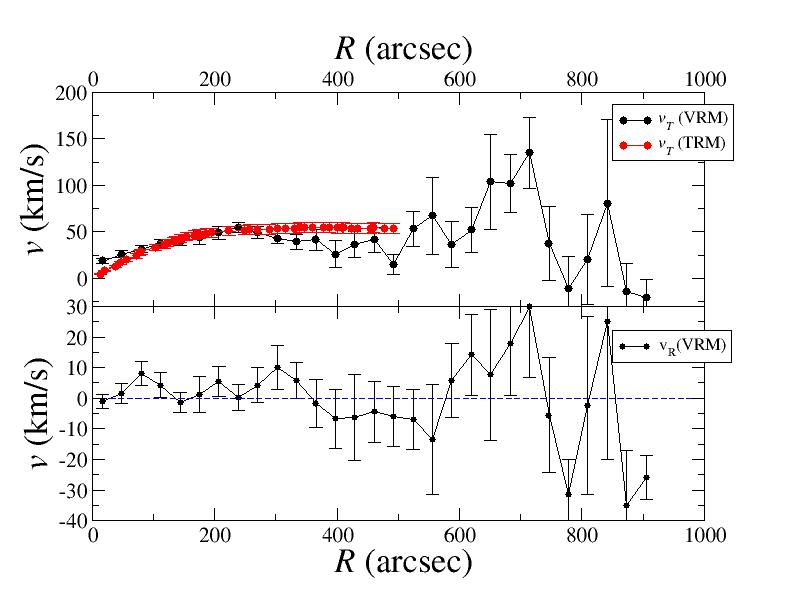}
\caption{}
\end{subfigure}
\quad
\caption{As Fig.\ref{fig:CVnIdwA3} but for NGC2366.  
} 
\label{fig:NGC2366_3}
\end{figure*}

NGC2366 exhibits an irregular pattern in its LOS velocity field, as depicted in Figures \ref{fig:NGC2366_2} to \ref{fig:NGC2366_3}. The orientation angles, in the inner part of the disc, display fluctuations without a clear radial trend. This behavior is consistent with the irregularities observed in the LOS velocity field and with the significant anisotropies detected by the maps reconstructed by the VRM. Due to the lack of a clear radial trend in the orientation angles, it is not surprising that the VRM and the TRM yield similar results for the rotation velocity. {  The two mass models provide similarly good fits to the data, with a reduced $\chi^2$ of 0.92. }

\subsection*{NGC3738} 
\begin{figure*}
 \quad
 \begin{subfigure}[a]{0.3\textwidth}
 \centering
\includegraphics[width=5.0cm,angle=0]{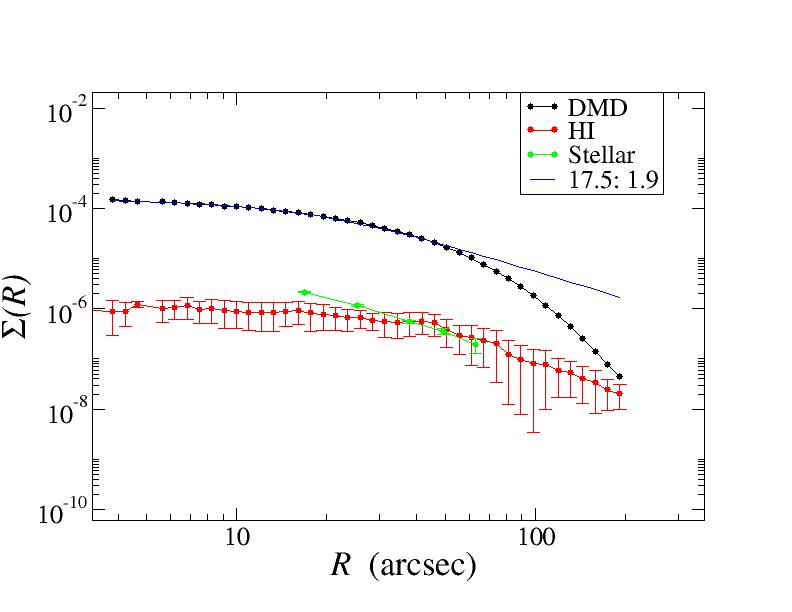}
\caption{}
\end{subfigure}
\quad
\begin{subfigure}[a]{0.3\textwidth}
\centering
\includegraphics[width=5.0cm,angle=0]{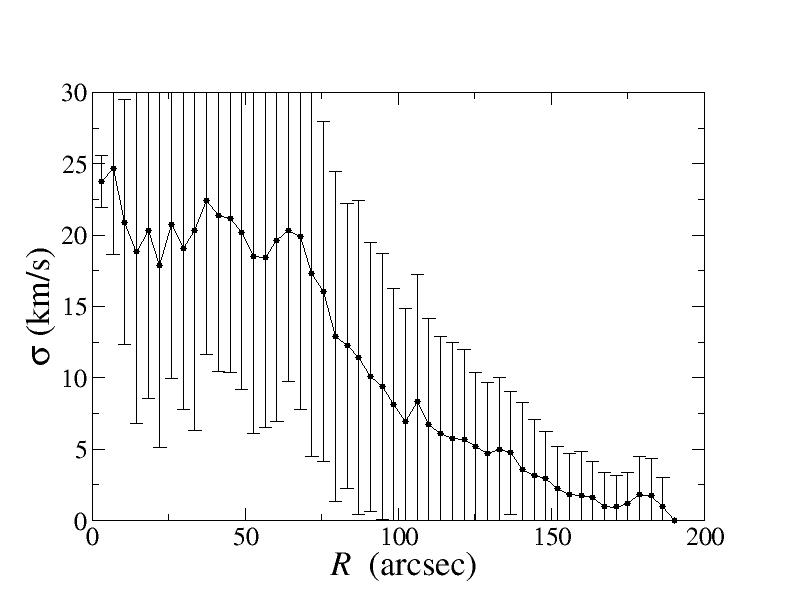}
\caption{}
\end{subfigure}
\quad
\begin{subfigure}[a]{0.3\textwidth}
\centering
\includegraphics[width=5.0cm,angle=0]{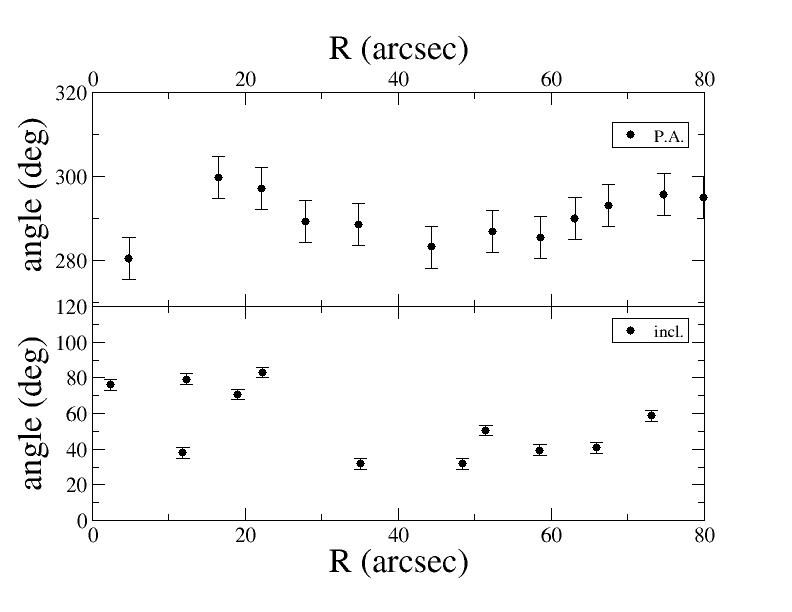}
\caption{}
\end{subfigure}
\quad
\caption{As Fig.\ref{fig:CVnIdwA2} but for  NGC3738.	
} 
\label{fig:NGC3738_2}
\end{figure*}

\begin{figure*}
\quad
\begin{subfigure}[a]{0.3\textwidth}
\centering
\includegraphics[width=5.0cm,angle=0]{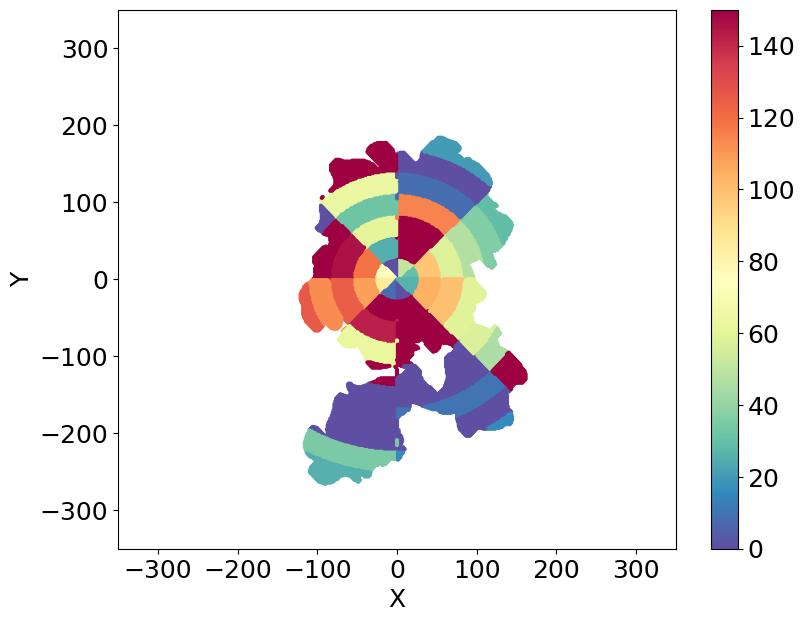}
\caption{}
\end{subfigure}
\quad
\begin{subfigure}[a]{0.3\textwidth}
\centering
\includegraphics[width=5.0cm,angle=0]{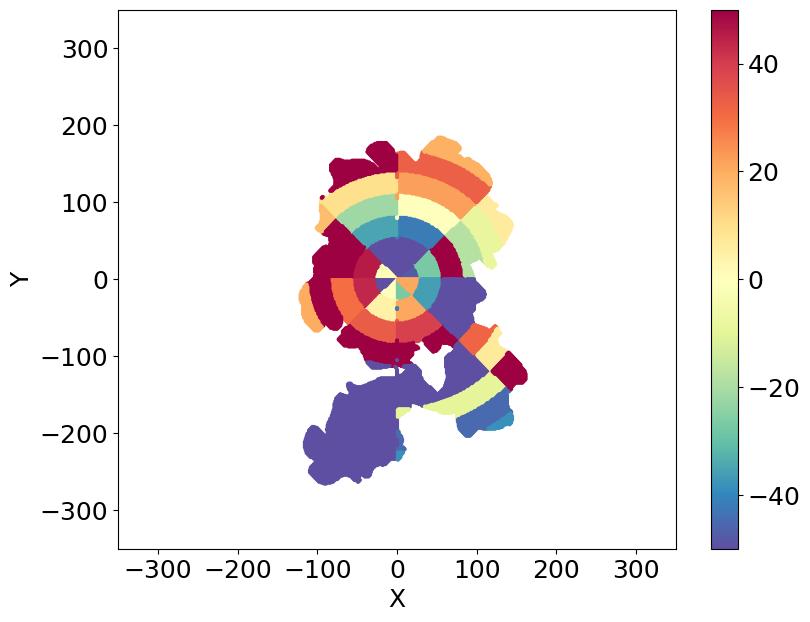}
\caption{}
\end{subfigure}
\quad
\begin{subfigure}[a]{0.3\textwidth}
\centering
\includegraphics[width=5.0cm,angle=0]{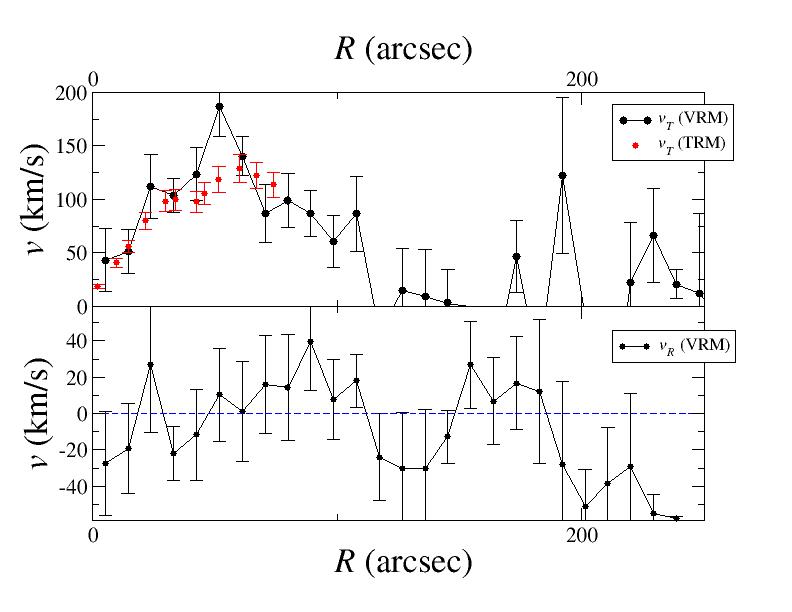}
\caption{}
\end{subfigure}
\quad
\caption{As Fig.\ref{fig:CVnIdwA3} but for  NGC3738.
} 
\label{fig:NGC3738_3}
\end{figure*}

NGC3738 exhibits an irregular shape and a LOS velocity field characterized by irregular and asymmetrical patterns, as evident from Figures \ref{fig:NGC3738_2} -\ref{fig:NGC3738_3}. The orientation angles display noticeable fluctuations, indicating the presence of irregularities in the galaxy's dynamics. In particular, the VRM analysis of the velocity components maps shows the presence of anisotropies throughout the disc, indicating variations in the velocity components in different directions within specific regions of the galaxy. However, in the inner disc, the rotational velocities obtained by both the TRM and VRM are similar. {  
For this reason, the fits with the theoretical model are similar, and their relatively high $\chi^2$ values are attributed to the large fluctuations in the observed velocity profile.}

\subsection*{WLM} 
\begin{figure*}
 \quad
 \begin{subfigure}[a]{0.3\textwidth}
 \centering
\includegraphics[width=5.0cm,angle=0]{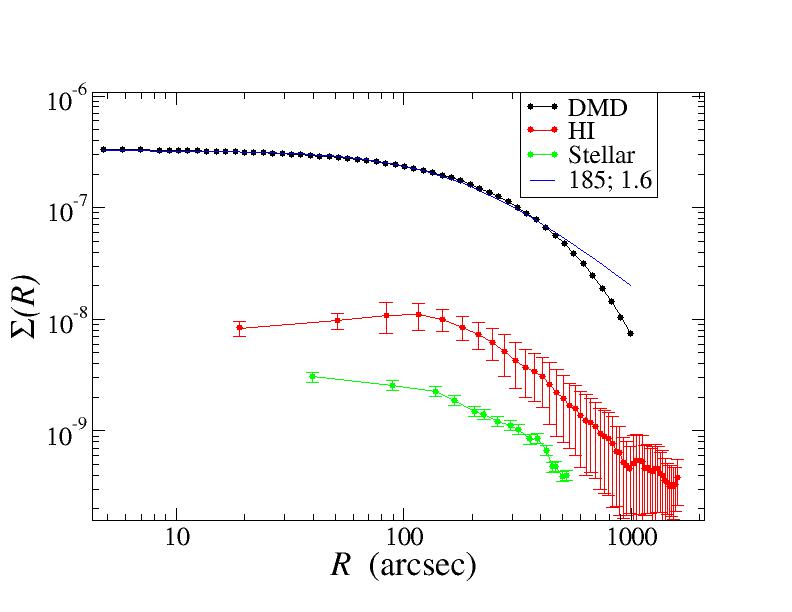}
\caption{}
\end{subfigure}
\quad
\begin{subfigure}[a]{0.3\textwidth}
\centering
\includegraphics[width=5.0cm,angle=0]{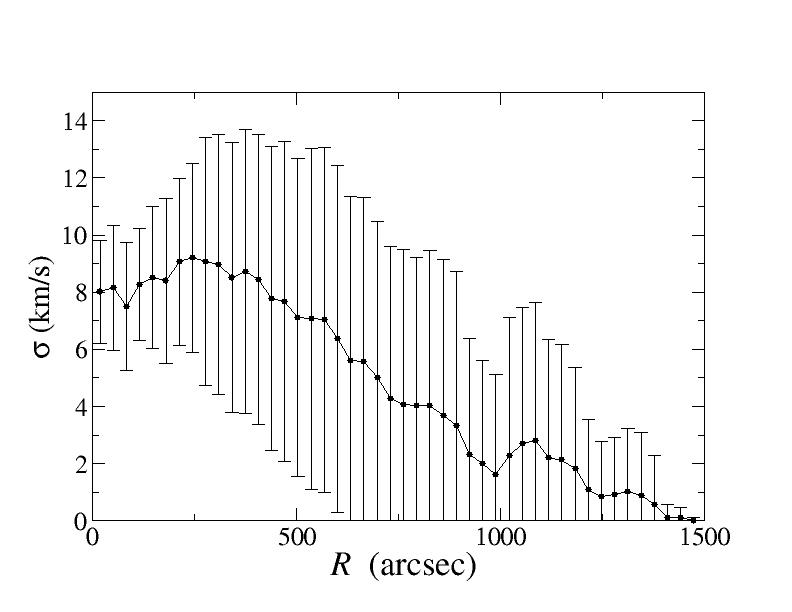}
\caption{}
\end{subfigure}
\quad
\begin{subfigure}[a]{0.3\textwidth}
\centering
\includegraphics[width=5.0cm,angle=0]{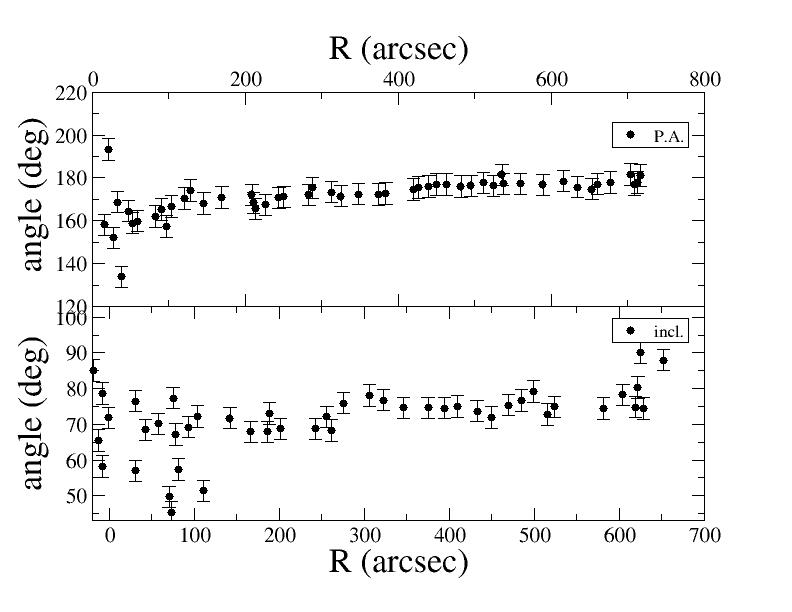}
\caption{}
\end{subfigure}
\quad
\caption{As Fig.\ref{fig:CVnIdwA2} but for  WLM.
} 
\label{fig:WLM_2}
\end{figure*}

\begin{figure*}
\quad
\begin{subfigure}[a]{0.3\textwidth}
\centering
\includegraphics[width=5.0cm,angle=0]{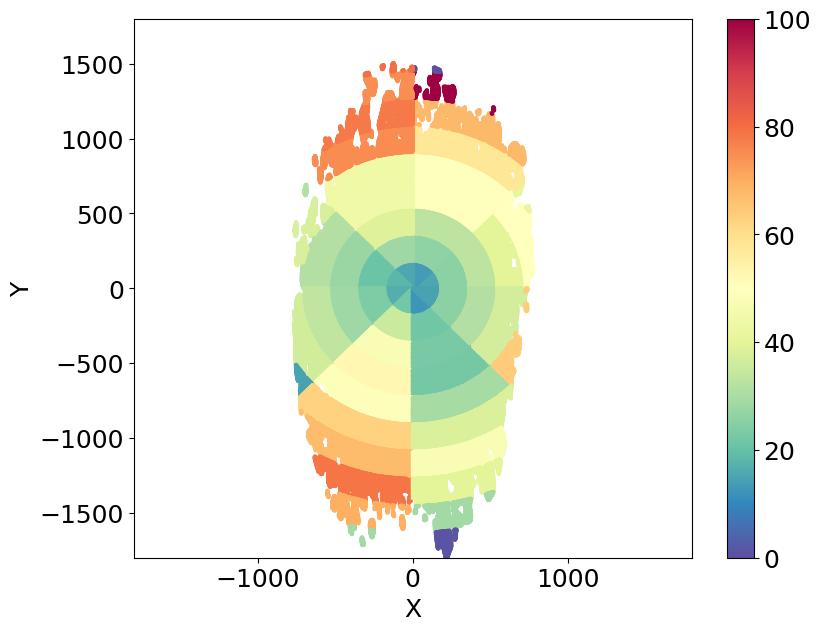}
\caption{}
\end{subfigure}
\quad
\begin{subfigure}[a]{0.3\textwidth}
\centering
\includegraphics[width=5.0cm,angle=0]{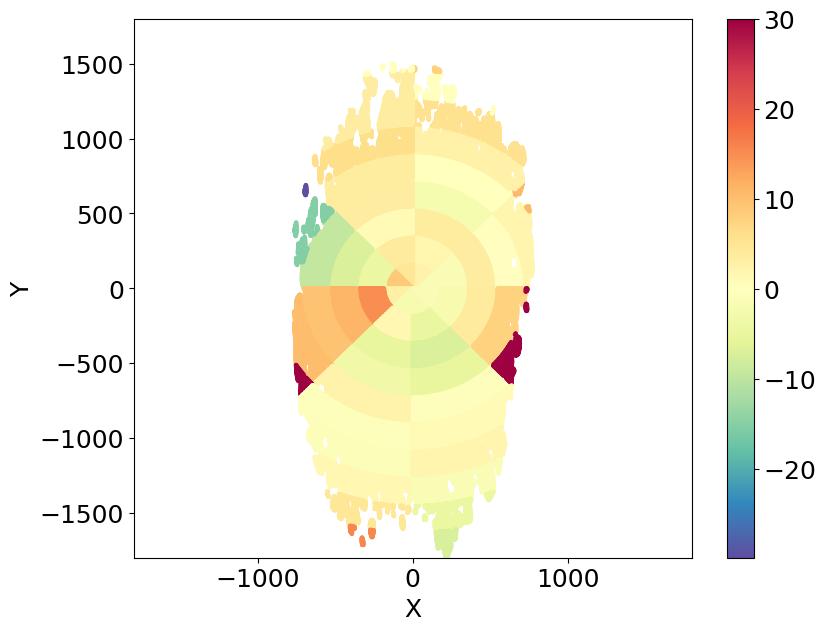}
\caption{}
\end{subfigure}
\quad
\begin{subfigure}[a]{0.3\textwidth}
\centering
\includegraphics[width=5.0cm,angle=0]{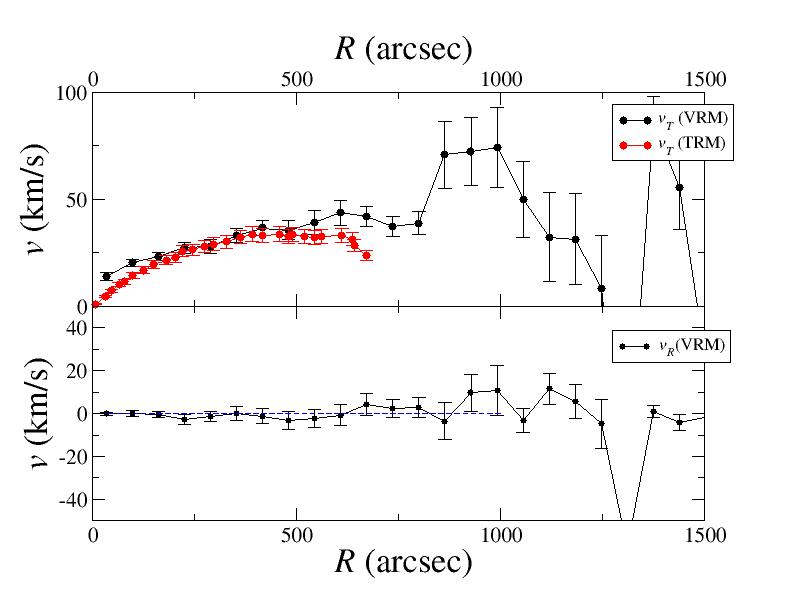}
\caption{}
\end{subfigure}
\quad
\caption{As Fig.\ref{fig:CVnIdwA3} but for WLM.  
} 
\label{fig:WLM_3}
\end{figure*}

In the case of WLM, the LOS velocity map is relatively more regular and exhibits a nearly symmetrical pattern, as depicted in Figures \ref{fig:WLM_2} - \ref{fig:WLM_3}. The orientation angles show a consistent behavior across all radii, indicating a certain level of uniformity in the galaxy's geometry, i.e.  the flat disc assuming is a reasonable one.
{  However, while there is a definitive radial trend, both orientation angles display a fluctuating behavior, which is reflected in the noisy radial profiles of both velocity components when analyzed using the VRM.  }
The velocity components maps obtained through the VRM analysis reveal moderate anisotropies in the inner disc and more significant anisotropies in the outermost regions. 
{  Both mass models provide a reasonable fit to the data, with comparable $\chi^2$ values.  } 

\end{document}